\documentclass[11pt,a4paper]{article}
\pdfoutput=1
\usepackage{jheppub}

\newcounter{comment}

{\refstepcounter{comment}%
\begin{quote}
\ttfamily\small$\blacksquare$ \textbf{\underline{Comment} $\sharp$\thecomment:}}%
{\end{quote}}

{
\begin{quote}
\ttfamily\small$\blacktriangleright$ \textbf{\underline{Reply} $\sharp$\thecomment:}}%
{\end{quote}}

\newcommand{\GeV}{\operatorname{GeV}}

\newcommand{\fm}{\operatorname{fm}}

\newcommand{\xB}{x_{\rm B}}
\newcommand{\Q}{{\cal Q}}
\newcommand{\KM}{{\it KM10 }}
\newcommand{\KMa}{{\it KM10a }}
\newcommand{\our}{{\it AFKM12 }}
\def\im{\Im{\rm m}}
\def\re{\Re{\rm e}}

\title{Deeply Virtual Compton Scattering at a Proposed High-Luminosity Electron-Ion Collider}

\author[a]{E.C.~Aschenauer}
\author[a]{S.~Fazio}
\author[b]{K.~Kumeri{\v c}ki}
\author[a,c]{and D.~M\"uller}

\affiliation[a]{Physics Department, Brookhaven National Lab, Upton, US}
\affiliation[b]{Department of Physics, University of Zagreb, Zagreb, Croatia}
\affiliation[c]{Institut f\"ur Theoretische Physik II, Ruhr-University Bochum,
                Bochum, Germany }

\abstract{
Several observables for the deeply virtual Compton scattering process have been simulated
in the kinematic regime of a proposed Electron-Ion Collider to explore the possible
impact of such measurements for the phenomenological access of generalized parton
distributions. In particular, emphasis is given to the transverse distribution of
sea quarks and gluons and how such measurements can provide information on the angular
momentum sum rule. The exact lepton energy loss dependence for the unpolarized $t$-differential
electroproduction cross section, needed for a Rosenbluth separation, is also reported.
}
\arxivnumber{}

\begin{document}
\maketitle

\toccontinuoustrue

\newpage
\section{Introduction}

During the last decade the collaborations at the Hadron Electron Ring Accelerator (HERA)
and the Thomas Jefferson National Accelerator Facility (JLAB)
spent lately significant effort to measure
exclusive processes such as the electroproduction of a real photon (a process known
as deeply virtual Compton scattering (DVCS)
\cite{Adloff:2001cn,Chekanov:2003ya,Aktas:2005ty,Aaron:2007ab,Chekanov:2008vy,Aaron:2009ac,
Airapetian:2006zr,Airapetian:2008aa,Airapetian:2009aa,Airapetian:2010ab,Airapetian:2011uq,Airapetian:2012pg,
Airapetian:2012mq,
Chen:2006na,Girod:2007aa,Gavalian:2008aa,Munoz_Camacho:2006hx,Mazouz:2007aa}), %
vector mesons (VM)
$\rho^0$ \cite{
Aidetal96a,Breetal98,Adletal99,Breetal99,Adletal02,Aaretal09,Cheetal07,
Airapetian:2000ni,Airapetian:2010dh,
Hadjidakis:2004zm,Morrow:2008ek},
$\phi$ \cite{
Deretal96a,Adletal97,Adletal00a,Cheetal05,Aaretal09,
Borissov:2001fq,
Santoro:2008ai},
$\omega$ \cite{
Breetal00,
Morand:2005ex}, $J/\psi$ \cite{
Breetal98,Adloff:1999zs,Cheetal04,Aktetal05},
$\Upsilon$
\cite{Adletal00,Chekanov:2009zz,Abramowicz:2011fa},
and the pseudoscalar meson $\pi^+$ \cite{Airapetian:2007aa,Horn:2007ug,Blok:2008jy}
in the deeply virtual region in which the virtuality $\Q^2 \gtrsim 1 \GeV^2$ of the
exchanged space-like photon allows to resolve the internal structure of the proton.
The HERA collider experiments
\cite{Adloff:2001cn,Chekanov:2003ya,Aktas:2005ty,Aaron:2007ab,Chekanov:2008vy,Aaron:2009ac}.
found that the exclusive cross sections grow with increasing energy $W$,
where the effective ``pomeron'' intercept is larger and the slope parameter smaller
than for the soft pomeron trajectory \cite{Donnachie:1983hf}, introduced to describe elastic
(anti-)proton-proton high energy scattering.  Moreover, the exponential $t$-slope parameter as
a function of the scale $\Q^2+M^2_{\rm VM}$ was determined by fitting the
$t$-dependence of the cross section
for exclusive vector meson production and DVCS \cite{Abramowicz:2011fa}, which makes loose
contact to the idea of imaging the proton content \cite{Ralston:2001xs}.

Various phenomenological and theoretical descriptions for these
exclusive processes have been proposed and utilized.
In the high-energy region it is popular to understand these processes in
terms of the pomeron picture \cite{DonLan98}, perturbative high-energy
QCD \cite{BalLip78,KurLipFad77}, the color dipole picture
\cite{MuePat94,Muea94}, or in terms of the color glass condensate
approach \cite{McLerran:1994vd,Iancu:2000hn}.
In the deeply virtual regime exclusive processes provide an important
tool in accessing the generalized parton distributions (GPDs)
\cite{Mueller:1998fv,Radyushkin:1996nd,Ji:1996nm}, bridging thereby the
high and medium energy regions.
GPDs also enter in the hand bag model approach
\cite{Radyushkin:1998rt,Diehl:1998kh,Goloskokov:2005sd}, which allows to
describe observables that in the perturbative GPD approach are
considered as non-factorizable contributions that cannot be
perturbatively treated.
In all these approaches the underlying mechanism is a $t$-channel
exchange with different degrees of freedom.

Based on factorization theorems \cite{Collins:1996fb,Collins:1998be}, GPDs offer
a partonic interpretation of these processes,
where unobserved transverse degrees of freedom are integrated out.
Thereby, these universal functions, defined in terms of matrix elements of quark
and gluon operators or, alternatively,
as a non-diagonal overlap of light-cone wave functions
\cite{Diehl:1998kh,Diehl:2000xz,Brodsky:2000xy},
encode the non-perturbative aspects of the nucleon. Because of their fundamental
QCD definition, a whole framework is built up around GPDs, various aspects of the GPD
framework are reviewed in \cite{Diehl:2003ny,Belitsky:2005qn}.
In particular, GPDs provide an access to the transverse spatial distribution of
patrons
\cite{Burkardt:2000za,Die02,KogSop70}, and appear in the gauge invariant
decomposition of the nucleon spin in
terms of quark and gluon degrees of freedom \cite{Ji:1996ek}.

\begin{figure}[ht]
\begin{center}
\includegraphics[width=0.95\textwidth]{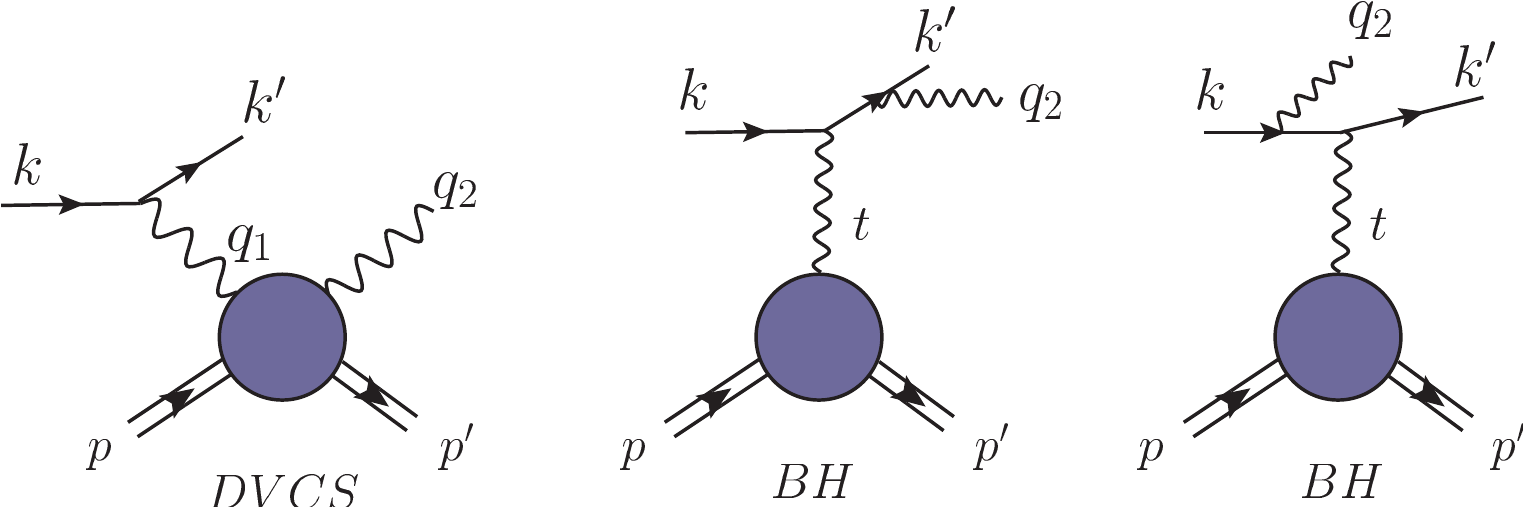}
\end{center}
\caption{
\label{fig:DVCS-BH}
\small Amplitudes contributing to the photon leptoproduction
cross section in leading order approximation of QED: the DVCS amplitude
(left) while the remaining two diagrams (middle and right) represent the
Bethe-Heitler amplitudes, parameterized by hadronic electromagnetic form factors.
}
\end{figure}
Phenomenologically, exclusive electroproduction of a real photon, DVCS,
diagrammatically depicted in Fig.~\ref{fig:DVCS-BH} (left),
is the golden channel to constrain GPDs as it is theoretically clean and
the phase of its amplitude can be measured using the interference with
the Bethe-Heitler (BH) amplitude
(see Fig.~\ref{fig:DVCS-BH} middle/right). Besides that, the measurement of
Compton scattering observables, even at rather low photon virtuality, is
important since it provides insight into the fundamental Compton scattering
process in the virtual regime. Since the virtual Compton process
contains {\em twelve} helicity amplitudes (or equivalently {\em twelve} complex
Compton form factors (CFFs) \cite{Belitsky:2001ns}), their disentanglement is
already an experimental challenge. The measurement of CFFs should
be considered a primary task, as important as the measurement of electromagnetic
nucleon form factors.
In return, the (partial) disentanglement of the various CFFs offers then a
phenomenologically much simpler and cleaner access to GPDs.

Based on present phenomenological GPD knowledge, Monte Carlo simulations, and
GPD fitting routines, we explore in our studies here
both the DVCS process and the access to the spatial transverse distribution of
quarks and gluons at a proposed Electron-Ion Collider (EIC).
The much more general physics case of this suggested high-luminosity collider
with a dedicated detector for exclusive channels in the medium to high energy
regime of lepton-nucleon and lepton-nuclei scattering is described in \cite{Accardi:2012hwp}.

The rest of this article is organized as follows:
in Sect.~\ref{sec:presendDVCS} we introduce the theory elements, needed for the
access of GPDs from DVCS observables, including also, for the unpolarized case,
the exact dependence of a (reduced) $t$-differential photon electroproduction
cross section on the electron energy loss variable $y$.
Furthermore, we give a short overview of existing DVCS measurements. In
Sect.~\ref{sec:simulation} we describe the planned EIC at its different
stages and the Monte Carlo simulation technique used in
the generation of EIC DVCS pseudo-data. In Sect.~\ref{sec-pred} we shortly
introduce three GPD models which are then utilized to provide
predictions for the $t$-differential DVCS cross section, single spin and lepton
charge DVCS asymmetries at different EIC kinematics.
In Sect.~\ref{sec:interpretation} we discuss the access of GPD $H$ and $E$ at the
final stage of EIC by using the DVCS cross section and single transverse proton spin
asymmetry. Furthermore, we quantify the implications of such measurements for the
imaging of the proton and comment on the qualitative aspects of
such measurements for the spin sum rule.
Finally,  we summarize and conclude in Sect.~\ref{sec:summary}.

\section{Deeply virtual Compton scattering}
\label{sec:presendDVCS}

The differential photon electroproduction cross section is five-fold and
consists of the sum of the BH amplitude squared, DVCS amplitude squared,
and the interference (INT) terms,  where the latter is charge odd:
\begin{equation}
\label{X-electroproduction}
\frac{d\sigma^{ep\to ep\gamma}}{d\xB dt d\Q^2  d\phi d\varphi} =
\frac{d\sigma^{ep\to ep\gamma,\mbox{\tiny BH}}(F_1,F_2)}{d\xB  dt d\Q^2  d\phi d\varphi} \pm
\frac{d\sigma^{ep\to ep\gamma,\mbox{\tiny INT}}(F_1,F_2,{\cal F})}{d\xB dt d\Q^2 d\phi d\varphi} +
\frac{d\sigma^{ep\to ep\gamma,\mbox{\tiny (D)VCS}}({\cal F},{\cal F}^\ast)}{d\xB dt  d\Q^2 d\phi d\varphi}\,.
\end{equation}
Here the $+ (-)$ sign is valid for electron (positron) beam,
$\xB$ is the common Bjorken scaling variable,
$\phi$ is the azimuthal angle between lepton and hadron scattering planes, and
$\varphi=\Phi-\phi$, where $\Phi (\equiv\phi_S)$
is the angle between the lepton scattering plane and a possible transverse spin
component
of the incoming proton at rest. We adopt in the following the frame conventions
of \cite{Belitsky:2001ns} (virtual photon momentum is counter-along the
$z$-direction and $x$-component of the incoming electron momentum is positive).
To the leading order (LO) in the electromagnetic
fine structure constant $\alpha_{\rm em} = \frac{e^2}{4\pi} \approx
\frac{1}{137},$ and neglecting the electron mass, the three terms on the
r.h.s.~of (\ref{X-electroproduction}) are exactly known in terms of
the electromagnetic Pauli form factor $F_1(t)$ and the Dirac form factor $F_2(t)$,
parameterizing the BH amplitude, and a set of twelve photon helicity dependent
CFFs ${\cal F}_{a b}(\xB,t,\Q^2)$, parameterizing the DVCS amplitude, see
Fig.~\ref{fig:DVCS-BH}. These CFFs are labeled by the helicities of the incoming
$a \in \{+,0,-\}$ and outgoing photon $b\in \{+,-\}$
and they are called
\begin{equation}
\label{cffF_{ab}}
{\cal F}_{a b} \in \{{\cal H}_{ab},{\cal E}_{ab}, \widetilde{\cal H}_{ab},
\widetilde{\cal E}_{ab} \} \quad\mbox{with}\quad
{\cal F}_{0 -} = {\cal F}_{0 +}\,, \quad  {\cal F}_{+ -}=  {\cal F}_{- +}\,,
\end{equation}
more details can be found in \cite{Belitsky:2010jw,Belitsky:2012ch}.
Analogously to the Dirac and Pauli form factor $F_1$ and $F_2$ (axial and pseudo-scalar
form factors $F_A$ and $F_P$), the CFFs ${\cal H} (\widetilde{\cal H})$  and
${\cal E} (\widetilde{\cal E}) $ ) are associated with conserved proton helicity
amplitudes and helicity flipped ones, respectively.

The three separate terms of the differential electroproduction cross section
(\ref{X-electroproduction}) can be expanded w.r.t.~harmonics of the azimuthal angle $\phi$,
\begin{equation}
\label{X-BH}
\frac{d\sigma^{ep\to ep\gamma,\mbox{\tiny BH}}(F_1,F_2)}{d\xB dt d\Q^2 d\phi d\varphi}
= \frac{\alpha_{\rm em}^3 } {16 \, \pi^2 \, {\Q}^4}\,
\frac{\xB^{-1}(1 + \epsilon^2)^{-5/2} }{t\, {\cal P}_1 (\phi,y) {\cal P}_2 (\phi,y)}
\left\{ \sum_{n = 0}^2 c^{\mbox{\tiny BH}}_n \cos{(n\phi)} + s^{\mbox{\tiny BH}}_1 \sin{(\phi)}
\!\right\}, \\
\end{equation}
\begin{equation}
\label{X-INT}
\frac{d\sigma^{ep\to ep\gamma,\mbox{\tiny INT}}(F_1,F_2)}{d\xB  dt d\Q^2  d\phi d\varphi}
 = \frac{\alpha_{\rm em}^3} {16 \, \pi^2 \,  {\Q}^4}\, \frac{y^{-1}(1 +
\epsilon^2)^{-1/2}} {t {\cal P}_1 (\phi,y) {\cal P}_2 (\phi,y)}
\left\{\! c^{\mbox{\tiny INT}}_0 + \sum_{n = 1}^3
\left[ c_{n}^{\mbox{\tiny INT}} \cos(n \phi) + s_{n}^{\mbox{\tiny INT}} \sin(n \phi)\right] \!\right\}, \\
\end{equation}
\begin{equation}
\label{X-VCS}
\frac{d\sigma^{ep\to ep\gamma,\mbox{\tiny VCS}}({\cal F},{\cal F}^\ast)}{d\xB  dt d\Q^2  d\phi d\varphi}
 =  \frac{\alpha_{\rm em}^3  } {16 \, \pi^2 \,  {\Q}^4}\,
\frac{\xB(1 + \epsilon^2)^{-1/2}}{{\cal Q}^2}\left\{\! c^{\mbox{\tiny VCS}}_0 + \sum_{n=1}^2
\left[ c^{\mbox{\tiny VCS}}_n \cos(n\phi) + s^{\mbox{\tiny VCS}}_n \sin(n \phi) \right]\! \right\}.
\end{equation}
Here $1/({\cal P}_1 (\phi,y) {\cal P}_2 (\phi,y))$ are (rescaled) BH propagators,
defined in (32) of \cite{Belitsky:2001ns}, the energy loss
\begin{equation}
\label{y(xB,Q2,s)}
y=\frac{1}{\xB} \frac{\Q^2}{s-M_p^2}
\end{equation}
of the electron depends for fixed $\xB$ and $\Q^2$ on the center-of-mass
(c.o.m.) energy squared $s$, and,
finally, we used the shorthand $\epsilon \equiv 2\xB M_p/\Q$. Moreover, all of
the Fourier coefficients $c_n^{\cdots}$ and $s_n^{\cdots}$ depend on
the polarization vectors of the protons. The explicit expressions for an
incoming polarized nucleon has been presented in \cite{Belitsky:2012ch}, where
for transverse polarization the
coefficients can be further decomposed in $\cos(\varphi)$ and $\sin(\varphi)$
harmonics. Note that if one reduces the five-fold cross section
(\ref{X-electroproduction}) to a four-fold one by integrating over $\varphi$,
the $\varphi$-harmonics drop out
and the remaining unpolarized and longitudinally polarized parts of the
expressions (\ref{X-BH}--\ref{X-VCS}) are multiplied by a factor $2\pi$.
The knowledge of the coefficients in the BH term (\ref{X-BH}) is
limited only by the knowledge of the proton form factors $F_1(t)$ and $F_2(t)$.
The coefficients of the interference (\ref{X-INT}) and (D)VCS term (\ref{X-VCS})
are linear and bi-linear in the CFFs, respectively.
We emphasize that electromagnetic corrections will enter in all three terms. So
far such $\alpha_{\rm em}/\pi$-proportional
corrections are only partially taken into account in radiative correction
procedures.

Adopting the discussion of \cite{Belitsky:2001ns}, we can state that an
over-complete
set of observables exist and that at least in principle their experimental
measurements would allow to extract the real and imaginary parts of all twelve
CFFs (\ref{cffF_{ab}}). Loosely speaking, in the deeply virtual regime the first
harmonics
in the interference term are dominant, i.e., proportional to $1/\Q^3$, and are
governed by twist-two associated CFFs (or GPDs), while the constant and second
harmonics are kinematical suppressed by $1/\Q$ and arise in leading order of
perturbative QCD from
both twist-two and twist-three associated CFFs (or GPDs). The third harmonics
are counted as leading twist contributions, however, they arise in
next-to-leading order (NLO) of perturbative QCD from gluon transversity GPDs. A
rather analogous counting scheme holds
for the zeroth, first, and second harmonics of the DVCS term, where, 
compared to the interference term, an additional kinematical factor $1/\Q$
appears. Because of this mismatch in twist and power counting, some care is
needed.

In the rest of this Sect.~\ref{sec:presendDVCS} we consider: in
Sect.~\ref{sec:Rosenbluth}, the $y$-dependence of the $\phi$-integrated
electroproduction cross section (\ref{X-electroproduction})
for an unpolarized proton and in Sect.~\ref{sec:GPD22CFFs} we point out that the
relation of helicity CFFs to GPDs can be
systematically improved. In Sect.~\ref{sec:phenomenology} we give a short
overview of existing DVCS measurements and make a loose
contact to CFF/GPD phenomenology.

\subsection{Rosenbluth separation of electroproduction cross section}
\label{sec:Rosenbluth}

It would be very desirable to decompose the photon electroproduction cross
section (\ref{X-electroproduction}) into its different parts
(\ref{X-BH}--\ref{X-VCS}).
In an experimental setup in which both electrons and positrons are available,
the charge-odd interference term (\ref{X-INT}) and
the charge-even part, given as sum of BH and DVCS cross sections
(\ref{X-BH},\ref{X-VCS}), can be obviously separated from each other by forming
the difference and sum of electron and positron cross sections.
Having only an electron beam at hand,
it remains so far unclear to what extent a variation of c.o.m.~energy (or
electron/proton beam energy)
allows for a Rosenbluth separation, which is expected to be
much more intricate than in the case of elastic form factors or deeply inelastic
scattering (DIS)
structure functions. We recall that in these cases two form factor combinations
(or structure functions) enter the unpolarized cross sections; however, both of
them arise from transversely or longitudinally polarized photon exchanges and
are thus accompanied with a different $y(\xB,\Q^2,s)$ dependence, which
varies for fixed $\xB$ and $\Q^2$ with the c.o.m.~energy $\sqrt{s}$, see
(\ref{y(xB,Q2,s)}), i.e., with the beam energy (or energies).

Having the exact analytic expressions of \cite{Belitsky:2012ch} in mind, it
looks hopeless to employ a Rosenbluth separation directly to the five-fold (or
four-fold) cross section (\ref{X-electroproduction}). Thus, it is more
appropriate
to project first on the azimuthal angle harmonics, where, however, the $\phi$-
and $y$-dependencies of the BH propagators should be treated in such a way
that the final result is most appropriate for the analyzes of experimental data.
Including these propagators in the
integral, as done in \cite{Belitsky:2001ns}, provides a truncated
Fourier series and allows for a rather simple power counting scheme; however,
these Fourier coefficients will not have a simple $y$-dependence. Alternatively,
one may stick to the standard Fourier coefficients, e.g., calculated from
\begin{equation}
\int_{-\pi}^{\pi}\! d\phi\, \cos{(n\phi)}\,
\frac{d\sigma^{ep\to ep\gamma}}{d\xB dt d\Q^2  d\phi} \quad\mbox{for}\quad n = 0,1,2,3,\cdots,
\end{equation}
where the DVCS cross section only contributes to the first three lowest coefficients.
In the following we will first consider only the lowest harmonic, i.e., $n=0$,
which has a surprisingly simple and obvious $y$-dependence.

Let us first introduce a formally defined $t$-differential ``photoproduction''
cross section.  It is obtained by integrating the four-fold electroproduction
cross section over the azimuthal angle $\phi$, multiplying it with an infinitesimal
electron phase space element, and dividing it by a flux factor,
\begin{eqnarray}
\label{dX-reduced}
\frac{d\sigma^{\mbox{\tiny TOT}}(\xB,t,\Q^2|y)}{dt} & \equiv & \frac{1}{\Gamma(\xB,\Q^2|y)}
\int_{-\pi}^\pi\! d\phi\,
\frac{d\sigma^{ep\to ep\gamma}(\xB,t,\Q^2|y)}{dt d\phi d\xB d\Q^2} \times  d\xB d\Q^2 \\
\nonumber\\
& = & \frac{d\sigma^{\mbox{\tiny\rm BH}}(\xB,t,\Q^2|y)}{dt} \pm \frac{d\sigma^{\mbox{\tiny\rm INT}}
(\xB,t,\Q^2|y)}{dt} + \frac{d\sigma^{\mbox{\tiny\rm DVCS}}(\xB,t,\Q^2|y)}{dt}\,,
\nonumber
\end{eqnarray}
where as before the positive (negative) sign of the
interference term refers to an electron (positron) beam. For the virtual
photon flux we adopt the Hand convention \cite{Hand:1963bb} by taking
\begin{eqnarray}
\label{Gamma_V}
\Gamma(\xB,\Q^2|y) = \frac{\alpha_{\rm em}}{2\pi} \frac{y^2}{1-\varepsilon(y)} \frac{1-\xB}{\xB \Q^2}
\quad\mbox{with}\quad  \varepsilon(y) =
\frac{1-y-\frac{\epsilon^2  y^2}{4}}{1-y+ \frac{y^2}{2} + \frac{\epsilon^2  y^2}{4}}
\end{eqnarray}
where $\varepsilon(y)$ is the  ratio of longitudinal and transverse photon flux.

The $y$-dependence of the three terms in (\ref{dX-reduced}) and the explicit expressions for
the Fourier coefficients can be evaluated from (\ref{X-BH}--\ref{X-VCS}) . Thereby, the DVCS
cross section is the most simplest
one and given by the constant harmonic in (\ref{X-VCS}), which is further
specified in (36) of \cite{Belitsky:2012ch}. The BH cross section
can also be analytically calculated, where due to the $\phi$-dependence of the
BH propagators also higher $\phi$ harmonics that arise from the interference of
photon helicity flip amplitudes, specified in (35-37) of \cite{Belitsky:2001ns},
enter. Thereby, the integration over the azimuthal angle $\phi$ generates a
characteristic $y$-dependent function that stems from the product $1/({\cal
P}_1(\phi,y){\cal P}_2(\phi,y))$ of BH propagators.
Consequently, this function inherits the $u$-channel pole of one BH propagator
at
$$y=y_{\rm col} \quad\mbox{with}\quad y_{\rm col} = \frac{\Q^2 + t}{\Q^2 + \xB t}\,,$$
where the real photon and incoming electron momenta are collinear. In the
following we present results for the region $y < y_{\rm col}$, in which this
characteristic function reads:
\begin{equation}
\label{BH-integrated}
\frac{1}{\left[1+\frac{t}{\Q^2}-y \frac{t}{\Q^2}(1-\xB) \right] \left[1+\frac{t}{\Q^2}-y\left(1+\frac{\xB t}{\Q^2}\right)\right]}
= \frac{\left(1+\frac{\xB t}{\Q^2}\right)^{-1}}{\left[1+\frac{t}{\Q^2}-y \frac{t}{\Q^2}(1-\xB) \right] \left[y_{\rm col}-y\right]} .
\end{equation}

The interference term is the most intricate one, since various CFF combinations,
which have different $y$-dependencies, enter in the harmonics and
due to the BH propagators all of them will contribute to the $\phi$-integrated
interference term. Utilizing the exact results, given in (66,67,69) and appendix
B.1 of \cite{Belitsky:2012ch}, it can be shown that due to the $\phi$
integration the transverse CFFs ${\cal F}_{-+}$ disappear. Moreover, the
$y$-dependent factor (\ref{BH-integrated}), arising from the BH propagators,
cancel exactly in all remaining expressions and we also find a unique
$y$-dependence for the net result. We also emphasize that the dominant first
harmonic gets suppressed by $1/\Q$ and cancels a contribution in the constant
term, yielding a result that is proportional to $\xB^2$. Finally, we add that
the CFFs $\widetilde{\cal E}_{++}$ and $\widetilde{\cal E}_{0+}$ are absent in
the unpolarized interference term.

Let us skip here further details and quote the new results for the moderate/small-$\xB$ region:
\begin{equation}
\label{XBH-red}
\frac{d\sigma^{\rm BH}}{dt} = \frac{4 \pi \alpha_{\rm em}^2 y^2 }{-t  \Q^2 (1+\epsilon^2)}\,
\frac{ \left[1-\frac{2\xB t}{\Q^2-t}\, \varepsilon(y) \right]
\frac{(-1)\widetilde{K}^2 (\Q^2-t)}{t(\Q^2+t)(1-\xB)}\left[F_1^2(t)-
\frac{t}{4 M_p^2}F_2^2(t) \right]+ {\cal O}(\xB^2) }
{\left[1+\frac{t}{\Q^2}-y \frac{t}{\Q^2}(1-\xB) \right]
\left[1+\frac{t}{\Q^2}-y\left(1+\frac{\xB t}{\Q^2}\right)\right] } ,
\\
\end{equation}
\begin{equation}
\label{XINT-red}
\frac{d\sigma^{\mbox{\tiny\rm INT}}}{dt} =
\frac{4 \pi \alpha_{\rm em}^2\, y(2-y)}{\Q^4(1+\epsilon^2)(2-2y+y^2+ \frac{\epsilon^2 y^2}{2})}\;
\frac{\xB^2}{1-\xB}  \Re{\rm e}\, {\cal C}({\cal F}_{++}|{\cal F}_{0+}) \,,
\phantom{\Bigg|}\\
\end{equation}
\begin{equation}
\label{XDVCS-red}
\frac{d\sigma^{\mbox{\tiny\rm VCS}}}{dt} =
\frac{\pi \alpha_{\rm em}^2}{\Q^4\sqrt{1+\epsilon^2}} \,
\frac{\xB^2}{1-\xB}\left[ {\cal C}({\cal F}_{++},
{\cal F}_{++}^\ast)+ {\cal C}({\cal F}_{-+},{\cal F}_{-+}^\ast ) +
\varepsilon(y)\, {\cal C}({\cal F}_{0+},{\cal F}^\ast_{0+} ) \right] ,
\end{equation}
where the bi-linear ${\cal C}$-coefficient of the (D)VCS term is given in (45)
of \cite{Belitsky:2012ch} and the linear ${\cal C}$-coefficient of the interference term reads
\begin{equation}
\label{XINT-red1}
{\cal C}({\cal F}_{++}|{\cal F}_{0+}) = \left[F_1(t) {\cal H}_{++} -  \frac{t}{4 M^2} F_2(t) {\cal E}_{++} -
\left\{F_1(t) + F_2(t)\right\}\widetilde{\cal H}_{++}\right]\!(\xB,t,\Q^2) + {\cal O}(\xB)\,.
\end{equation}
Note that here the longitudinal helicity CFFs are suppressed by an additional
$\xB$ factor. From the equations (\ref{XBH-red}-\ref{XDVCS-red}) one immediately
reads off the well-known canonical scaling and the characteristic $y$ hierarchy
of the BH, interference, and DVCS term, given by
$$\frac{d\sigma^{\mbox{\tiny BH}}}{dt} \propto \frac{y^2}{-t \Q^2}\,, \qquad
\frac{d\sigma^{\mbox{\tiny INT}}}{dt} \propto \frac{y}{\Q^4}\,,
\quad\mbox{and}\quad \frac{d\sigma^{\mbox{\tiny DVCS}}}{dt} \propto
\frac{1}{\Q^4}\,, $$
respectively.
A few further comments about the variable dependencies are in order.

\begin{itemize}
\item $y$-dependencies
\end{itemize}
The  power behavior in $y$ of the  BH, interference, and DVCS term is modified.
More precisely, we have for these three terms the hierarchy
\begin{equation}
\frac{y^2}{\left[1+\frac{t}{\Q^2}-y \frac{t}{\Q^2}(1-\xB) \right]
\left[1+\frac{t}{\Q^2}-y\left(1+\frac{\xB t}{\Q^2}\right)\right] }\,, \qquad
\frac{y(2-y)}{2-2y+y^2+ \frac{\epsilon^2 y^2}{2} }\,, \qquad 1,
\end{equation}
where both the BH and DVCS term is further separated into transverse and longitudinal parts.
The latter is proportional to the polarization parameter $\varepsilon(y)$, which, however,
in  DVCS kinematics appears to be power suppressed. The additional $y$-dependence of the BH term,
cf.~(\ref{BH-integrated}), depends on both the $t/\Q^2$ ratio and $\xB$.  At
$y=0$ and in the vicinity of $y_{\rm col}$ it has the values
\begin{eqnarray}
\frac{1}{\left(1+\frac{t}{\Q^2}\right)^2} \sim 1 \quad  \mbox{and}\quad
\frac{1}{\left(1+\frac{t}{\Q^2}\right) \left(1-\frac{t}{\Q^2}+\frac{2 \xB t }{\Q^2}\right)}  \,
\frac{1 }{y_{\rm col} -y} \sim \frac{1}{1-y}\,,
\end{eqnarray}
respectively.  In the DVCS kinematics this function can be approximated by $1/(1-y)$.
The additional $y$-dependence of the interference term (\ref{XINT-red})
is given by the rather mild concave function
$$\frac{2-y}{2-2y+y^2+ \frac{\epsilon^2 y^2}{2} } \approx \frac{2-y}{2-2y+y^2}\,,$$
which takes the value one at both endpoints $y\in \{0,
\approx\!\!1\}$ and has a maximum of $\approx 1.21$ at $y\approx 0.59$.

\begin{itemize}
\item  $t$-dependence of the BH cross section
\end{itemize}
The kinematical factor $\widetilde{K}^2/(-t)$ in the BH cross section (\ref{XBH-red}) is proportional
to $(t-t_{\rm min})/t$ and, hence, it vanishes at the phase space boundary $t\to t_{\rm min}$. Thereby,
the BH cross section (\ref{XBH-red}) remains finite and is proportional to $\xB^2/(-t_{\rm min})$.
If we have  the region $-t_{\rm min} \ll -t$ in mind, where
$-t_{\rm min} \sim \xB^2 M_p^2$ vanishes at small $\xB$,
we will loosely say that the BH cross section is proportional to $1/(-t)$.
We add that in this $t$-region and for $\xB \lesssim 0.05$  the approximation
(\ref{XBH-red}) works on the level of one percent and better.

\begin{itemize}
\item small-$\xB$ region
\end{itemize}
At small $\xB$ the CFF behavior is governed by a possible ``pomeron'' exchange, which
yields that even $\xB\times {\cal F}(\xB,t,\Q^2)$ may grow for decreasing $\xB$ values.
Taking the limit $\xB\to 0$ for the kinematical factors of (\ref{XBH-red}--\ref{XDVCS-red})
yields for $-t_{\rm min} \ll -t$ the rather accurate kinematic expressions
\begin{eqnarray}
\label{XBH-approx}
\frac{d\sigma^{\rm BH}}{dt} &\! \approx \! & \frac{4 \pi \alpha_{\rm em}^2 }{-t  \Q^2 }\; \frac{y^2
\left[F_1^2(t)-\frac{t}{4 M_p^2} F_2^2(t) \right]}{\left[y_{\rm col}-y\right]\left[1+(1-y)\frac{t}{\Q^2}\right]} \quad\mbox{for}\quad y < y_{\rm col} \approx 1+\frac{t}{\Q^2}\sim 1,
\\
\label{XDVCS-approx}
\frac{d\sigma^{\rm DVCS}}{dt} & \!\approx \!&  \frac{\pi \alpha_{\rm em}^2}{\Q^4 }
\xB^2\left[ {\cal C}({\cal F}_{++},{\cal F}_{++}^\ast)+ {\cal C}({\cal F}_{-+},{\cal F}_{-+}^\ast ) +
\varepsilon(y)\, {\cal C}({\cal F}_{0+},{\cal F}^\ast_{0+} ) \right],
\end{eqnarray}
where
\begin{eqnarray}
\xB^2\,  {\cal C}({\cal F},{\cal F}^\ast) \approx
\left[\left|\xB {\cal H} \right|^2  - \frac{t}{4 M^2_{p}} \left| \xB{\cal E} \right|^2 +
\left|\xB \widetilde {\cal H}\right|^2 - \frac{t}{4 M^2_{p}} \left|\xB \overline{\cal E} \right|^2
\right]\!\left(\xB,t,{\cal Q}^2\right)
\end{eqnarray}
with the new notation%
\footnote{This redefinition absorbs a common prefactor $\xB/(2-\xB+\xB t/\Q^2)$ of $\widetilde{\cal E}$
that appears in all $\cal C$-coefficients and it ensures that $\overline{\cal E}$ has the same
phenomenological Regge counting in the small-$\xB$ region as the other CFFs. It cancels the
$1/\xi\approx (2-\xB)/\xB$ factor that appears in the form factor in front of $\cal E$, used
for the decomposition of the DVCS amplitude (analogously for GPD $\widetilde E$).}%
\begin{eqnarray}
\label{bE}
\overline{\cal E}(\xB,t,\Q^2) \approx \frac{\xB}{2-\xB} \widetilde{\cal E}(\xB,t,\Q^2) \,.
\end{eqnarray}
The interference term is suppressed w.r.t.~DVCS cross section  by an additional factor
$\xB\,y$ and can be safely neglected. As one immediately realizes from these rather accurate kinematic
approximations, the BH cross section (\ref{XBH-approx}) is kinematically
enhanced at small $-t$ and suppressed at small $y$ values.
However, most important is that the DVCS cross section (\ref{XDVCS-approx}) in the small-$\xB$
region grows with decreasing $\xB$, caused by an effective ``pomeron'' exchange in the $t$-channel.
Thus, even the relative kinematical  $-t(1-y)/\Q^2 y^2$ suppression of the DVCS cross section
w.r.t.~BH one can
be overcome. Moreover, the DVCS signal can be further experimentally enhanced by an upper $y$ cut.
However, it should be kept in mind that the ratio of DVCS cross section to the BH one depends on
the competing interplay of $\xB$, $\Q^2$, and $t$ dependencies.
In particular, if the DVCS cross section falls off much
faster with increasing $-t$ than the electromagnetic form factor $F_1(t)$,
like in the case of the often assumed exponential $t$-dependence, the ratio
of DVCS cross section to BH one can become very small at larger $-t$ values.

Finally, let us quote the $y$-dependence of the $t$-differential cross section
(\ref{dX-reduced}) in the most obvious manner for general DVCS kinematics $\Q^2 > -t$ and $y < y_{\rm col}$:
\begin{equation}
\label{XTOT-y}
\frac{d\sigma^{\mbox{\tiny TOT}}}{dt} =
\frac{y^2  \left[ \frac{d\sigma_{\rm T}^{\mbox{\tiny BH}}}{dt}  + \varepsilon(y)
\frac{d\sigma_{\rm L}^{\mbox{\tiny BH}}}{dt} \right] }{\left(1-y \frac{(1-\xB) t}{\Q^2+t} \right)
\left(\frac{\Q^2+t}{\Q^2+\xB t}-y\right)} \pm  \frac{y\left(1-\frac{y}{2}\right)\sqrt{1+\epsilon^2}}{1-y+
\frac{y^2}{2}+ \frac{\epsilon^2  y^2}{4}}\, \frac{d\sigma_{\rm T}^{\mbox{\tiny INT}}}{dt} +
\frac{d\sigma_{\rm T}^{\mbox{\tiny DVCS}}}{dt} + \varepsilon(y)\frac{d\sigma_{\rm L}^{\mbox{\tiny DVCS}}}{dt}\,.
\end{equation}
The reduced  BH cross section
$d\sigma_{\rm T}^{\mbox{\tiny BH}}/dt  + \varepsilon(y)d\sigma_{\rm L}^{\mbox{\tiny BH}}/dt$
for the smaller-$\xB$ region, the reduced  interference term
$d\sigma_{\rm T}^{\mbox{\tiny INT}}/dt$,  and the DVCS
cross section $d\sigma_{\rm T}^{\mbox{\tiny DVCS}}/dt + \varepsilon(y)d\sigma_{\rm L}^{\mbox{\tiny DVCS}}/dt$
can be read off from (\ref{XBH-red}), (\ref{XINT-red}), and (\ref{XDVCS-red}), respectively.
Note that the $y$-dependent factor in front of the interference term is given by a transverse
photon flux asymmetry,
$$
\frac{y\left(1-\frac{y}{2}\right)\sqrt{1+\epsilon^2}}{1-y+\frac{y^2}{2}+ \frac{\epsilon^2  y^2}{4}}=\frac{{\cal L}_{--}-{\cal L}_{++}}{{\cal L}_{--}+{\cal L}_{++}}.
$$
Depending on the kinematics, the application of the formula (\ref{XTOT-y}) is two-fold.
In the case that the subtraction of the BH cross section can be reliably done, the
measurement of this subtracted cross section at three different beam energies allows in
principle to separate the longitudinal DVCS cross section,  transverse DVCS  cross section,
and the interference term. One may also utilize the $y$-dependence
to cross-check experimentally if a BH-subtraction procedure is well understood.

\subsection{Relating DVCS observables to GPDs}
\label{sec:GPD22CFFs}
\begin{figure}[ht]
\begin{center}
\includegraphics[width=0.95\textwidth]{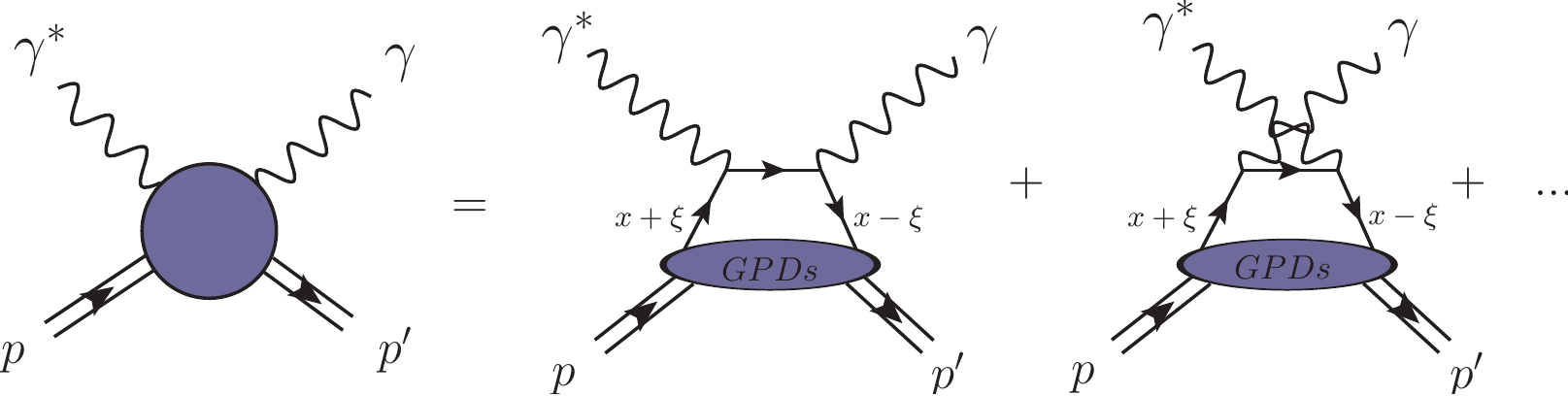}
\end{center}
\caption{
\small
Factorization of the DVCS amplitude to leading order in perturbative QCD and to leading
twist-two accuracy.
This yields equation (\ref{cffs}) that expresses CFFs (filled circle) in terms of GPDs
(filled ellipse).
\label{fig:DVCS2GPD} }
\end{figure}
GPDs, denoted here generically as
$$
F(x,\eta=\xi, t, \mu^2) \quad\mbox{with}\quad F\in\{H,E,\widetilde{H},\widetilde{E}\}\,,
$$
are intricate functions that, besides depending on the partonic momentum fraction $x$
and the momentum transfer squared $t$, depend
also on the $t$-channel longitudinal momentum fraction $\eta$, called \emph{skewness}
(often denoted by $\xi$ in the literature\footnote{$\xi$ stands
for a Bjorken-like scaling variable while $\eta$ is a second scaling
variable appearing, e.g., in doubly virtual Compton scattering. In deeply
virtual production of photon and mesons one has $\eta\approx \xi$. Note that
below, in Sect.~\ref{sec:simulation} only, the
symbol $\eta$ will be used also to denote rapidity.}),
and on the factorization scale $\mu^2$.
The unpolarized parton GPDs are called $H$ and $E$ \cite{Ji:1998pc}, where the former (latter)
GPD can be loosely associated with a proton helicity (non)conserved distribution.
Analogous nomenclature is used for the polarized parton GPDs $\widetilde H$ and $\widetilde E$
\cite{Ji:1998pc}. GPDs have certain spectral properties
\cite{Mueller:1998fv,Radyushkin:1997ki}
and so their $x$-moments are polynomials of certain order in $\eta$, with lowest moments being
equal to elastic nucleon form factors. In the forward limit ($t\to 0$, $\eta\to 0$)
$H$ ($\widetilde H$) reduce to the unpolarized (polarized) Parton Distribution Functions (PDFs),
commonly called $(\Delta)q$ and $(\Delta)g$ for quarks and gluon, respectively. Furthermore,
in the region $|x| > \eta$, where a parton
is exchanged in the $s$-channel, GPDs are constrained by positivity conditions 
\cite{Martin:1997wy,Pire:1998nw,Radyushkin:1998es,Ji:1998pc,Diehl:2000xz},
which can be viewed in their most general form as a consequence of a wave function overlap representation
\cite{Pobylitsa:2002iu,Pobylitsa:2002vw,Pobylitsa:2002vi}.
However, this GPD property is exact only to LO accuracy.
To our best knowledge,
no attempt has been undertaken to derive positivity constraints for the $\eta=x$ case.
This implies that existing positivity constraints mostly do not apply
for the phenomenological description of deeply virtual processes.
However, as we will see below, they are important constraints for GPD models, e.g., as used by us in
Sect.~\ref{sec:imaging} for the purpose of extrapolation from the $\eta=x$ to the $\eta=0$ case.

DVCS observables can be exactly evaluated in terms of the helicity CFFs (\ref{cffF_{ab}}).
To express them in terms of GPDs in a systematically improvable manner, it is maybe
appropriate to utilize a conventionally defined GPD-inspired
CFF basis, such as the one introduced in \cite{Belitsky:2001ns}\footnote{To simplify
notation we set here ${\cal F}_3  = 2 \xi \left({\cal F}^{\rm{tw}-3}_+ - {\cal F}^{\rm{tw}-3}_-\right)$. Note
that the prefactor $\xi$ does not imply that ${\cal F}_3$ vanishes in the $\xi\to 0$ limit.}:
\begin{equation}
\label{cffF}
{\cal F} \in \{{\cal H},{\cal E}, \widetilde{\cal H}, \widetilde{\cal E}, {\cal H}_3,{\cal E}_3,
\widetilde{\cal H}_3, \widetilde{\cal E}_3,
{\cal H}_{\rm T},{\cal E}_{\rm T}, \widetilde{\cal H}_{\rm T}, \widetilde{\cal E}_{\rm T} \}\,.
\end{equation}
Here, the CFFs ${\cal H},{\cal E}, \widetilde{\cal H}$, and $\widetilde{\cal E}$ are
associated with twist-two GPDs $F \in \{H,E,\widetilde{H}, \widetilde{E}\}$ and govern
the photon helicity non-flip DVCS amplitude, i.e., at leading twist-two accuracy we have
\begin{eqnarray}
{\cal F}_{++}(\xB,t,\Q^2) = {\cal F}(\xB,t,\Q^2) + {\cal O}(1/\Q^2)\quad\mbox{for}\quad{\cal F}
\in \{{\cal H},{\cal E}, \widetilde{\cal H}, \widetilde{\cal E}\}.
\end{eqnarray}
It is ensured by the factorization theorem \cite{Collins:1998be,Radyushkin:1997ki,Ji:1998xh}
that these four dominant CFFs arise from the convolution of twist-two GPDs with hard coefficients,
which are perturbatively calculable as a series in the strong coupling constant $\alpha_s$.
Presently, these coefficients are known to NLO accuracy in the standard minimal subtraction
scheme \cite{Belitsky:1997rh,Mankiewicz:1997bk,Ji:1997nk,Ji:1998xh,Belitsky:1999hf,Pire:2011st} and to
next-to-next-to-leading order (NNLO) accuracy in a special scheme \cite{Mue05a,KumMuePasSch06}.
To LO they are calculated from the  handbag diagram, depicted in Fig.~\ref{fig:DVCS2GPD}, yielding
the convolution formula
\begin{equation}
\label{cffs}
{\cal F}(\xB,t,\Q^2)\stackrel{\rm LO}{=}\sum\limits_{i}\int_{-1}^1 \! dx \left[ \frac{e^2_i}{\xi-x-i\epsilon} \mp \{ x \rightarrow -x \} \right] F_i(x,\xi,t,\mu^2)\;\; \mbox{for}\;\;
{\cal F} \in \left\{ {{\cal H},{\cal E}  \atop \widetilde{\cal H}, \widetilde{\cal E}}
\right\},
\end{equation}
where $e_i$  are the fractional quark charges. The variable
$\xi\sim \xB/(2-\xB)$ is a conventionally defined Bjorken-like scaling variable, equated to
the longitudinal momentum fraction in the $t$-channel,
and $\mu^2 \sim \Q^2$ being the factorization scale. Note the conventional dependence as
function of this scaling variable is in the orders of $O(1/\Q^2)$. It further reduces
if one takes into account kinematic corrections, evaluated to twist-four
accuracy at LO in $\alpha_s$ \cite{Braun:2011zr,Braun:2011dg,Braun:2012bg,Braun:2012hq}.
As is well known, the ambiguity in setting the factorization scale diminishes
in higher orders of perturbation theory as long as the perturbative corrections to the GPD
evolution are consistently taken into account. Moreover, as long as we consider only
the DVCS process, the perturbative order to which we describe its amplitude can be mainly
understood as a convention (in a DVCS scheme, like in the DIS scheme,
only the perturbatively predicted evolution would alter, if we would switch, e.g., from LO to NLO).
In the minimal subtraction scheme the evolution kernels are known to NLO accuracy
\cite{Belitsky:1999hf}.  We also recall the fact, well known from unpolarized DIS, that
the absence of gluon GPDs in the LO convolution equations (\ref{cffs})
does not imply that these GPDs are absent from a LO  description;
they drive the evolution of the sea quarks.

The CFFs  ${\cal H}_3,{\cal E}_3, \widetilde{\cal H}_3$, and $\widetilde{\cal E}_3$ are
expressed by twist-three GPDs, containing information on three-parton correlation functions,
and enter into the photon helicity longitudinal-transversal flip amplitude, which reads
to twist-three and LO in $\alpha_s$ accuracy as
\begin{equation}
\label{F3toF0+}
{\cal F}_{0+}(\xB,t,\Q^2) = -\frac{\sqrt{2}\widetilde{K}}
{\Q \sqrt{1+\epsilon^2}\left(2-\xB+\frac{\xB t}{\Q^2}\right)} \left[
\xB {\cal F} +  {\cal F}_3 \right](\xB,t,\Q^2) + {\cal O}\left(1/{\cal Q}^2\right) +{\cal O}(\alpha_s)\,, \quad
\end{equation}
where
\begin{eqnarray}
\label{tK}
\widetilde{K} =  \sqrt{-(1-\xB) \left(\! 1+\frac{\xB t}{\Q^2}\!\right) t -\left(\!1+\frac{t}{\Q^2}\!\right)^2
\xB^2  M_p^2}
\end{eqnarray}
is a kinematical factor that vanishes at the minimal value of $-t$.
The CFFs ${\cal H}_{\rm T},{\cal E}_{\rm T}, \widetilde{\cal H}_{\rm T}$, and $\widetilde{\cal E}_{\rm T}$
are the dominant contributions to the transverse helicity flip DVCS amplitude, which, at
leading twist accuracy, arises from the transversely polarized gluon GPDs that are perturbatively
\cite{Belitsky:2000jk}
and power suppressed \cite{Kivel:2001rw,Braun:2012bg,Braun:2012hq}. Consequently, we have
\begin{eqnarray}
{\cal F}_{-+}(\xB,t,\Q^2) = {\cal F}_{\rm T}(\xB,t,\Q^2)  +
{\cal O}\left(1/{\cal Q}^2\right)\quad \mbox{with}\quad
{\cal F}_{\rm T}(\xB,t,\Q^2) =  {\cal O}(\alpha_s)\,.
\end{eqnarray}

If not stated otherwise, in the following we work for
convenience to twist-two and LO accuracy, where we take four
light quarks and we adopt the conventions
\begin{eqnarray}
\xi= \frac{\xB}{2-\xB} \quad\mbox{and} \quad \mu^2= \Q^2\,.
\end{eqnarray}
With these approximations GPD phenomenology can be drastically simplified. Namely,
the convolution formula (\ref{cffs}) tells us that the imaginary parts of the four
dominant CFFs are given by the GPDs on the cross-over line $x=\xi$,
\begin{eqnarray}
\label{KLMSPM-DR-Im}
\Im{\rm m}  {\cal F}(x_{\rm Bj},t,{\cal Q}^2)  & \stackrel{\rm LO}{=} &
\pi  F (\xi,\xi,t,{\cal Q}^2)\,, \quad F \in \{H, E, \widetilde H, \widetilde E\}\,.
\end{eqnarray}
Furthermore, by means of the GPD spectral property one obtains from (\ref{cffs}) a
dispersion integral representation for the real parts of these CFFs \cite{Teryaev:2005uj},
\begin{eqnarray}
\label{KLMSPM-DR-Re}
\Re{\rm e}  \!
\left\{\! {\cal H}  \atop {\cal E} \!\right\}\!(x_{\rm Bj},t,{\cal Q}^2) & \stackrel{\rm LO}{=} &
{\rm PV}\! \int_{0}^1\!dx\, \frac{2x}{\xi^2-x^2}\!  \left\{\! H \atop E \! \right\}\! (x,x,t,{\cal Q}^2)
\mp {\cal C}(t,\Q^2)\,,
\\
\label{KLMSPM-DR-Re1}
\Re{\rm e}  \!
\left\{\! \widetilde{\cal H}  \atop\widetilde {\cal E} \!\right\}\!(x_{\rm Bj},t,{\cal Q}^2) &
\stackrel{\rm LO}{=} &
{\rm PV}\! \int_{0}^1\!dx\, \frac{2}{\xi^2-x^2}\!  \left\{\! \xi \widetilde H \atop x^2 \widetilde E \! \right\}\! (x,x,t,{\cal Q}^2)
+ \frac{1}{\xi}\left\{\! 0 \atop \widetilde{\cal C}(t,\Q^2) \!\right\}\,.
\end{eqnarray}
Here ${\cal D}= - {\cal C}$, entering in (\ref{KLMSPM-DR-Re}) as subtraction term,
is given as convolution of the so called $D$-term contribution (introduced in
\cite{Polyakov:1999gs} to complete GPD polynomiality in one possible manner),
which can be extracted for a given GPD.  Note that  the  dispersion relation for
$\widetilde{\cal E}$ is over-subtracted and that the subtraction constant
$\widetilde{\cal C}(t,\Q^2)$ contains a pion pole contribution.
This pole contribution can be calculated rather analogously to the $D$-term,
e.g., from the suggested parameterizations \cite{Mankiewicz:1998kg,Frankfurt:1999xe} or
from extraction using a Regge-inspired GPD parametrization \cite{Bechler:2009me}.
Since in this approximated framework at fixed photon virtuality only the GPDs at the cross-over line
$x=\xi$ and two subtraction constants enter,  GPD phenomenology is drastically simplified.

\subsection{Present status of DVCS measurements and GPD analyzes}
\label{sec:phenomenology}
Let us first consider  experiments which have only an electron beam available. The three
parts of the electroproduction cross section (\ref{X-electroproduction})
contribute, depending on the kinematics, with different strength to the various harmonics.
One can remove the  BH cross section (\ref{X-BH}), taken to LO accuracy in $\alpha_{\rm em}$, by measuring
the cross section differences for single spin flip observables, e.g., the beam-helicity
difference $\Delta_{\rm LU}$
\begin{eqnarray}
\label{DX-electroproduction}
\frac{d\Delta_{\rm LU}}{d\xB dt d\Q^2  d\phi} = \frac{1}{2} \left[
\frac{d\sigma^\rightarrow}{d\xB dt d\Q^2  d\phi} -
\frac{d\sigma^\leftarrow}{d\xB dt d\Q^2  d\phi}
\right],
\end{eqnarray}
and analogously for a longitudinally ($\Delta_{\rm UL}$) and transversely
($\Delta_{\rm UT}$) polarized proton target. These observables are expanded
in terms of odd harmonics%
\footnote{Here and in the following
$\sin(n\phi)$, $\cos(\varphi)\sin(n\phi)$, and also $\sin(\varphi)\cos(n\phi)$
are called odd harmonics, while $\cos(n\phi)$, $\cos(\varphi)\cos(n\phi)$, and
also $\sin(\varphi)\sin(n\phi)$ are called even harmonics.}.
In fixed target kinematics they are mainly dominated by the $\sin(\phi)$ and/or
$\sin(\varphi)\cos(\phi)$ harmonics of the interference term (\ref{X-INT}),
giving access to the imaginary part of four twist-two associated CFF combinations, see
(\ref{A_{LU}-approx}--\ref{A_UL}) below.
However, the DVCS term (\ref{X-VCS}), suppressed in these observables by
$1/\Q^2$, may also contribute to some extent.
We add that in double spin flip experiments the BH cross section (\ref{X-BH})
also enters, however its $\cos(\phi)$ harmonic can be
quite small, which may allow the access to the $\cos(\phi)$ harmonic of the
interference term, i.e., three combinations of
twist-two associated CFF combinations.

Unpolarized electroproduction and electron-helicity dependent cross section
measurements at rather large $\xB$ and small $-t$ have been performed with small
uncertainties by the Hall A collaboration at JLAB \cite{Munoz_Camacho:2006hx}.
The measured cross section differences (\ref{DX-electroproduction}) is compatible
with various GPD model predictions, see
\cite{Munoz_Camacho:2006hx,Polyakov:2008xm,Kroll:2012sm}. In the unpolarized
case, however, the measurements at four different $-t$ values,
at $\Q^2=2.3\,\GeV^2$ and rather large $\xB =0.36$ indicate that the DVCS cross
section at these kinematics is much larger and drops much faster with growing
$-t$ than expected from common GPD models.
As explained in Sect.~\ref{sec:Rosenbluth}, at small $\xB$ (large $W$) the
``pomeron'' behavior
leads to the DVCS amplitude outgrowing the BH amplitude and
as a result of the $\phi$-integration, the interference term is negligibly small
in this region.
Therefore, at the H1 \cite{Adloff:2001cn,Aktas:2005ty,Aaron:2007ab,Aaron:2009ac}
and ZEUS \cite{Chekanov:2003ya,Chekanov:2008vy} collider experiments the
DVCS cross section has been accessed by subtracting the BH cross section.
Thereby, the subtraction method has been checked experimentally,
since in some parts of the kinematic phase space the BH cross section dominates
and Monte Carlo simulations can be directly confronted with measurements. The
size of the cross section was predicted by a simple model \cite{FraFreStr98}
and can be at NLO also described with standard%
\footnote{We distinguish here between standard and flexible GPD models. Former,
e.g., set up in
\cite{GoePolVan01,Belitsky:2001ns,GuzTec06,Goloskokov:2007nt,Goloskokov:2009ia},
rely on a more or less fixed skewness prescription and are used in model
predictions the latter allow for a flexible adjustment of the skewness effect
and a consistent GPD description of present DVCS data.}
GPD models, however, not at LO \cite{Freund:2001hd,Guzey:2008ys}. A simple
flexible GPD model allows to describe the HERA collider data at LO, NLO, and
NNLO, which allows to quantify GPD reparametrization effects \cite{Kumericki:2009uq}.

In some experiments only asymmetries, less affected by possible normalization
problems, are measurable. Having only an electron beam at hand one can access the
interference term with single spin flip experiments by polarizing the electron
beam longitudinally (electron beam-helicity asymmetry)
\begin{eqnarray}
\label{A_{LU}}
A_{\rm LU} = \left(\frac{d\sigma^{\rightarrow}}{d\xB dt d\Q^2 d\phi}-
\frac{d\sigma^{\leftarrow}}{d\xB dt d\Q^2 d\phi}\right)\Big/
\left(\frac{d\sigma^{\rightarrow}}{d\xB dt d\Q^2
d\phi}+\frac{d\sigma^{\leftarrow}}{d\xB dt d\Q^2 d\phi}\right),
\end{eqnarray}
and analogous equations hold true for single spin flip asymmetries with
longitudinally ($A_{\rm  UL}$) or transversely ($A_{\rm  UT}$) polarized
nucleons and unpolarized electron beams.
Here, the squared BH term in the numerator will drop out again at LO accuracy in
$\alpha_{\rm em}$ and the squared DVCS term will yield some contamination,
while the normalization is governed by all three terms of the unpolarized cross
section (\ref{X-electroproduction}).
In addition to longitudinal proton spin asymmetry measurements at HERMES
\cite{Airapetian:2010ab} and CLAS \cite{Chen:2006na}, electron beam-helicity
asymmetries were measured at CLAS \cite{Girod:2007aa,Gavalian:2008aa}.

The HERA experiments had both electrons and positrons beams available, which allowed
to access the interference term via the beam charge asymmetry
\begin{eqnarray}
\label{A_C}
A_{\rm C} = \left(\frac{d\sigma^{+}}{d\xB dt d\Q^2  d\phi}- \frac{d\sigma^{-}}{d\xB dt d\Q^2  d\phi}\right)\Big/
\left(\frac{d\sigma^{+}}{d\xB dt d\Q^2  d\phi}+\frac{d\sigma^{-}}{d\xB dt d\Q^2  d\phi}\right),
\end{eqnarray}
where the numerator is entirely given by the interference term, however, the normalization
depends also on the DVCS squared term.
This asymmetry has been measured by the HERMES collaboration \cite{Airapetian:2006zr},
where the correlation between the lowest and first harmonics, predicted in
\cite{Belitsky:2001ns}, was confirmed.
The beam charge asymmetry was measured also by the H1 collaboration \cite{Aaron:2009ac} at
large $W$ (small $\xB$) where, however, this observable (as well as the $t$-differential cross
section and the longitudinal spin asymmetry) is dominated by the CFF $\cal H$ and
uncertainties are large. Hence, the  CFF $\cal E$, giving access to sea quark and gluon
GPD $E$ that enters Ji`s angular momentum sum rule, could not be revealed at small $x$.

The HERMES collaboration provided the most complete
measurement of {\em thirty-four} DVCS asymmetries, where a missing-mass event
selection method was employed. This includes also
a partial interference/DVCS decomposition for asymmetries measured
with a transversely polarized  \cite{Airapetian:2008aa} and unpolarized
\cite{Airapetian:2009aa} proton target. However, since the normalization depends on the
unpolarized DVCS cross section and  both statistical and systematical uncertainties are
rather large, a full disentanglement of  twist-two related CFFs and an access to the
twist-three sector could not be achieved. In particular, GPD $E$ cannot
be accessed from these measurements in a GPD model unbiased manner.

\begin{figure}[ht]
\begin{center}
\includegraphics[width=0.95\textwidth]{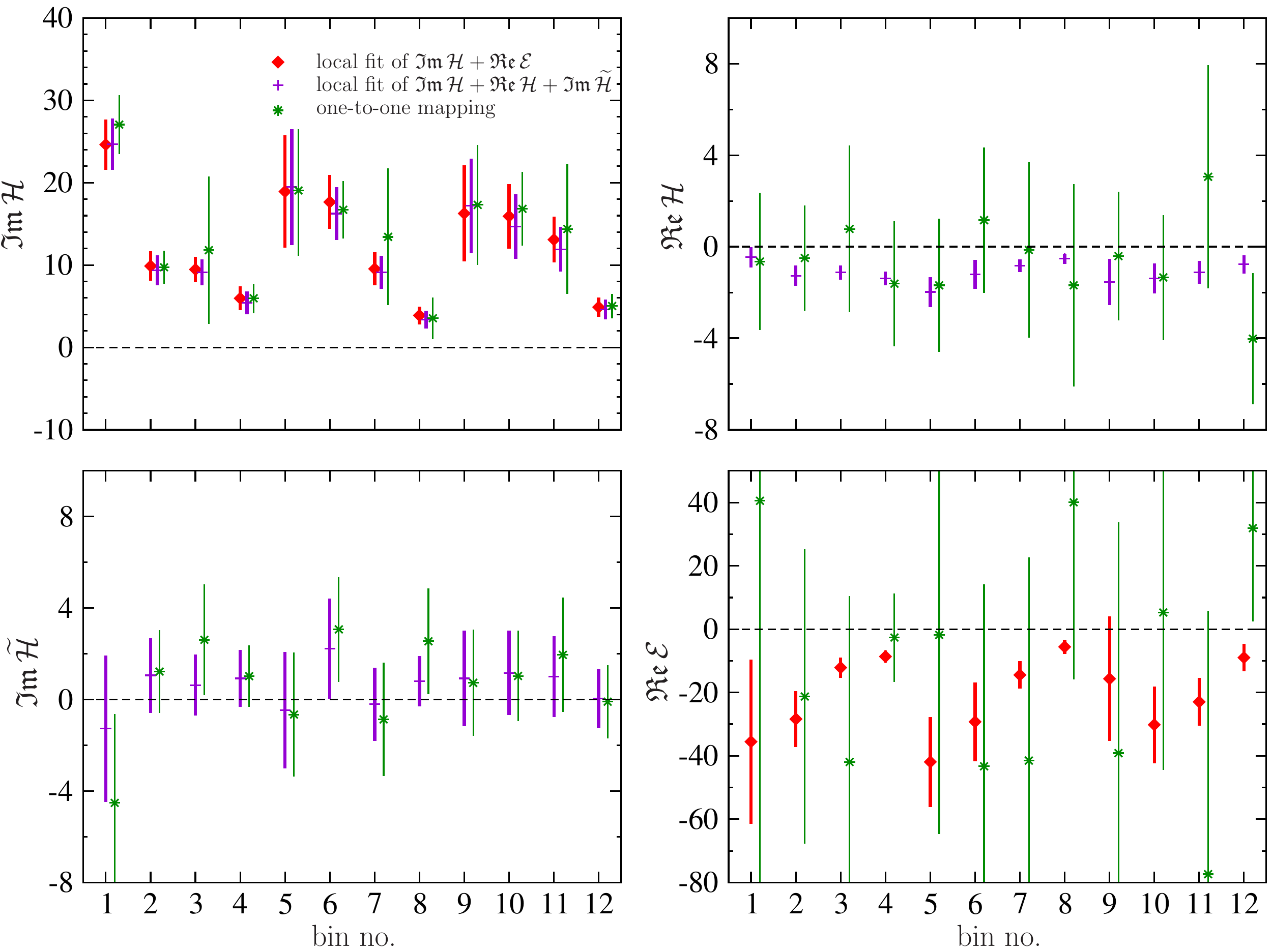}
\end{center}
\caption{\small Results of least-squares fits in two scenarios with only a small
number of CFFs locally fitted to data separately for each of 12 HERMES bins. First,
with only $\Im{\rm m}\mathcal{H}$ and $\Re{\rm e}\mathcal{E}$ fitted  (red diamonds) and,
second, with $\Im{\rm m}\mathcal{H}$, $\Re{\rm e}\mathcal{H}$ and $\Im{\rm m}\widetilde{\mathcal{H}}$
(purple pluses).  For comparison, result of a one-to-one mapping procedure is also shown (green stars).
\label{fig:regr}}
\end{figure}
This large set of DVCS observables, measured by HERMES in twelve kinematical bins
(some are measured in 18 bins), allows for a local extraction of CFFs. Since experimental uncertainties
are rather large for most of the observables, one may still rely on the hypothesis
of twist-two dominance and extract the twist-two associated CFFs by maps \cite{Belitsky:2001ns}, by
least-squares fits \cite{Guidal:2008ie, Guidal:2009aa,Guidal:2010ig,Guidal:2010de}, or neural networks
\cite{Kumericki:2011rz}. To avoid a misinterpretation of experimental measurements,
these local methods should be utilized with care. In particular, differences exist between the view
points of random variable map and regression methods, see Fig.~\ref{fig:regr}. A one-to-one map of eight
twist-two dominated asymmetries into the space of CFFs reveals that only the imaginary part of the
CFF $\cal H$ significantly differs from zero while its real part and the imaginary part of CFF
$\widetilde{\cal H}$ are relative small.
All other twist-two dominated CFFs have large uncertainties and are compatible with zero
\cite{Kumericki:2013br}.
By means of the LO approximation (\ref{KLMSPM-DR-Im}) the results for the imaginary parts can
now be viewed as GPDs on the cross-over line, while the dispersion relations
(\ref{KLMSPM-DR-Re}, \ref{KLMSPM-DR-Re1}) may in principle be utilized as sum rules to constrain
the GPDs on parts of the cross-over line that are outside of the accessible kinematics
\cite{Kumericki:2008di}.  We add that so far no attempt has been made to access photon
helicity flip contributions, related to twist-three and transversity GPDs. However, the
smallness of higher harmonics is compatible with the hypothesis of twist-two dominance.

Certainly, the partonic interpretation of DVCS measurements, the inclusion of the
$\Q^2$ evolution, perturbative corrections, kinematic corrections
\cite{Braun:2011zr,Braun:2011dg}, and the access to three-parton correlations
\cite{Belitsky:2000vx,Belitsky:2001ns} requires a global analysis with flexible GPD models.
Having measurements over a wide $\Q^2$ range allows, through evolution, to reveal the
GPD away from the cross-over line.
This is used for the description of the DVCS cross section measurements at small $\xB$,
whereas for fixed target kinematics the $\Q^2$ lever arm is small and evolution effects
are relatively weak (for an example study see \cite{Kumericki:2011zc}).

\begin{figure}[ht]
\begin{center}
\includegraphics[width=0.95\textwidth]{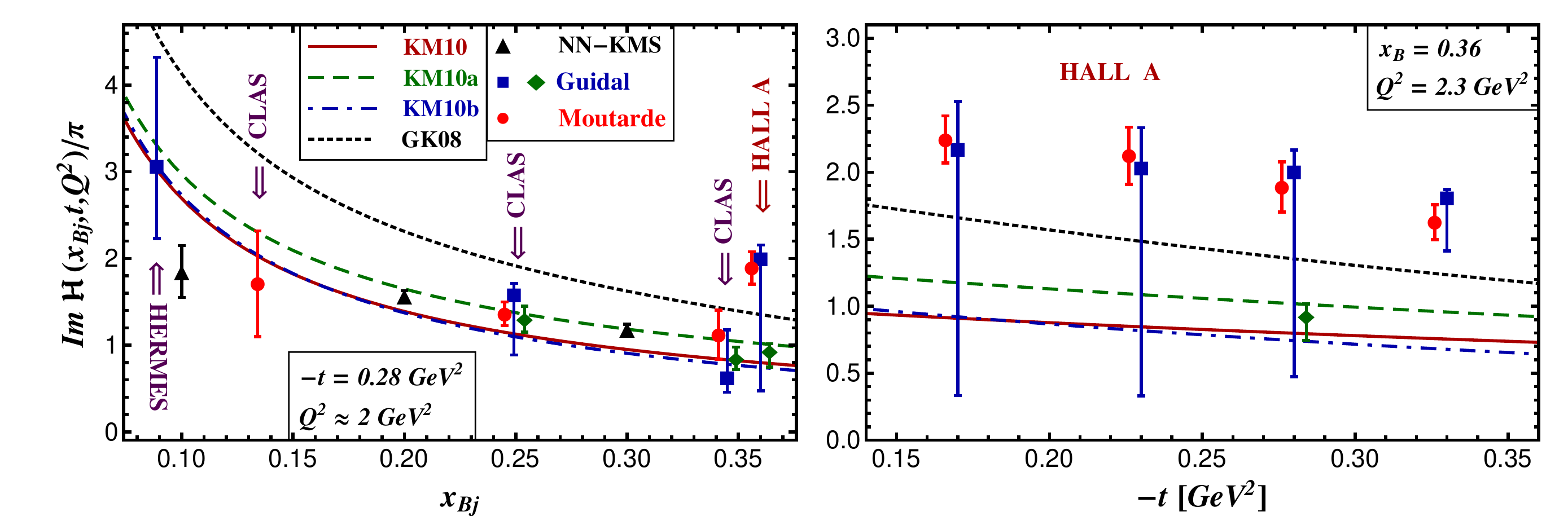}
\end{center}
\vspace{-8 mm}
\caption{\label{plot-comparision}\small
$\Im{\rm m} {\cal H}/\pi$ obtained from DVCS observables with different strategies:
hybrid model fits {\em KM10} (solid) {\em KM10a} (dashed), {\em KM10b} (dash-dotted)
[Hall A cross section data are neglected],
{\em GK07} model from DVEM (dotted) \cite{Goloskokov:2007nt},
seven-fold CFF fit~\cite{Guidal:2008ie,Guidal:2009aa} with boundary conditions (squares),
$\cal H$, $\widetilde {\cal H}$ CFF fit~\cite{Guidal:2010ig} (diamonds),
smeared conformal partial wave model fit~\cite{Moutarde:2009fg} within $H$ GPD (circles).
The triangles result from neural network fit~\cite{Kumericki:2011rz}.
}
\end{figure}

In a first step of a global DVCS analysis, unpolarized proton data were employed in GPD
fits \cite{Kumericki:2009uq,KumMueWeb,Kumericki:2011zc}. Thereby, the world data set could
be described with $\chi^2/{\rm d.o.f.} \approx 1$, using the {\em KM10} model.
Nevertheless, in such a fit the four CFFs ${\cal H}, {\cal E}, \widetilde{\cal H},
\widetilde{\cal E}$ cannot be disentangled and, partially for this reason, even the dominant
${\cal H}$ suffers from larger uncertainties, see Fig.~\ref{plot-comparision}. Below we
will also employ the model {\em KM10a}, which has
also a good $\chi^2/{\rm d.o.f.} \approx 1$ fit to the data set, but ignores the
Hall A cross section measurements. Including polarized proton data in a global fit could
certainly help to disentangle CFFs even better. In a more recent fit, given in
\cite{Kumericki:2013br}, we found that the {\em KM} model, designed for the unpolarized case,
describes even such a set of DVCS data with $\chi^2/{\rm d.o.f.} \approx 1.6$,
where most of the tension is due to the four unpolarized cross section
measurements of Hall A collaboration. We emphasize that this tension can have
different origins, e.g., it is maybe prudent to still consider the possibility that
the experimental issue of exclusivity plays a role in most of the world data. For instance,
beam spin asymmetry measurements from the
HERMES collaboration with a complete event reconstruction yields an increase of their size,
softening, thereby, the tension between measurements
and standard GPD predictions \cite{Airapetian:2012pg,Kroll:2012sm}.
>From present DVCS data, we can certainly state
that GPD $H$ plays the dominant role, some phenomenological constraints for GPD $\widetilde H$
can be obtained, and proton helicity flip GPDs $E$ and $\widetilde{E}$ remain unconstrained.
Let us add that present GPD phenomenology includes also
deeply virtual meson production, in first place in the hand-bag model approach
\cite{Goloskokov:2007nt,Goloskokov:2008ib,Goloskokov:2009ia,Goloskokov:2011rd}
and was started in the perturbative factorization framework with flexible GPD models
\cite{Bechler:2009me,Meskauskas:2011aa}.
So far a reasonable description of the considered deeply virtual meson production channels and
DVCS, currently explored on the level of LO accuracy,
can be reached \cite{Kumericki:2011zc,Meskauskas:2011aa,Kroll:2012sm} except for the large-$\xB$ region.

\begin{figure}[ht]
\begin{center}
\includegraphics[width=0.95\textwidth]{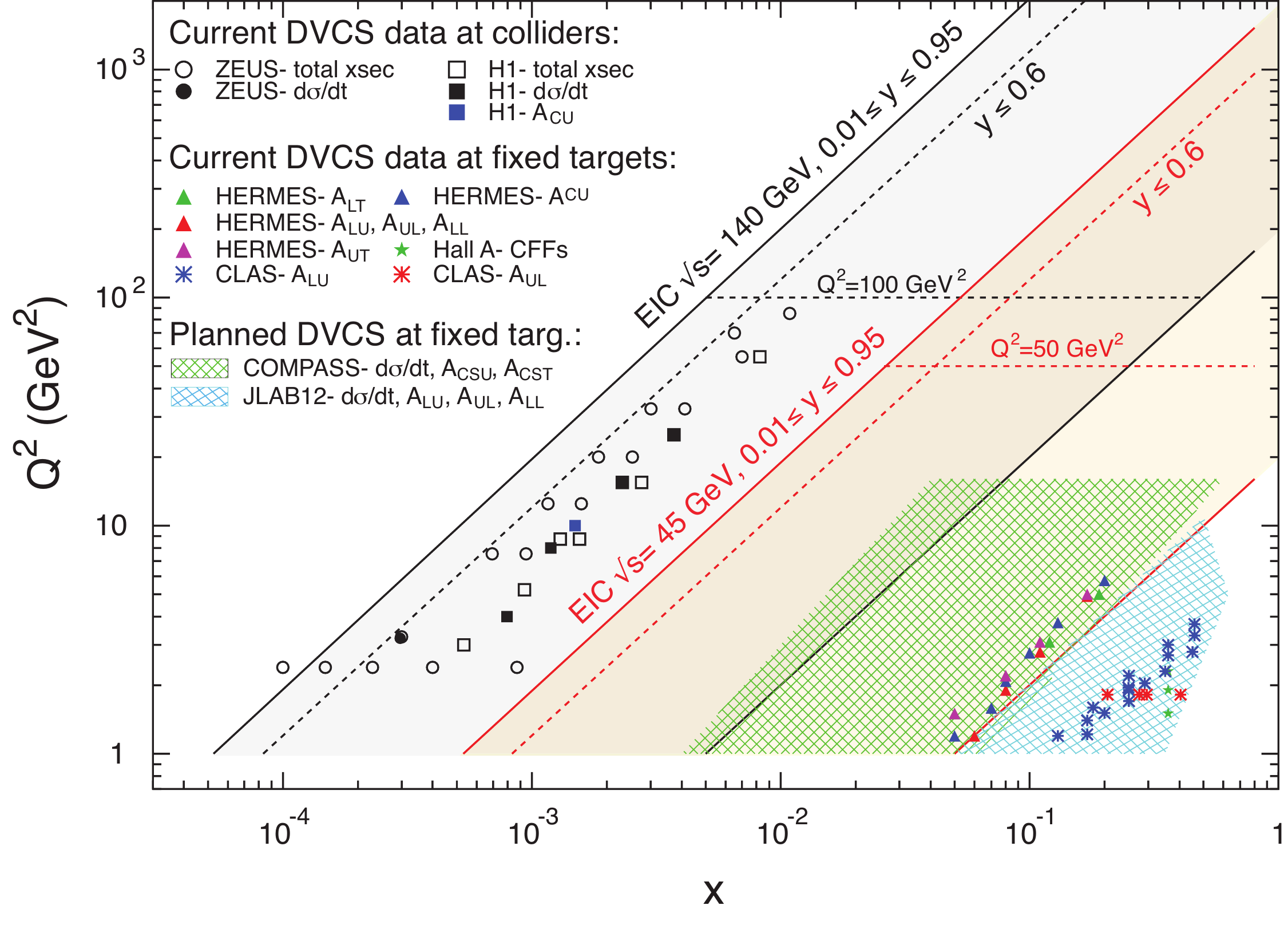}
\end{center}
\vspace{-8 mm}
\caption{\label{x-q2-dvcs}\small
The kinematic reach in $x-\Q^2$ for existing DVCS measurements from H1/ZEUS, HERMES,
CLAS and Hall A, as well as planned ones at COMPASS II and JLAB@12GeV, and the
proposed EIC.}
\end{figure}

As pointed out and illustrated in Figs.~\ref{fig:regr} and \ref{plot-comparision},
present DVCS measurements provide some limited information on GPDs and future precision
measurements are required to pin them down.
New fixed target experiments are planned at COMPASS-II with a polarized muon beam,
extending the HERMES kinematics to lower $\xB$, and JLAB-12~GeV will bridge the gap between the
kinematics of the present JLAB experiments' to the HERMES experiment, see Fig.~\ref{x-q2-dvcs}.
Moreover, a high luminosity machine in the collider mode with polarized electron and proton
or ion beams has been proposed \cite{Accardi:2012hwp} and will be introduced in the next Section.

\section{The EIC project and Monte Carlo simulation}
\label{sec:simulation}
\begin{figure}[ht]
\begin{center}
\includegraphics[width=0.45\textwidth]{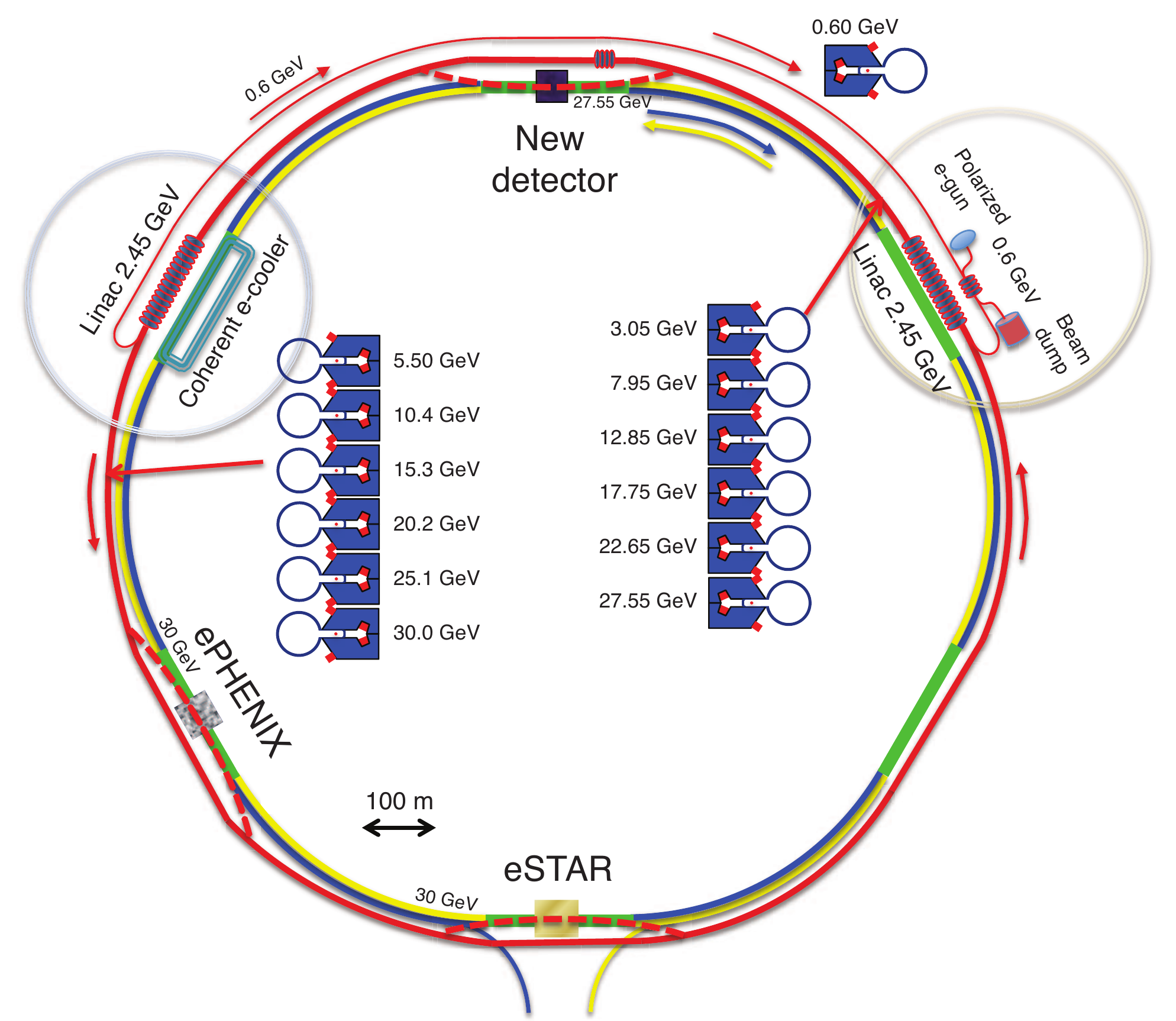}
\includegraphics[width=0.45\textwidth]{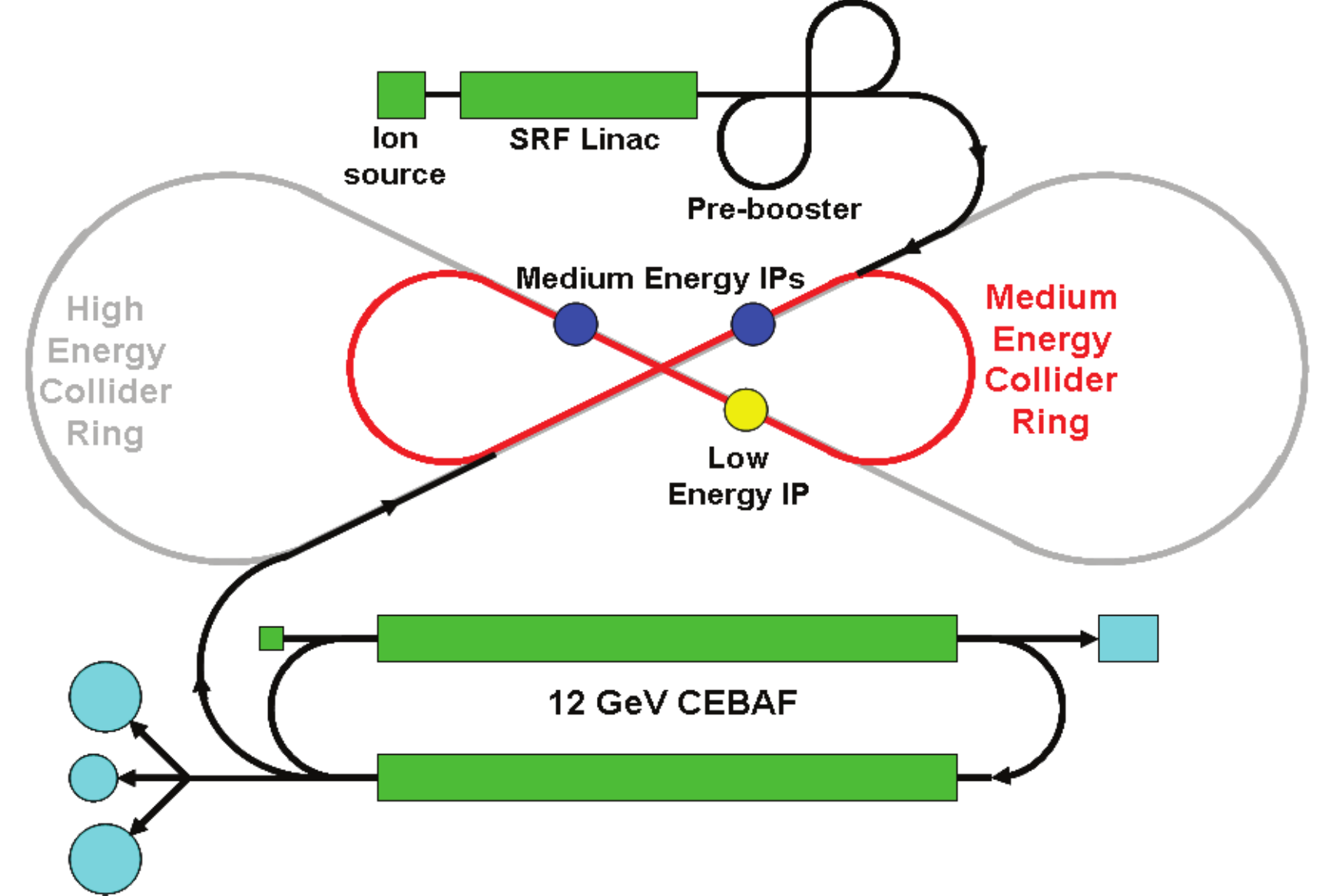}
\end{center}
\vspace{-8 mm}
\caption{\label{fig:EIC}\small
The Layout of the two proposed EIC machines: eRHIC (left) and ELIC (right).
}
\end{figure}

In order to open a new window into a kinematic regime that allows the systematic study of
quarks and gluons,
EIC is designed to provide a wide range in c.o.m.~energies, polarized lepton
and light ions beams and heavy ion beams, all at a very high luminosity \cite{Accardi:2012hwp}.
This creates an unprecedented opportunity for discovery and precision measurements, and
would allow us to study the momentum and space-time distribution of gluons and sea quarks
in nucleons and nuclei  \cite{EIC11,Accardi:2012hwp}. The main requirements for an EIC machine are:

\begin{itemize}
\item Highly polarized ($>70 \%$) electron and proton/light ion  beams;
\item Ion beams from deuteron to heaviest nuclei (uranium, lead);
\item Variable center of mass energy, ranging from about $20 \GeV$ up to $150 \GeV$;
\item Collision luminosity $\sim 10^{33-34}\,{\rm cm}^{-2}{\rm s}^{-1}$.
\end{itemize}

Two independent designs for a future EIC have evolved, eRHIC and ELIC, both using part of
already available infrastructure and facilities (see chapter~5 in \cite{Accardi:2012hwp}).
At Brookhaven National Laboratory (BNL) the eRHIC design (Figure~\ref{fig:EIC} left)
utilizes a new electron beam facility based on an Energy Recovery LINAC (ERL) to be built
inside the RHIC tunnel to collide with RHICs high-energy polarized proton and nuclear beams.
At JLAB the ELIC design (Figure~\ref{fig:EIC} right) employs a new
electron and ion collider ring complex together with the $12\GeV$ upgraded CEBAF, now
under construction, to achieve similar collision parameters.
The kinematic phase space achievable at an EIC for
electron-proton collisions is shown in Fig.~\ref{x-q2-dvcs} and compared to existing
DVCS data and planned future experiments. At an EIC it will be possible to study DVCS
measuring, for the first time simultaneously and with high accuracy, both differential
cross section and spin and charge asymmetries in a kinematic range that extends from large
$\xB$, typical for fixed target experiments, down to small $\xB$, typical for the HERA collider
experiments.

The present study is based on the eRHIC version of an EIC and its new dedicated
detector, designed to fulfill the requirements for the golden experiments at an EIC and
thus being simultaneously highly efficient for inclusive, semi-inclusive and exclusive
reactions. The eRHIC expected luminosity for $e p$ collisions as a function of the
beam-energy is shown in Figure~\ref{fig:eRHIC_lumi}. At eRHIC the full range of proton-beam
energies will be at hand from the early beginning of operations, whereas the energy of
the new electron-beam will be initially at $5-10 \GeV$ (stage I) and will be later upgraded
to higher energies up to $20-30 \GeV$ (stage II).
The newly designed eRHIC detector, shown in Fig.~\ref{fig:eRHIC_lumi},  will have the
following properties:

\begin{figure}[htbp]
\begin{center}
\includegraphics[width=0.35\textwidth]{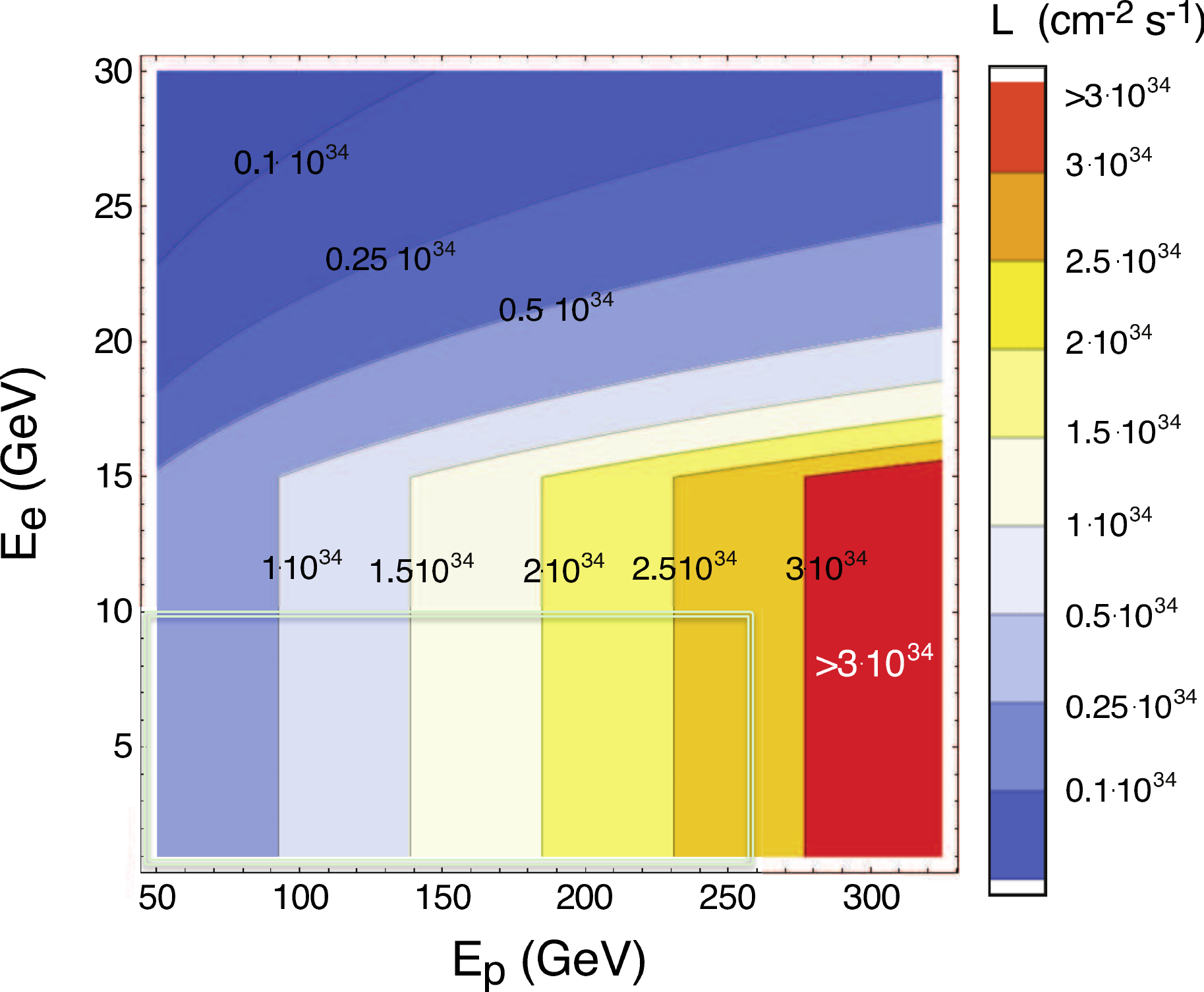}
\includegraphics[width=0.58\textwidth]{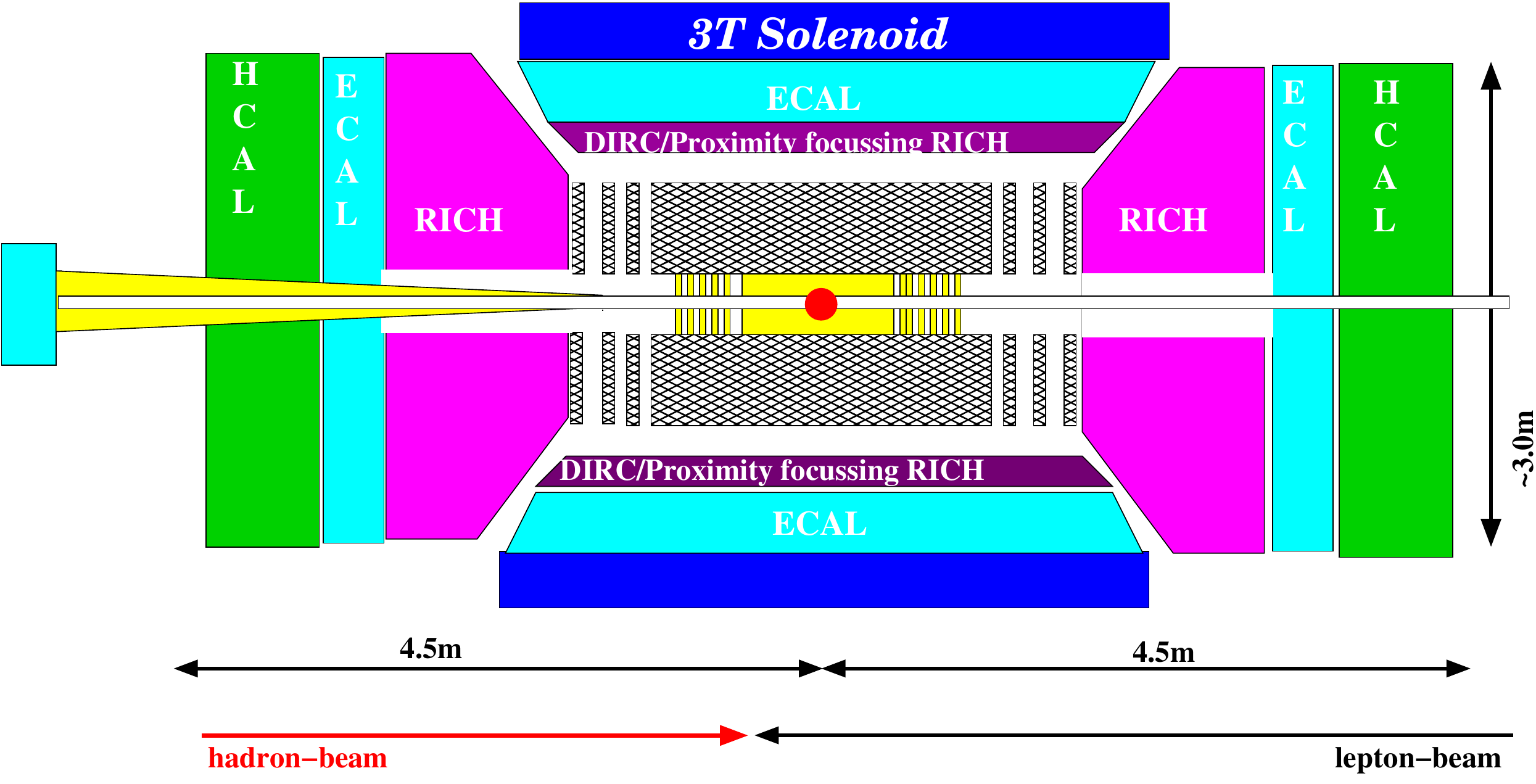}
\end{center}
\caption{\label{fig:eRHIC_lumi}
\small
Left: the expected luminosity at the eRHIC collider as a function of the
beam-energy configuration for $e p$ collisions. Right: a sketch of the eRHIC detector.  }
\end{figure}

\begin{itemize}
\item Wide acceptance $-5 < \eta < 5$ for both the scattered lepton and the produced hadrons;

\item The same rapidity coverage in electromagnetic calorimetry and tracking;

\item High electron track finding/reconstruction efficiency, capability to discriminate
two electromagnetic clusters down to a difference of $1$~degree of polar angle in the rear
endcap electromagnetic calorimeter and good precision for momentum (energy) reconstruction;

\item Particle identification to separate electrons and hadrons as well as pions, kaons
and protons over a momentum range of $0.5 \GeV$ to $10 \GeV$ for rapidities between
-1 to 1 and $0.5 \GeV$ to $80 \GeV$ for $1 < |\eta| < 3$;

\item Good vertex resolution;

\item High acceptance for forward going protons and neutrons from exclusive reactions
as well as from heavy ion breakup (Roman Pots and Zero Degree Calorimeter will be part
of the detector).

\item Low material budget to reduce electron bremsstrahlung and to achieve good resolution
in the reconstruction of all the kinematic variables.

\item Very small low scattering angle forward scattered electron tagger
      ($Q^2 < 0.1$ GeV$^2$)
\end{itemize}

The Monte Carlo (MC) generator used in the present study is MILOU~\cite{PerSchFav04}, which
simulates both the DVCS and the BH (initial and final state radiation) processes together
with their interference term.
It is explicitly noted that the case of the incoming electron radiating a photon
before the actual DVCS process can also be simulated.
It is based on the code by Freund/McDermott \cite{Freund:2002qf,Freund:2003qs}, which
utilizes the approximations described in \cite{Belitsky:2001ns}, and is tuned to H1 and ZEUS
measurements. The DVCS amplitude is evaluated in a GPD-inspired
framework to NLO accuracy \cite{Belitsky:1997rh,Mankiewicz:1997bk,Ji:1997nk,Ji:1998xh},
including the NLO GPD evolution \cite{Belitsky:1999hf}, by a routine, which provides tables
of CFFs.
The real and imaginary parts of CFFs then are used to calculate the cross sections for
DVCS, BH and their interference term.
The $t$-dependence of the DVCS amplitude is introduced as an exponential, i.e., the DVCS
cross section reads
$$\frac{d\sigma^{\rm DVCS}(W,t,\Q^2)}{dt}\propto \exp\left\{B(\Q^2) t\right\},$$ with the
exponential $t$-slope parameter $B(\Q^2)$ being constant or having a logarithmic $\Q^2$-dependence%
\footnote{If the $t$-dependence of flavor singlet quark and gluon GPDs is chosen
differently at the input scale, perturbative evolution will alter the $t$-dependence
for the resulting DVCS cross section.
This should not be confused with the MILOU option to alter additionally the $\Q^2$-dependence of
the exponential $t$-slope by hand for a given GPD model.  \label{foot:MILOU} }.
The MILOU code has been slightly modified from its original version as described in
Appendix~\ref{appendix:MILOU}.

The simulations used for our studies are based on the following MILOU options:
\begin{itemize}
\item The slope $B(\Q^2)= 5.6\,\GeV^{-2}$ is set to be constant.

\item CFFs tables are generated from a GPD model to NLO and twist-two accuracy.

\item Proton dissociation background, $ep\rightarrow e\gamma Y$, has not been included
in the simulation.
\end{itemize}
To our best knowledge, the first two choices guarantee that a pure and consistent
GPD framework is utilized in the MILOU simulations, see \cite{Diehl:2007jb} and
footnote \ref{foot:MILOU}.

The DVCS and BH processes have been simulated according to the following selection
criteria:

\begin{itemize}
\item $\Q^2 \ge 1 \GeV^2$;  $10^{-5}<\xB<10^{-1}$; binned logarithmically in 4
$\Q^2$- and 5 $\xB$-bins per decade and in several $|t|$-bins;
the bins in $\Q^2$ are: $1.0< \Q^2 < 1.78 \GeV^2$; $1.78 < \Q^2 < 3.16 \GeV^2$;
$3.16 < \Q^2 < 5.62 \GeV^2$; $5.62 < \Q^2 < 10 \GeV^2$; $10 < \Q^2 < 17.78 \GeV^2$.

\item Detector acceptance criteria: $0.01<y<0.85$ for the asymmetries, 
$0.01<y<0.60$ for the cross sections and $|\eta| < 5.0$ for the scattered
electron and produced photon, and the scattered proton acceptance: $0.03<|t|<1.5 \GeV^2$
(proton detected in the roman pots);

\item BH rejection criteria applied for the cross section measurement: em-clusters-energy
$> 1 \GeV$; $\theta_{el}-\theta_\gamma > 0\,{\rm rad}$.
In the case of a DVCS event with initial state radiation, the radiated photon is emitted
collinear to the incoming lepton beam, which means it remains undetected and leads
to a mis-reconstruction of the kinematic variables, i.e. $Q^2$ and $x$, of the process.
Thus, the ISR has been taken into account in the simulation and it can be shown that
only 15\% of the events radiate a photon carrying more than 2\% of the incoming
electron energy. These events can be nicely corrected to Born level using MC
simulations.

\end{itemize}
The $\Q^2$ and $\xB$ range is within the phase space reachable with an EIC/eRHIC.
The electron and proton beam-energy configuration considered for the present study are:
$5\times100 \GeV^2$, $5\times250 \GeV^2$  (for stage I) and $20\times250 \GeV^2$
(an example for stage II).

\begin{figure}[ht]
\centering
\includegraphics[width=0.32\textwidth,angle=0]{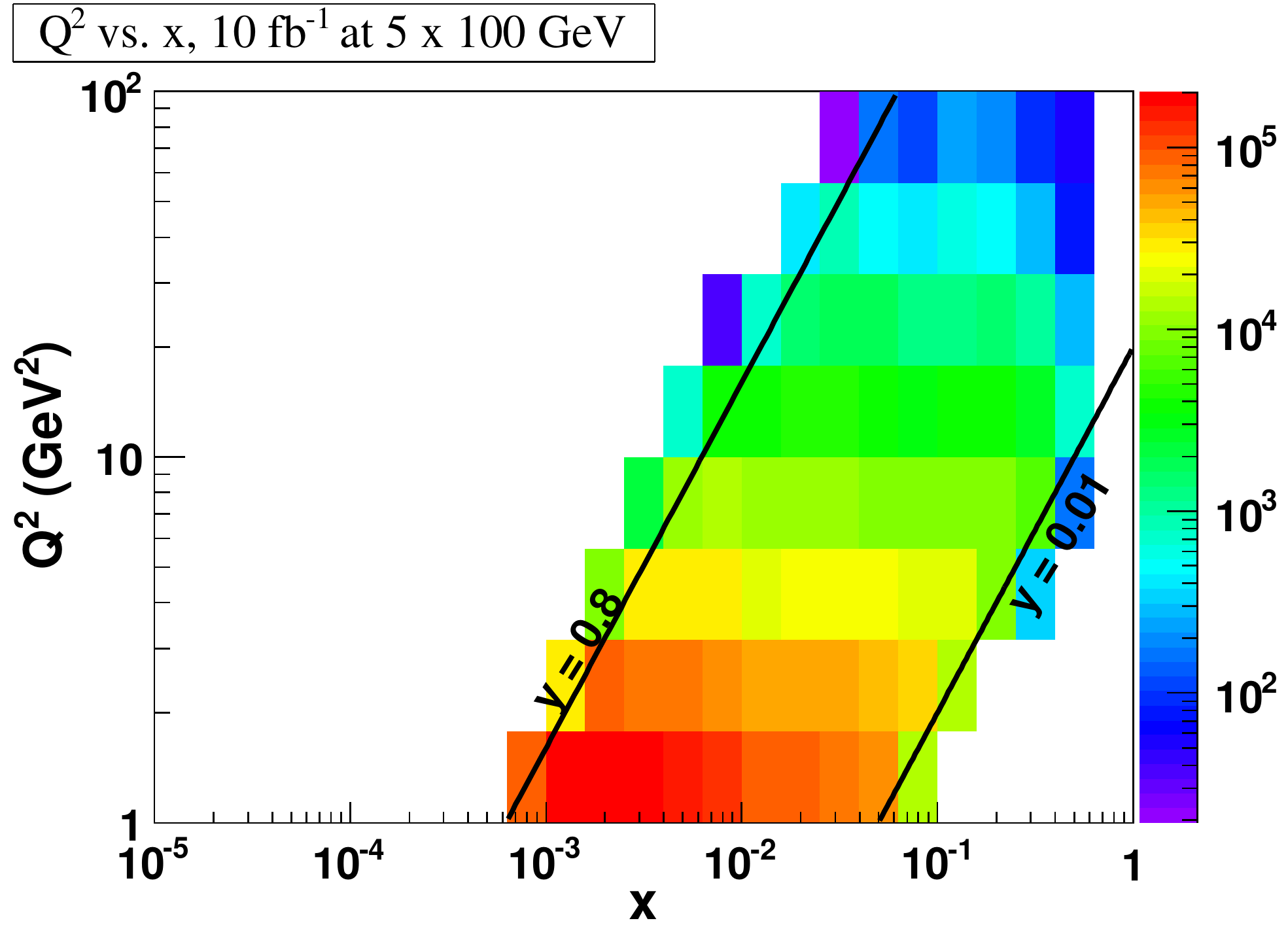}
\includegraphics[width=0.32\textwidth,angle=0]{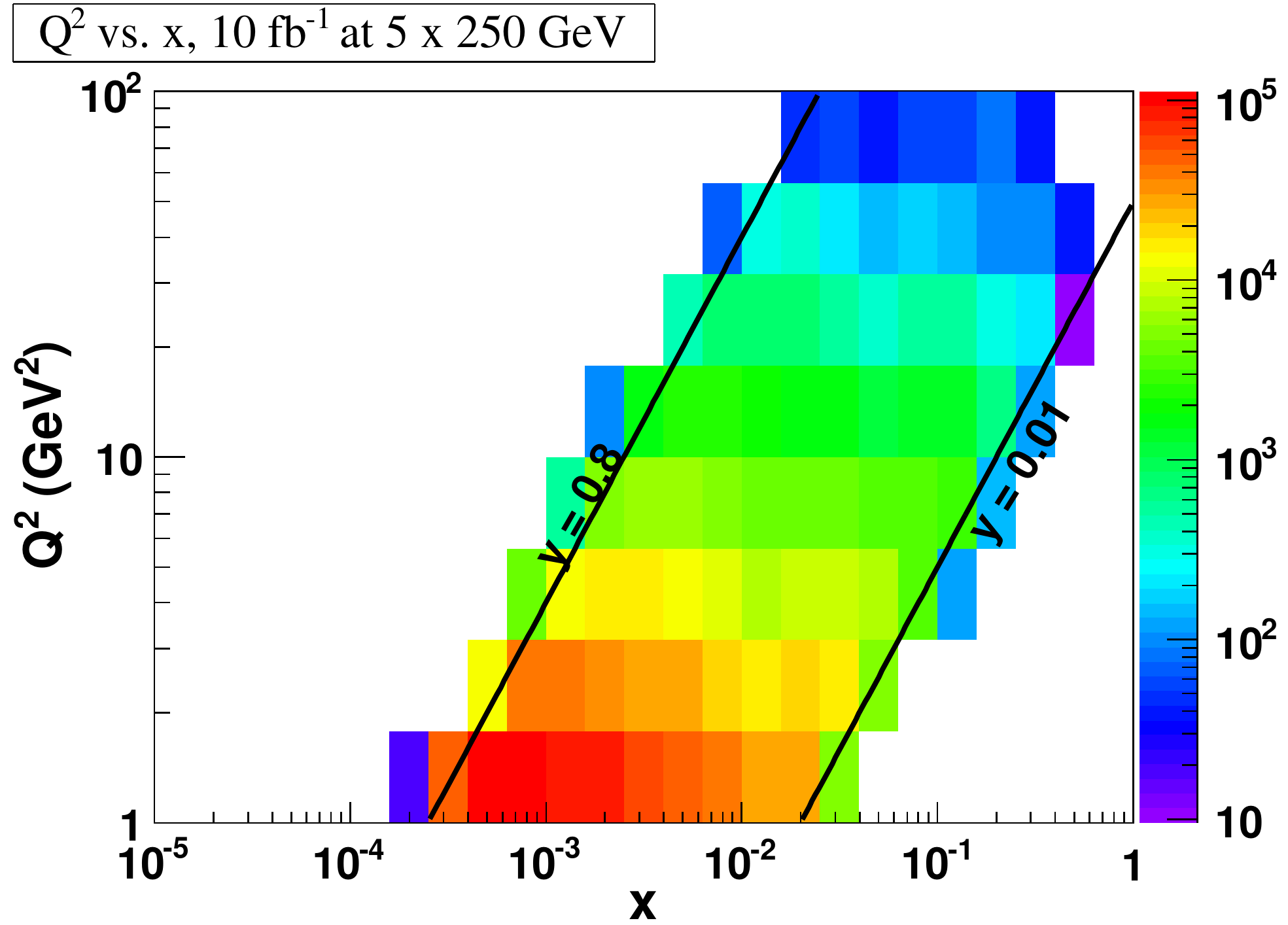}
\includegraphics[width=0.32\textwidth,angle=0]{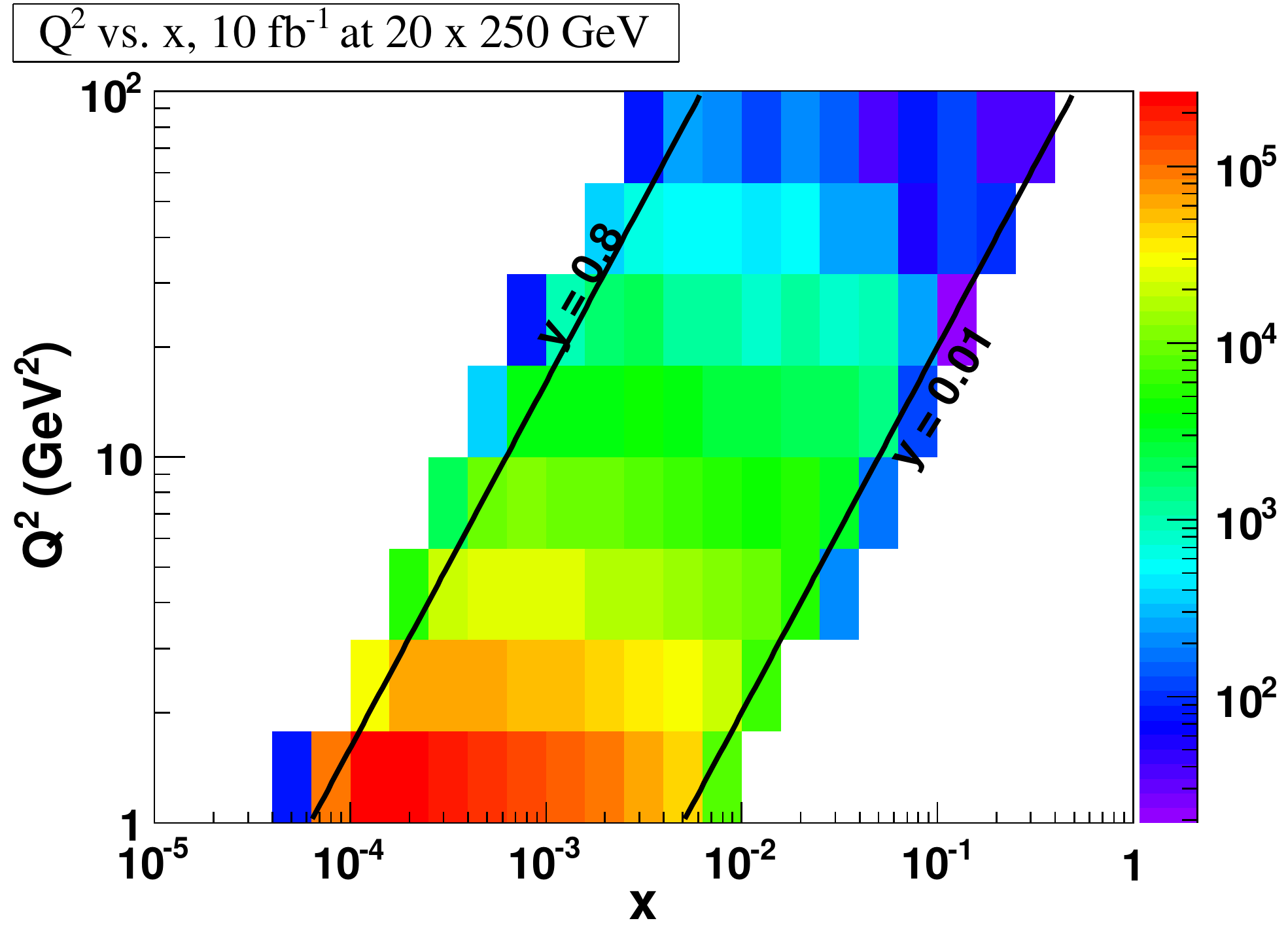}
\caption{\small The distribution of statistics in each $\{\Q^2,\xB\}$ bin for
eRHIC stage I (left and middle) and stage II (right) at a luminosity of $10\,{\rm fb}^{-1}$.
}
\label{phase_space}
\end{figure}

For the purpose of  DVCS cross section measurements it is
important to remove from the signal the background coming from the BH
events. The latter is a QED process, well known to an uncertainty of the order of
3\% coming from the uncertainty on the proton form factors. It can be subtracted from the
signal by means of a MC technique. Thus, especially at a high luminosity machine like
eRHIC where systematic uncertainties will dominate the measurements, it is important to
minimize the BH contribution, particularly at low c.o.m.~energy, where BH tends
to dominate over the DVCS (see Sect.~\ref{sec:Rosenbluth}).
The fraction of BH events has been estimated using a MC sample containing both DVCS and
BH processes.
The BH contamination was investigated for each $\{\Q^2, \xB, t\}$ bin as a function of the
electron energy loss $y$. After all BH suppression
criteria have been applied it was found that at large c.o.m.~energies the
BH contamination grows from negligible (at low-$y$) to about 70\% at $y\sim0.6$ allowing
for a safe BH subtraction, whereas  for lower c.o.m.~energies the BH contamination
grows faster with $y$ and can be dominant depending on the bin; nevertheless most of the
statistics at this low c.o.m.~energy is contained in the safe region $y<0.3$.

Figure~\ref{phase_space} compares the distribution of the statistics per bin for the
eRHIC beam-energy configurations $5\times100 \GeV^2$,  $5\times250 \GeV^2$
(both reachable at a stage I) and $20\times250 \GeV^2$ (available at a stage II),
considering an integrated luminosity of $\sim 10\,{\rm fb}^{-1}$.
The results shown in the present paper are based on simulated data samples corresponding
to an integrated luminosity of $100\,{\rm fb}^{-1}$ for the $20\times250 \GeV^2$ configuration and
$10\,{\rm fb}^{-1}$ for the $5\times100 \GeV^2$ configuration, both corresponding to
approximately 1 year of data taking at eRHIC assuming a 50\% operational efficiency.
The data samples generated for the propose of measuring the differential cross section
only contain the DVCS process whereas samples containing DVCS, BH, and their
interference term have been generated for measurements of different single spin asymmetries.

All the generated events have been smeared according to expected momentum and angular
resolutions.
The statistical uncertainty for the differential cross section can be at small values of
$-t$ as low as few percent; the same is true for the uncertainty for the extracted
slope parameter $B$. This implies that the measurement is actually limited by systematics.
For the purposes of the present work, a systematic uncertainty of 5\% has been assumed,
based on the experience at HERA and the expected coverage and technology
improvements of the new detector at eRHIC. The overall systematic uncertainty, due to the
uncertainty on the measurement of luminosity, is not considered for this paper as it
simply affects the normalization of the cross section measurement.

\section{Selected DVCS observables at EIC}
\label{sec-pred}

As explained in Sect.~\ref{sec:presendDVCS}, the isolation of CFFs is a rather intricate
task, which can be only achieved by measuring a complete set of observables. However,
we have also seen that photon helicity flip contributions, which are suppressed in DVCS
kinematics, are not traceable in the present world data set.
Hence, we restrict ourselves to four twist-two associated  CFFs to study
the physics case of DVCS measurements at a suggested eRHIC, giving emphasis to twist-two
dominated observables.  As motivated in  Sect.~\ref{sec:simulation}, we choose two scenarios: one with
a relatively low and another with a high c.o.m.~energy, corresponding to the beam configurations
$$E_e\times E_p = 5\times100 \GeV^2\quad\mbox{and}\quad E_e\times E_p = 20\times250 \GeV^2.$$

For future DVCS measurements at $5\times100 \GeV^2$ it is maybe expected that
the description of precise data in this region of transition to the small-$\xB$ physics requires
rather complex GPD models, which are not needed for the description of the present  DVCS
data. For the higher energy case it is expected that valence quark contributions are negligibly
small and non-negligible CFFs
$$\xB\times {\cal H}\quad \mbox{and potentially}\quad \xB\times {\cal E} $$
are governed by an effective ``pomeron'' exchange in the $t$-channel, associated with
both sea quarks and gluon contributions, and  that they  (moderately) grow with decreasing $\xB$.
Thereby, almost  nothing is known about the CFF $\cal E$, which, as pointed out in
Sect.~\ref{sec:phenomenology},
is not accessible from present DVCS measurements in neither
the collider nor the fixed target mode. The available theoretical/phenomenological guidance
is not yet fully trustworthy. On one hand a ``pomeron'' coupling to
proton helicity non-conserved quantities such as the CFF $\cal E$ is phenomenologically not
established, see Ref.~\cite{Don05} and references therein. On the other hand a pomeron
like behavior for the
CFF $\cal E$  is perturbatively predicted by GPD evolution%
\footnote{As for the perturbative evolution of unpolarized PDFs in the flavor singlet
sector, the evolution of both GPD $H$ and $E$ in this sector is at small $x$ driven by
gluons, which generate an effective `pomeron' like behavior. The solution of the
evolution equation yields in fact an essential singularity rather a pole. Such a
behavior can be only avoided if both the quark singlet and gluon GPDs vanish simultaneously.}.
A separate study on the access of GPD $E$ as well as the transverse spatial distribution of sea
quarks and gluons at stage II  will be presented in Sect.~\ref{sec:interpretation}, which
without
additional information or assumptions is hard to achieve for EIC measurements at
low beam energies.

We expect that the remaining two twist-two associated CFFs if multiplied with $\xB$,
$$\xB\times \widetilde {\cal H} \quad \mbox{and}\quad
\xB\times\overline{\cal E} \approx \xB\times \frac{\xB}{2-\xB} \widetilde{\cal E},$$
go to zero in the small-$\xB$ region. Note, however, that in contrast to the CFF ${\cal H}$,
Regge phenomenology provides no clear guidance for their small-$\xB$ behavior. The
phenomenological situation is analogous to the polarized DIS function $g_1$ (GPD $\widetilde H$
embeds the polarized PDF $\Delta q$). We emphasize that these essentially
unknown contributions may play a role at the stage I kinematics.

To cover possible scenarios, we employ  in our studies three hybrid models, where sea quark
and gluonic components of CFFs $\cal H$ and $\cal E$ are based on GPD models that include the
perturbative evolution, while their valence quarks and remaining GPDs are treated with
dispersion relations as described in Sect.~\ref{sec:GPD22CFFs} . Two
of the models are pinned down from global fits to the world data of unpolarized DVCS measurements,
which are described very well, despite having rather different partonic content.
We now list the models and describe their main properties.
\begin{itemize}
\item \KM describes the world data set of DVCS measurements using an unpolarized proton
 target. It contains the twist-two GPDs $H$ and $\widetilde H$, while the real part of
 helicity-flip CFFs $\cal E$ and $\widetilde{\cal E}$ are only given by subtraction
 constants in the dispersion relation (related to so-called $D$-term and pion pole
 contribution, respectively). Both GPD $\widetilde H$ and the (real) CFF
 $\widetilde{\cal E}$ are rather large and they are considered as effective degrees of
 freedom that allow to describe the unpolarized cross section measurements from
 the Hall A collaboration \cite{Munoz_Camacho:2006hx}.

\item \KMa is analogous to the \KM model; however, the Hall A cross section measurements
 are not well described. In this model the GPD $H$ is the dominant one, $\widetilde H$
 is set to zero, and $\widetilde{\cal E}$ contains only the pion pole, which is accounted in
 the standard way \cite{Mankiewicz:1998kg,Frankfurt:1999xe}.

\item \our is a flexible GPD model for the small-$x$ region, specifically designed for the
 present study. It contains besides the sea quark and gluon GPDs $H^{\rm sea}$ and
 $H^{\rm G}$ also a flexible small-$x$ parametrization of GPDs $E^{\rm sea}$ and
 $E^{\rm G}$.  All of these GPDs include a ``pomeron'' behavior, which can be individually
 adjusted at the input scale. The normalization of $E$-type GPDs is controlled by the
 anomalous magnetic moment of sea quarks $\kappa^{\rm sea}=1.5$, which is fixed to
 be positive and rather large. The parton polarized GPDs $\widetilde H$ and $\widetilde E$
 are set to zero.
\end{itemize}

Our small-$x$ GPD models are set up in terms of (conformal) GPD moments
rather than in $x$-space, at the input scale $\Q^2=4 \GeV^2$ for four light quarks.
They yield, similarly to other GPD models, the following {\it effective} functional form%
\footnote{This form arises exactly in the small-$x$ limit of standard GPD models at the input scale;
however, strictly spoken it is not stable under perturbative evolution. Nevertheless,
the resulting CFF output of a GPD model can be reparameterized for a given $\Q^2$ value
and put in a Regge-inspired form. Thereby, the ``pomeron'' trajectory is altered,
indicated by its $\Q^2$-dependence.}
of CFFs:
\begin{eqnarray}
\label{cffHcffE-smallx}
\left\{ {\cal H} \atop {\cal E} \right\} (\xB,t,\Q^2) \sim \pi
\left[i-\cot\left(\frac{\pi \alpha(t,\Q^2)}{2}\right) \right]
\xi^{-\alpha(t,\Q^2)} \left\{ h_\alpha \atop e_\alpha \right\}(t,\Q^2)\,,
\end{eqnarray}
which resembles a Regge phenomenological ansatz with a linear ``pomeron'' trajectory
\begin{eqnarray}
\alpha(t) = \alpha(t=0) + \alpha^\prime t.
\end{eqnarray}
In the \KM and \KMa models a dipole parametrization $\left(1-\frac{t}{M^2}\right)^{-2}$
for the residual $t$-dependency was taken, while the \our model alternatively relies,
as in the MILOU simulation, on an exponential ansatz $e^{b t}$.
The boundary value of the residue $h_\alpha$ at $t=0$ depends on both the momentum
fractions $N^i(\Q^2)$, carried by the unpolarized parton type $i$, and the skewness effect,
parameterized in terms of two model parameters $s^i_2$ and $s^i_4$, which control both the
normalization of the CFFs and their $\Q^2$ evolution; a detailed discussion is given in
\cite{Kumericki:2007mh,Kumericki:2009uq,doi:10.1142/S201019451100167X}. Analogously, we
parameterize in the \our model the GPD $E$ with an independent set of parameters, however,
here the momentum fractions $N^i(\Q^2)$ are replaced by the partonic gravitomagnetic
moments ${\cal B}^i= N^i \kappa^i$, parameterized at the input scale by the product of the momentum
fraction $N^i$ and the anomalous magnetic moments $\kappa^i$. The momentum and gravitomagnetic
sum rules are utilized to fix the gluonic momentum fraction and gravitomagnetic moment,
respectively. From a DIS fit the PDF-related parameters were found \cite{Kumericki:2009uq},
\begin{eqnarray}
N^{\rm sea}=0.152\,, \quad \alpha^{\rm sea}=\alpha_E^{\rm sea}=1.158\,, \quad
\alpha^{\rm G}=\alpha_E^{\rm G}=1.247\,,
\end{eqnarray}
which we, for simplicity, also adopt for GPD $E$ in the \our model.
Some other relevant model parameters are listed in Tab.~\ref{tab:parameters},
where, again for simplicity, we equate the Regge slope parameters of GPD $H$
and the residue slope parameter for GPD $E$ with those of GPD $H$,
\begin{eqnarray}
\label{cff-ansatz_Regge}
\alpha^{\prime~{\rm G}}=\alpha^{\prime~{\rm sea}}\,
\quad
b^{\rm sea}_E=b^{\rm sea}\,,
\quad
b^{\rm G}_E= b^{\rm G}\,.
\end{eqnarray}

Finally, we specify the remaining GPDs on the cross-over line and the form of subtraction constants,
where the CFFs are calculated from
 (\ref{KLMSPM-DR-Im}, \ref{KLMSPM-DR-Re}, \ref{KLMSPM-DR-Re1}).
\begin{table}
\begin{center}
\begin{tabular}{|c||c|c|c|c|c|c|c|c|c|c|c|}
  \hline
\phantom{A}model\phantom{A} &
$\alpha^{\prime~{\rm sea}}$ &
$\kappa^{\rm sea}$ &  $\alpha_{E}^{\prime~{\rm sea}}$ &
$\alpha_{E}^{\prime~{\rm G}}$  & $(M^{\rm sea})^2$ & $(M^{\rm G})^2$ &
$b^{\rm sea}$ & $b^{\rm G}$ \\ \hline  \hline
  \KM\!\!(a)  & 0.15& 0.0& -- & --& 0.51(0.52)& 0.7& --& -- \\
  \our & 0.10& 1.5& 0.02& 0.05& --& --        & 2.8  & 2.0 \\
  \hline \hline
\end{tabular}
\hspace{29mm}\phantom{end}
\end{center}
\vspace{-8mm}
\begin{center}
\begin{tabular}{|c||c|c|c|c|c|c|c|c|c|c|c|c|c|}
  \hline\hline
  \phantom{A}model\phantom{A} & $r$ & $b$ &  $M$ &  $c$  & $M_c$ & $\widetilde r$ & $\widetilde b$ & $\widetilde M$ & $r_\pi$ & $M_{\pi}$
  \\ \hline  \hline
  \KM\phantom{i} & 0.620& 0.404 & 4. & 8.777& 0.975 & 7.759	& 2.050& 0.884  & 3.536 & 4.020\\
  \KMa & 0.884& 0.400& 1.5& 1.722& 2.000 & 0.000  &   --  &  -- &  {\small cf.~\cite{PenPolGoe99}} & {\small cf.~\cite{PenPolGoe99}} \\
  \hline
\end{tabular}
\end{center}
\vspace{-3mm}
\caption{\small Some selected model parameters for unpolarized sea quark and gluon GPDs (upper table),
valence $H$ and $\widetilde{H}$ GPDs as well as for subtraction constants
(lower table), where squared mass parameters are given in $\GeV^2$ and slope parameters
$\alpha^\prime$ and $B$ in $\GeV^{-2}$.  }
\label{tab:parameters}
\end{table}
Only the target helicity conserved GPDs on the cross-over line are modeled
\begin{eqnarray}
\label{KLMSPM-ansHval}
H^{\rm val}(x,x,t) & = &
\frac{1.35\,  r}{1+x} \left(\frac{2 x}{1+x}\right)^{-\alpha(t)}
\left(\frac{1-x}{1+x}\right)^{b}
\left(1-  \frac{1-x}{1+x} \frac{t}{M^{\rm val}}\right)^{-1}\,, \\
\widetilde H(x,x,t)  & = & \frac{0.6\,  \widetilde r}{1+x}
\left(\frac{2 x}{1+x}\right)^{-\alpha(t)}
\left(\frac{1-x}{1+x}\right)^{ \widetilde b}
\left(1-  \frac{1-x}{1+x} \frac{t}{ \widetilde M}\right)^{-1}\,.
\end{eqnarray}
Here, the skewness effect is parameterized by the ratios
$$
r=\lim_{x\to 0}\frac{H(x,x,0)}{H(x,0,0)} \quad\mbox{and}\quad  \widetilde r=\lim_{x\to 0}\frac{\widetilde H(x,x,0)}{\widetilde H(x,0,0)}\,,
$$
$\alpha(t) =  0.43 + 0.85\, t/{\rm GeV}^2$,  $b$ ($\widetilde b$)  controls the
$x\to1$ limit,  $M^{\rm val}$ ( $\widetilde M$) the residual $t$-dependence,
where $q(x)=H(x,0,0)$ ($\Delta q(x)=\widetilde H(x,0,0)$) are unpolarized (polarized)
reference PDFs, e.g., the LO parametrization of \cite{Alekhin:2002fv} (\cite{GehSti95}).
The subtraction constant is normalized
by $c$ ($r_{\pi}$) and the cut-off mass $M_c$ ($M_\pi$) controls the $t$-dependence:
\begin{eqnarray}
\label{KLMSPM-ansD}
{\cal C}(t) =  \frac{c}{\left(1- \frac{t}{M_c^2}\right)^{2}}\,, \qquad
\widetilde{\cal C}(t) =\frac{2.164\, r_{\pi}}{(m_{\pi}^2-t) \left(1-\frac{t}{M_\pi^2}\right)^2}\,,
\end{eqnarray}
where $m_{\pi} \approx 0.14 \GeV$ is the pion mass and the normalization factor $2.164$
in the pion pole contribution matches the residue of the $t=m_\pi^2$ pole from the pseudo
scalar form factor $2 g_A M_p^2/(1+m_{\pi}^2/M_{\pi}^2) $ with $M_{\pi}=1.17 \GeV$. Note,
however, that in the GPD framework the normalization of the pion pole contribution
remains unknown.
In the \KMa model we use the pion pole parametrization of \cite{PenPolGoe99}.
More explanations on these simple  parameterizations can be found in \cite{Kumericki:2009uq}.
The parameters of the \KM and \KMa models are listed in Tab.~\ref{tab:parameters}.

In the remainder we illuminate the richness of a possible experimental DVCS program at an
suggested EIC.
Thereby, we will concentrate on  observables that are dominated by
twist-two associated CFFs.  In Sect.~\ref{sec-pred-XDVCS}
we restrict ourselves to the unpolarized cross
section  and in Sect.~\ref{sec:meas-SA} to single spin asymmetry measurements.
In Sect.~\ref{sec:opportunities} we will comment on further DVCS related measurements,
which are interesting on their own, and we shortly discuss the use of an unpolarized positron
beam to disentangle photon helicity non-flip contributions from longitudinal-transverse
helicity ones.

\subsection{Cross section measurements at stage I}
\label{sec-pred-XDVCS}

As emphasized in Sect.~\ref{sec:Rosenbluth},  the separation of the measurable
electroproduction cross section (\ref{dX-reduced})
into its three parts in the most model independent way and/or with a minimal
set of assumptions is an important goal.
So far the extraction of the $t$-differential DVCS cross section, entering
in (\ref{dX-reduced}),
has been only reached in the small-$\xB$ and $0.1 \GeV^2 \leq  -t \leq 0.8 \GeV^2$
region by the H1 and ZEUS collaborations. Thereby, the subtraction method
\begin{eqnarray}
\label{dsigma^{DVCS}-sub}
\frac{d\sigma^{\mbox{\tiny DVCS}}(\xB,t,\Q^2) }{dt}  \simeq  \frac{d\sigma^{\mbox{\tiny TOT}} (\xB,t,\Q^2)}{dt} -\frac{d\sigma^{\mbox{\tiny BH}}(\xB,t,\Q^2) }{dt}, \end{eqnarray}
was utilized,  where the interference term could be safely neglected and the BH
cross section was simulated. The latter was cross-checked experimentally
in the BH dominated phase space region.

To understand whether such a subtraction procedure would be also reliable in the
EIC kinematics and whether one can improve this method by utilizing the variable
beam energy option,
we consider first the generic dependence of the $t$-differential cross
section (\ref{dX-reduced}) on its variables. According to what was
explained in  Sect.~\ref{sec:Rosenbluth}, for smaller value of
$-t \gg -t_{\rm min} \approx \xB^2 M_p^2$  and large $y$ the BH cross section
dominates, since it is enhanced by the kinematical prefactor $y^2/(-t\Q^2)$. On
the other hand in the limit $y\to 0$ both the BH cross section and the
interference term drop out, where $\varepsilon(y=0)=1$ and,
thus, the sum of the transverse and longitudinal DVCS cross sections can be
accessed, see (\ref{XTOT-y}).
Moreover, the interference term (\ref{XINT-red}) has the same canonical
$1/\Q^4$ scaling as the DVCS cross section (\ref{XDVCS-red}), however, it has an
additional prefactor $\xB\, y$.
Restricting ourselves to the dominant CFF ${\cal H}$, we find that
the ratio of interference term (\ref{XINT-red}, \ref{XINT-red1}) to the sum of
BH (\ref{XBH-approx}) and DVCS (\ref{XDVCS-approx}) cross sections is estimated,
for smaller-$\xB$ values, to be
\begin{equation}
\label{XINT2XDVCS}
\frac{d\sigma^{\mbox{\tiny\rm INT}}}{d\sigma^{\mbox{\tiny\rm BH}}+d\sigma^{\mbox{\tiny\rm DVCS}}} \sim  2 \xB \sqrt{\frac{-(1-y) t}{\Q^2}}\,
\frac{F_1(t)\, \sqrt{\frac{-t(1-y)}{4y^2\Q^2}}\, \Re{\rm e}\, \xB {\cal H}(\xB,t,\Q^2)}{F_1^2(t)-\frac{t}{4 M_p^2} F_2^2(t) +
\frac{-t(1-y)}{4y^2\Q^2} \left|\xB {\cal H}(\xB,t,\Q^2)\right|^2}\,.
\end{equation}
Obviously, the suppression factor $2\xB \sqrt{-(1-y) t/\Q^2} \lesssim \xB$ (DVCS
requires $-t \ll \Q^2$ ) makes this ratio small for EIC kinematics. Moreover, we
expect from Regge arguments, consistent with phenomenological findings, that the
real part of the dominant CFF ${\cal H}$ is in the small- and even
moderate-$\xB$ region
much smaller than its imaginary part (at least for smaller values of $-t$, see
the results from HERMES in Fig.~\ref{fig:regr}). We conclude that
in most of the stage I bins, given in Sect.~\ref{sec:simulation}, the
interference term is negligible and we can simplify the $t$-differential cross
section (\ref{XTOT-y}) to
\begin{eqnarray}
\label{XTOT-EIC}
\frac{d\sigma^{\mbox{\tiny TOT}}}{dt} \approx
\frac{y^2  \left[ \frac{d\sigma_{\rm T}^{\mbox{\tiny BH,red}}}{dt}  + \varepsilon(y) \frac{d\sigma_{\rm L}^{\mbox{\tiny BH,red}}}{dt} \right] }{\left(1-y \frac{(1-\xB) t}{\Q^2+t} \right) \left(\frac{\Q^2+t}{\Q^2+\xB t}-y\right)}
+ \frac{d\sigma^{\mbox{\tiny\rm DVCS}}(y)}{dt}
\end{eqnarray}
with $\frac{d\sigma^{DVCS}(y)}{dt} =
\frac{d\sigma_{T}^{DVCS}}{dt} + \varepsilon(y)\frac{d\sigma_{L}^{DVCS}}{dt}.$
The smallness of the interference term has been also seen in numerical GPD model
calculations. Thereby, the use of
the approximate equations in \cite{Belitsky:2001ns} naturally yields only incomplete
cancelations in the $\phi$-integrated interference term. This causes the ratio (\ref{XINT2XDVCS})
to appear proportional
to $(-t/\Q^2)^{3/2}$ rather than to $\xB\times (-t/\Q^2)^{1/2}$. Nevertheless, also
in the MILOU simulations, based on the approximate equations in \cite{Belitsky:2001ns},
the interference term turns out to be negligibly small.

For an EIC experiment the equation (\ref{XTOT-EIC}) provides a further handle to
cross-check experimentally the BH subtraction procedure. However,
we expect that a  Rosenbluth separation of the transverse and longitudinal
DVCS cross section will be difficult to achieve in the small $-t$ region.
To suppress the BH contribution a relatively small $y$ is needed, which
also means that the variation of $\varepsilon(y)$, which functional
dependence can be mimicked by a truncated Taylor expansion
$ \varepsilon(y) \approx 1-\frac{y^2}{2}-\frac{y^3}{2},$
is only small. Moreover, if we stick to the twist-two expansion of the DVCS amplitude,
the longitudinal DVCS cross section in the small-$\xB$ region will be expressed by
twist-three associated CFFs and this
cross section will be kinematical suppressed by a factor
$\widetilde{K}^2/\Q^2 \approx -t/\Q^2$, see (\ref{XDVCS-red}), (\ref{F3toF0+}) and (\ref{F3toF0+}).
On the other hand these behaviors  may offer the possibility of access to the
twist-three contribution at larger values of $-t$,
which, in turn, allows the variation of $y$ over a larger region.
However, such an access may  only be possible if the $t$-dependence of CFFs, as compared to that
of electromagnetic form factors, is rather flat.

The transverse DVCS cross section contains both non-flip and transverse flip
helicity amplitudes, where the latter would be
perturbatively suppressed by $(\alpha_s/2\pi)^2$ or $1/\Q^2$ corrections.
Neglecting the suppressed photon helicity flip contributions and switching
to the GPD-inspired CFF basis (\ref{cffF}),
we can approximately write the DVCS cross section  for stage I kinematics as
\begin{equation}
\label{XDVCS}
\frac{d\sigma^{\mbox{\tiny\rm DVCS}}}{dt} \approx
\frac{\pi \alpha_{\rm em}^2}{{\cal Q}^4}
\left[\frac{\big|\xB {\cal H} \big|^2}{\left(1-\frac{\xB}{2}\right)^2}  -
\frac{t\, \big| \xB{\cal E} \big|^2}{4 M^2_{p}}  +
\frac{\big|\xB \widetilde {\cal H} \big|^2}{\left(1-\frac{\xB}{2}\right)^2} -
\frac{t\,\big|\xB \overline{\cal E} \big|^2}{4 M^2_{p}}  -
\frac{\xB\, \Re{\rm e}\,\widetilde{\cal H}\, \overline{\cal E}^\ast }{1-\frac{\xB}{2}} \right]\!
\left(\xB,t,{\cal Q}^2\right), \nonumber\\
\end{equation}
where the functional form arises from the exact $\cal C$-coefficients by neglecting
kinematically suppressed contributions of order ${\cal O}(\xB^2)$ and ${\cal O}(\xB t/\Q^2)$.
As somehow expected, in our numerical studies it turned out that the  DVCS cross section
(\ref{XDVCS}) for the $5\times 100 \GeV^2$ beam configuration is rather
sensitive to the choice of GPD model. To some extent this is also true for higher c.o.m.~energies
in the large $-t$ region. In other words, a definite conclusion whether the subtraction
method in these kinematics will be possible, cannot be taken without actual data.

As mentioned in Sect.~\ref{sec:simulation}, the eRHIC option allows also at stage
I to increase the proton beam energy, for kinematical coverage see Fig.~\ref{phase_space}.
To illustrate the energy dependence of the DVCS cross section, we consider its ratio
to the measurable electroproduction cross section
\begin{eqnarray}
\label{R^{DVCS} R^{INT}}
\frac{d\sigma^{\mbox{\tiny\rm DVCS}}}{d\sigma^{\mbox{\tiny\rm TOT}}} = \frac{
\int_{-\pi}^\pi\! d\phi\, \frac{d\sigma^{ep\to ep\gamma,\mbox{\tiny\rm DVCS}}}{d\xB dt d\Q^2 d\phi}
}{
\int_{-\pi}^\pi\! d\phi\, \frac{d\sigma^{ep\to ep\gamma}}{d\xB dt d\Q^2  d\phi} }
\,.
\end{eqnarray}
Considering again the CFF $\cal H$ as the dominant one and sticking to the small-$\xB$
approximation with $-t\gg t_{\rm min}$ and $y < y_{\rm col}\approx 1$, we can estimate this ratio as
\begin{equation}
\label{XDVCS2XTOT}
\frac{d\sigma^{\mbox{\tiny\rm DVCS}}}{d\sigma^{\mbox{\tiny\rm TOT}}} \sim
\frac{\frac{-t(1-y)}{4y^2\Q^2} \left|\xB {\cal H}(\xB,t,\Q^2)\right|^2}{F_1^2(t)-\frac{t}{4 M_p^2} F_2^2(t) +
\frac{-t(1-y)}{4y^2\Q^2} \left|\xB {\cal H}(\xB,t,\Q^2)\right|^2}\,.
\end{equation}
Clearly, as long as we stay away from $-t_{\rm min}$, which is at EIC not reachable in the considered bins,
this ratio will get very small at low $-t$ and its behavior at large $-t < \Q^2$
depends on the $-t$ drop-off of CFFs.
\begin{figure}[tt]
\vspace{-0.8cm}
\begin{center}
\includegraphics[width=0.48\textwidth,angle=0]{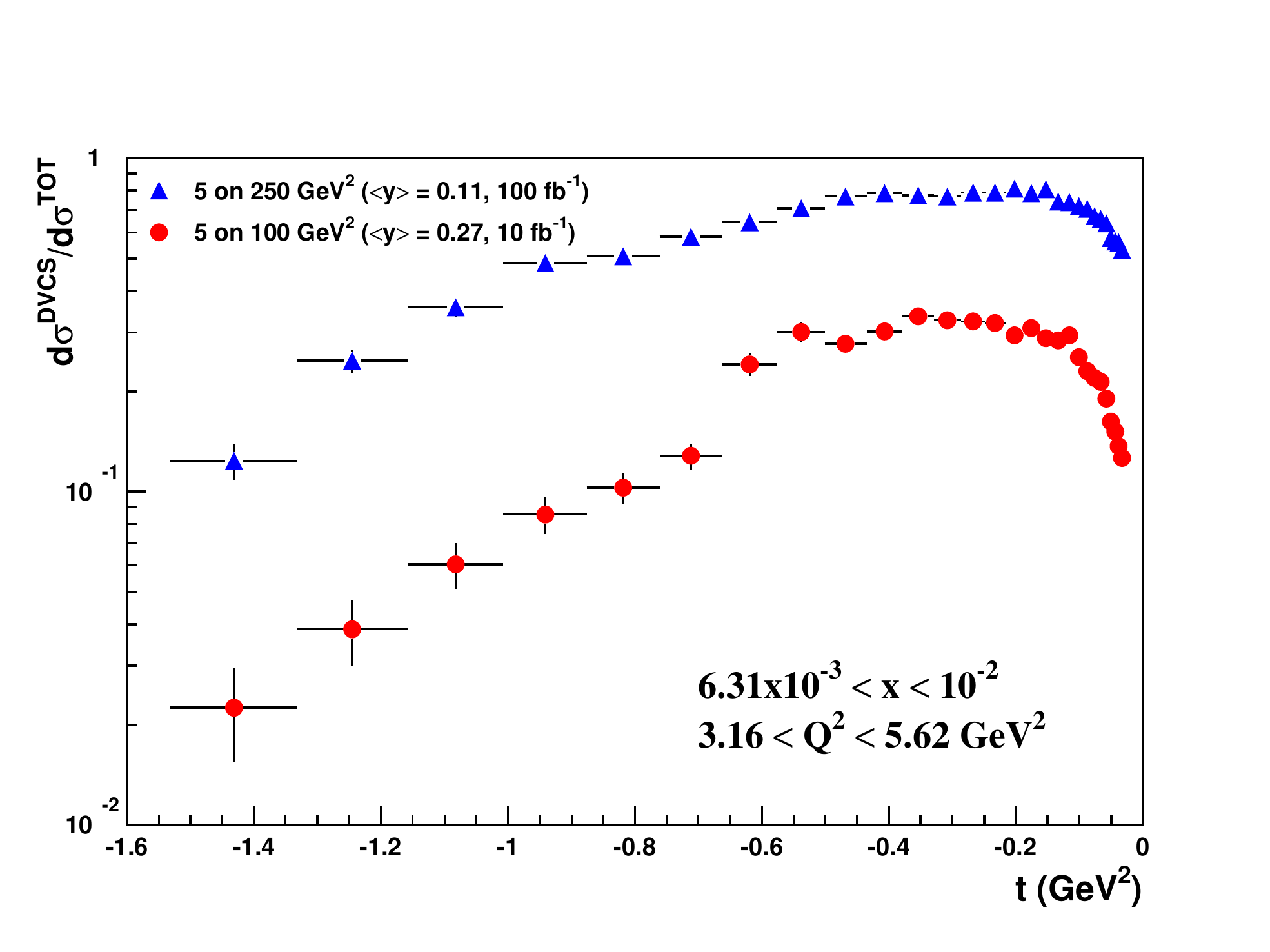}
\includegraphics[width=0.48\textwidth,angle=0]{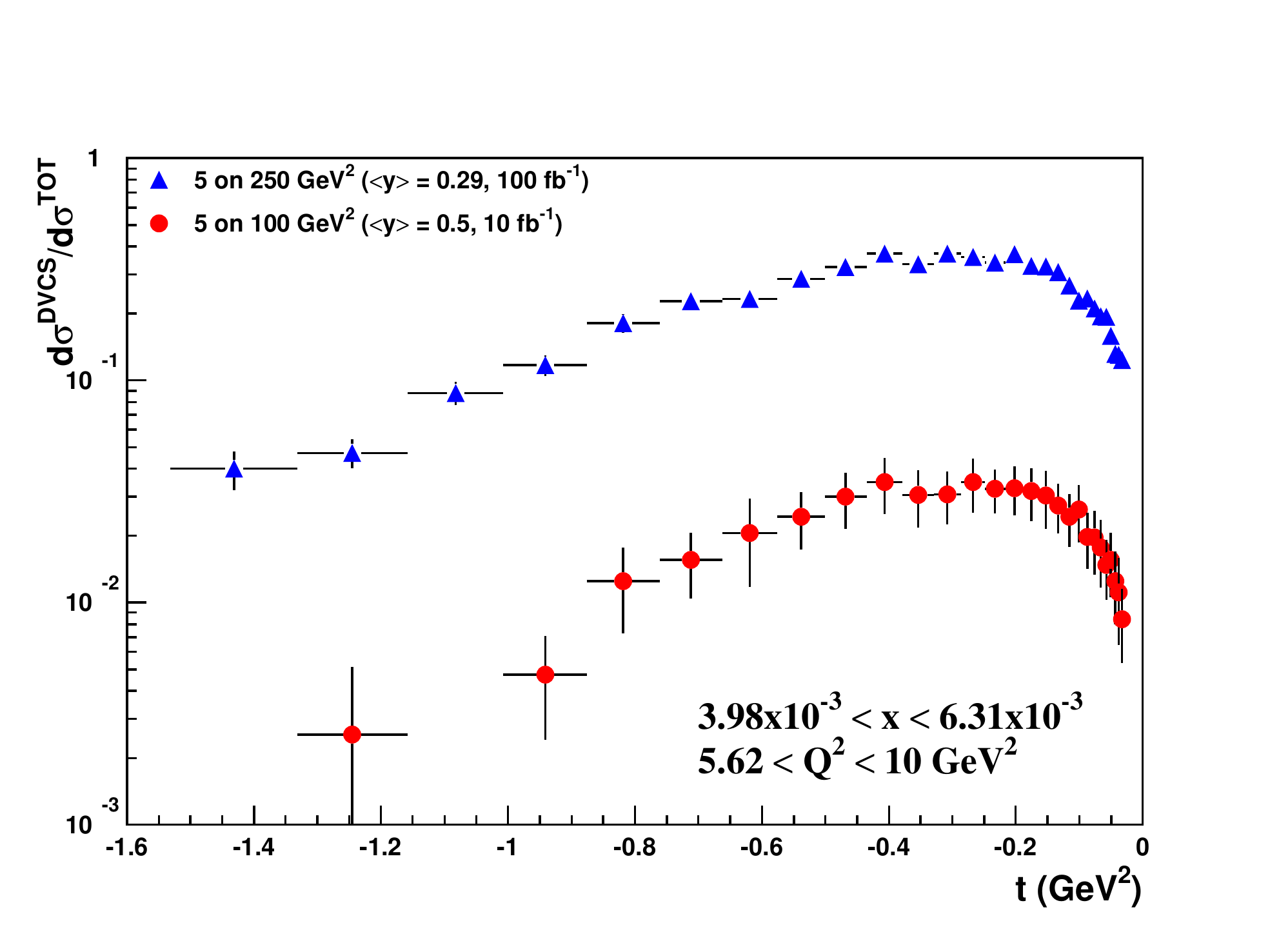}
\end{center}
\vspace{-0.8cm}
\caption{\small Cross section ratio  (\ref{R^{DVCS} R^{INT}}) of the DVCS
cross section to the photon electroproduction  cross section (\ref{XTOT-y}) as function of $-t$
for two $\xB-\Q^2$-bins and two different beam energy combinations.
\label{fig:X-ratios}
}
\end{figure}
In Fig.~\ref{fig:X-ratios} we show the typical $t$-shape of this ratio for an exponential $t$-dependence
at $5\times 100 \GeV^2$ (circles) and $5\times 250 \GeV^2$ (triangles) for  two $\{\xB,\Q^2\}$-bins,
\begin{eqnarray}
6.31\times 10^{-3} < \xB < 1.00 \times 10^{-2} &\mbox{and}& 3.16 \GeV^2 <\Q^2 < 5.62\GeV^2 \mbox{ (left)}\,,
\nonumber\\
3.98\times 10^{-3} < \xB < 6.31 \times 10^{-3} &\mbox{and}& 5.62 \GeV^2 <\Q^2 < 10.0\GeV^2 \mbox{ (right)}\,.
\nonumber
\end{eqnarray}
These results were simulated by MILOU, as described in
Sect.~\ref{sec:simulation}. The statistical uncertainties are obtained including
all the selection criteria to suppress the BH cross section also in the region where the DVCS
cross section is extremely small. Clearly, the functional multi-variable dependencies, that are
expected from the approximation (\ref{XDVCS2XTOT}), can be easily seen in the plots.
In this specific GPD model, utilized in MILOU, the DVCS cross
section is only accessible in a smaller set of
$\{\xB,\Q^2,t\}$-bins. However, as is clearly illustrated in Fig. \ref{fig:X-ratios}, an increase
of the proton beam energy from $100 \GeV$ to $250 \GeV$ allows to overcome such a
potential limitation.

\begin{figure}[ht]
\begin{center}
\includegraphics[width=1.00\textwidth]{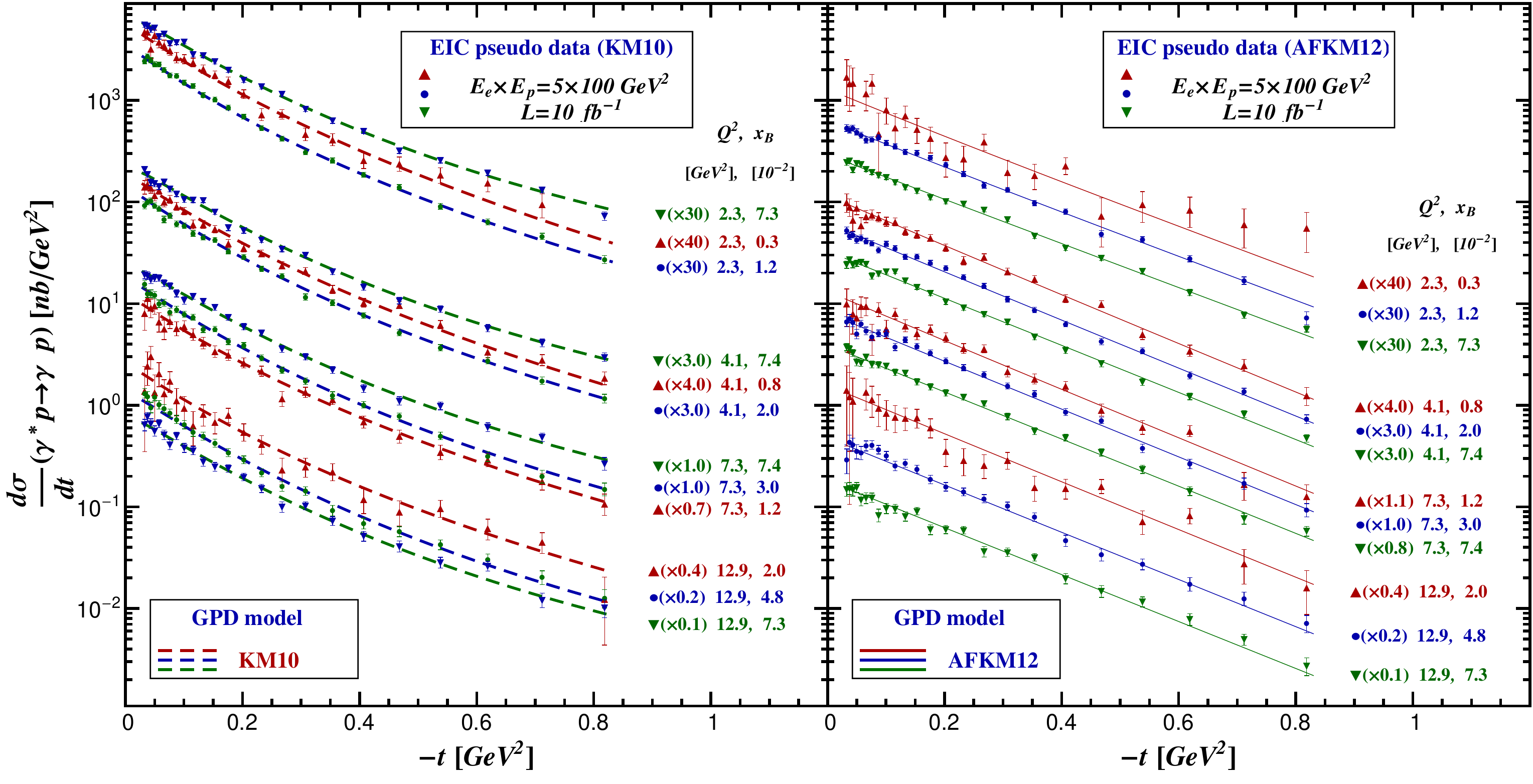}
\end{center}
\caption{
\small
\KM (left panel) and \our (right panel) model predictions of the differential DVCS cross
section versus $-t$ for unpolarized beams with energies
$E_e\times E_p = 5\times 100 \GeV^2$ and a luminosity of $10~{\rm fb}^{-1}$. The
uncertainties of the EIC pseudo data contain statistical, 5\% systematical, and
uncertainties due to BH cross section subtraction,
where for the latter a 3\% uncertainty of the BH cross section has been assumed.
\label{Fig-X5x100}}
\end{figure}
We take now the \KM and \our predictions to illustrate that the DVCS cross section can
be possibly obtained by a subtraction
procedure (\ref{dsigma^{DVCS}-sub}) even at the low beam energy configuration
$5\times100 \GeV^2$. Generally, these DVCS cross section
predictions overshoot those of the MILOU simulations, on the other hand the \KMa model
predictions are in agreement%
\footnote{
In the majority of bins the \KMa and MILOU cross sections are comparable to each other,
while in some low $\Q^2$ and large-$\xB$ bins the \KMa model prediction overshoots the
MILOU prediction up to 100\%, which could be attributed to model differences.}%
.
Based on the MILOU simulations, described in Sect.~\ref{sec:simulation},
we obtain the statistical uncertainties for the model predictions by rescaling according to
the ratios of the DVCS cross sections.
All uncertainties (statistical, $5\%$ systematical, and subtraction uncertainty from a 3\%
error of the BH cross section) were added in quadrature and the predicted cross
section for a kinematical point, given by the center of a three dimensional
$\{\xB,\Q^2,t\}$-bin, was assumed to be normally distributed.

In Fig.~\ref{Fig-X5x100} we show the \KM (left panel) and \our (right panel) model
predictions for the DVCS cross section versus $-t$ for the $5\times100 \GeV^2$ beam energy
configuration for four $\Q^2$ and three $\xB$ bins.
Apart from the different $t$-behavior, one notices model differences in the normalization
at lower $-t$ values, in particular for the largest $\xB$ values. One also realizes that
in the \KM model the cross section does not necessarily grow with decreasing $\xB$ as
it is the case for \our model (solid curves), containing only the sea quarks and gluon
components of the CFFs $\cal H$ and $\cal E$.  Both of these observations indicate that
valence-like contributions to $\cal H$ and/or non-dominant CFFs can play a certain role
at lower c.o.m.~energies. In both panels the sizable uncertainties arise from the
uncertainty of the BH cross section and, as expected, they appear for the low $\xB$ bins,
essentially, in the small $-t$ region and large $-t$ region. Note in Fig.~\ref{Fig-X5x100}
bins are not shown in which the DVCS cross section is entirely
dominated by the subtraction uncertainties, i.e., we ignored bins with
$y \gtrsim 0.25\cdots 0.4$. For values $-t >0.8\GeV^2$ (not shown) the uncertainties
associated with the BH subtraction become also large, particularly for \our model
which possesses an exponential $t$-dependence.  We remind that all models, including MILOU,
describe the H1/ZEUS DVCS cross sections measurements very well (see left panel
of Fig.~\ref{fig:X}) for which the aforementioned contributions play a minor role.

Let us summarize the lessons for an unpolarized DVCS cross section measurement at rather
low EIC energies. Certainly, it is safe to expect that the electroproduction cross
sections, i.e., containing all three terms, are large enough to provide precise data,
at present not available in this kinematical region of transition to small $\xB$. Such
data can be immediately included in global GPD fits; however, model assumptions will
affect the partonic interpretation of such measurements. The isolation of the DVCS
cross section is probably only feasible in a limited phase space (lower $y$ values,
limited $-t$ values).  Even in the case that this problem can be overcome by a (partial)
Rosenbluth separation, the measurements would only
provide a very qualitative insight in the transverse distribution of partons, since
the separation of different CFF contributions is based on assumptions. Hence, a
measurement of further observables is needed, which allows for a separation of the
various CFFs contributions.

\subsection{Single spin asymmetry measurements}
\label{sec:meas-SA}

Measuring the differences of spin-dependent cross sections (\ref{DX-electroproduction})
for unpolarized, longitudinally and transversely polarized protons allows the access
of the imaginary part of CFFs in a much cleaner manner than utilizing asymmetries. In such
measurements one may use harmonic analysis to access the imaginary parts of twist-two associated CFFs
from the first odd harmonics,  occurring from the interference of the BH and DVCS amplitudes.
However, even these observables are contaminated by power-suppressed helicity flip contributions that
stem from both the interference and DVCS squared term.
The latter contamination can be eliminated if lepton beams of both charges are available, see
discussion in the next section. This allows then for a harmonic analysis,
aiming to isolate the imaginary parts of twist-two associated CFFs from the remaining ones.
In this way one can separate to some extent twist-two, twist-three, and gluon
transversity contributions. What is the best  strategy to analyze a high quality data set,
measured in an experiment where only an electron beam is available,
is not so obvious at present. One may hope that, as in the case of unpolarized
electroproduction cross section, considered in Sect.~\ref{sec:Rosenbluth}, a
common Fourier analysis will finally yield some simplifications and may even allow to
employ the  Rosenbluth separation method to some extent.

\begin{figure}[htbp]
\begin{center}
\includegraphics[width=1.00\textwidth]{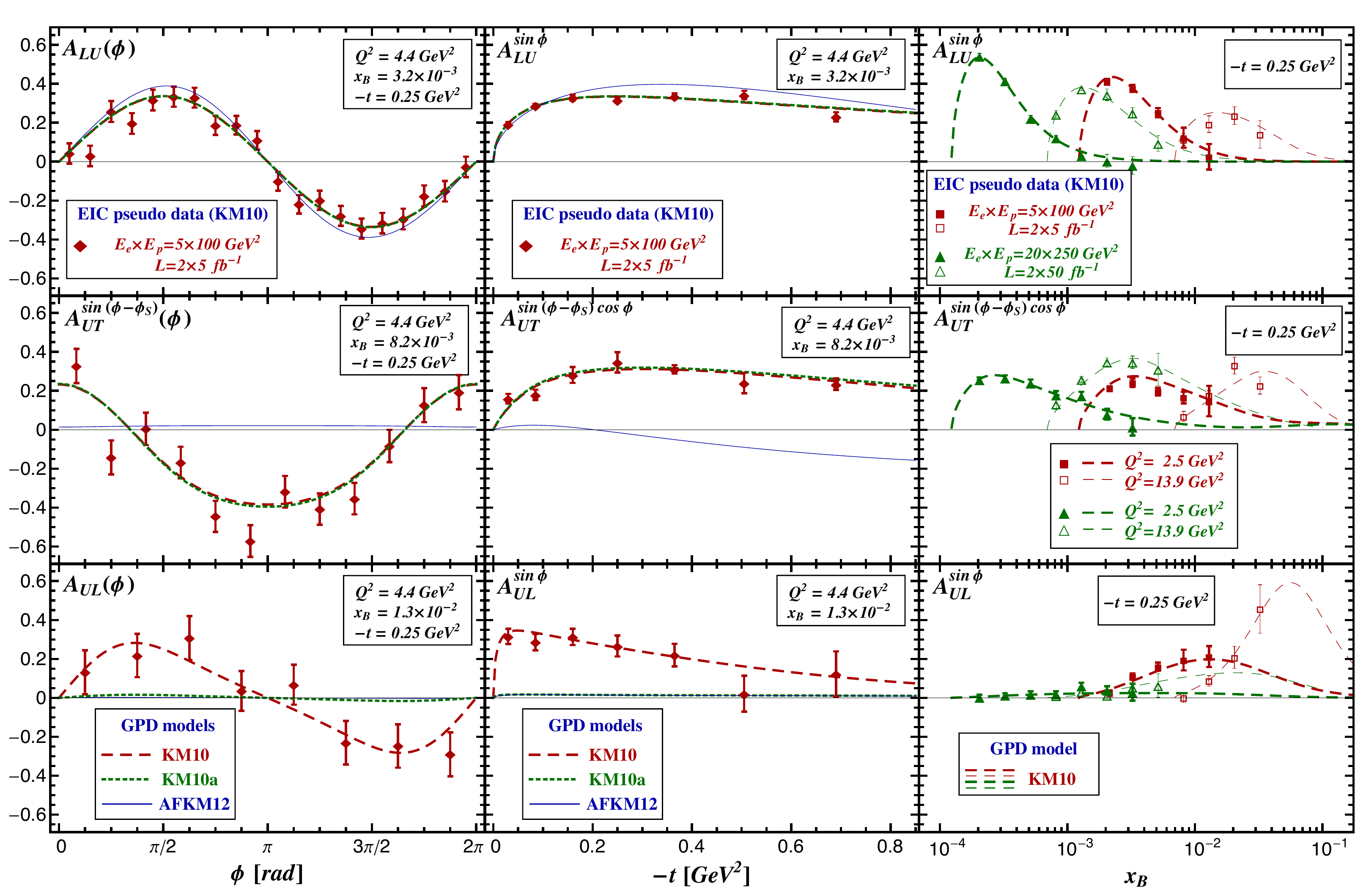}
\end{center}
\vspace{-0.7cm}
\caption{\small {\em KM10} model predictions for DVCS single spin asymmetries: electron
(\ref{A_{LU}-approx}) [upper], transverse proton (\ref{A^sin_UT}) [middle] and longitudinal
proton (\ref{A_UL}) [lower] with $E_e\times E_p = 5\times 100 \GeV^2$ (diamonds,squares)
and $E_e\times E_p =20\times 250 \GeV^2$ (triangles) EIC settings. Asymmetries versus
azimuthal angle $\phi$ for one selected bin ($\xB=8.2\times 10^{-3}, -t=0.25\GeV^2,
\Q^2=4.4 \GeV^2$) at $5\times 100 \GeV^2$ are shown in the left column for various
GPD models: \KM (dashed), \KMa (dotted), and \our (solid). In the middle column
the $t$-dependence for the projection on the first non-vanishing harmonic is displayed
for the same $\xB$, $\Q^2$ and beam energy values. In the right column the
$\xB$-dependencies is shown for the \KM model prediction at $-t=0.25\GeV^2$, two
different $\Q^2$ values, $\Q^2=2.5 \GeV^2$ (filled squares and triangles, thick curves)
and $\Q^2=13.9 \GeV^2$ (empty squares and triangles, thin curves), and two different
choices of beam energies, $5\times 100 \GeV^2$ (red) and $20\times250 \GeV^2$ (green).
\label{Fig-ALU_AUT_AUL} }
\end{figure}

For purpose of illustration we focus
in the following on twist-two GPD model predictions for single spin asymmetries rather
than on spin-dependent cross section differences (\ref{DX-electroproduction}). In
Fig.~\ref{Fig-ALU_AUT_AUL} we show pseudo data that are generated using the \KM model,
and randomized according to the uncertainties as specified in Sect.~\ref{sec:simulation}
(rescaled statistical errors from MILOU simulations,
$5\%$ systematical uncertainty on cross section level, $5\%$ normalization uncertainty for
the polarization measurement).
The error propagation from the $\phi$-dependent cross section to harmonic amplitudes was
simply done by fitting. Note that the uncertainty for the projection asymptotically scales for
the $N$ $\phi$-bins as $1/\sqrt{N}$, except for the zeroth harmonic for which scaling is
$1/\sqrt{2 N}$.  We note that the polarization error should be treated
as an overall normalization uncertainty, which, however, was not done here. Hence, the
projections on the first harmonic in Fig.~\ref{Fig-ALU_AUT_AUL} have an additional
normalization uncertainty, essentially given by the polarization uncertainty.

The upper panels in Fig.~\ref{Fig-ALU_AUT_AUL} show for a proton beam the
electron beam spin asymmetry (\ref{A_{LU}}) as function of the azimuthal angle $\phi$
for one selected bin with $5\times 100 \GeV^2$ beam energies (left panel), its projection
on the dominant first $\sin \phi$ harmonic,
\begin{equation}
\label{A_{LU}-approx}
A_{\rm LU}^{\sin\phi}\propto \frac{y\sqrt{1-y}}{2-2y+y^2}\sqrt{\frac{-t}{y^2 \Q^2}}\times \xB\,\im \left[
 F_1 {\cal H} -\frac{t}{4M_p^2} F_2 {\cal E} + \frac{\xB}{2} (F_1 + F_2) \widetilde {\cal H}
\right](\xB,t,\Q^2) +\cdots,
\end{equation}
as function of $-t$ (middle panel), and versus $\xB$ for a low $\Q^2=2.5\GeV^2$ and a
high $\Q^2=13.9\GeV^2$ value (right panel).  The asymmetry is dominated by helicity
conserved CFF ${\cal H}$ and proportional to the electron energy loss $y$.  Consequently,
if $y$ is not too small, the asymmetry might be rather sizable over a large kinematical
region, shown for $5\times 100 \GeV^2$ (squares, thick curves) and
$20\times 250 \GeV^2$ (triangles, thin curves).
The CFF $\cal E$ appears with a kinematic suppression factor $t/4 M_p^2$, induced by a proton
helicity flip, and remaining CFFs also contribute, which is in (\ref{A_{LU}-approx})
indicated by the ellipsis that include also further kinematically suppressed contributions.
Comparing the different predictions of the \KM (dashed), \KMa (dotted), and \our (solid)
models, one realizes that the contaminations of this asymmetry by other CFFs are in
fact small. Our $\cal E$ enhanced model prediction only slightly differs from
the other ones at $t\sim 0.5\, \GeV^2$.
It is noted that for a neutron target the ${\cal H}$ contribution is suppressed by the
accompanying Dirac form factor $F^n_1$ ($F_1^n(t=0)=0$), making this asymmetry
sensitive to the CFF ${\cal E}$. However, in this case one expects a smaller single
beam spin asymmetry that is also contaminated by other non-dominant CFF contributions.

A single spin asymmetry measurement with a transversely polarized proton beam,
cf.~(\ref{A_{LU}}), provides another handle on the imaginary part of the
helicity-flip CFF ${\cal E}$. This asymmetry has in addition to the $\phi$ dependence
a $\phi-\phi_S$ modulation. If the target spin in such a frame is perpendicular
to the reaction plane (e.g., $\phi-\phi_S=\pi/2$), the asymmetry
\begin{equation}
\label{A^sin_UT}
A^{\sin(\phi-\phi_S)\cos\phi}_{\rm UT} \propto
\frac{\sqrt{1-y}}{2-y} \frac{-t }{2y\,M_p\Q} \times \xB\im \left[F_2 {\cal H}-  F_1 {\cal E} +
\frac{\xB}{2} (F_1+F_2)\overline{\cal E} \right](\xB,t,\Q^2)  + \cdots
\end{equation}
is dominated by a linear combination of ${\cal H}$ and ${\cal E}$  CFFs. In the case
that the target spin is aligned with the reaction plane (e.g., $\phi-\phi_S=0$) the asymmetry
\begin{eqnarray}
\label{A^cos_UT}
A^{\cos(\phi-\phi_S)\sin\phi}_{\rm UT} \propto
\frac{\sqrt{1-y}}{2-y}  \frac{-t}{2y\, M_p \Q} \times \xB\im
\left[ F_2 \widetilde{\cal H} - F_1 \overline{\cal E}\right](\xB,t,\Q^2)  +\cdots
\end{eqnarray}
is formally dominated by a linear combination of CFFs $\widetilde{\cal H}$ and
$\overline{\cal E}$, cf.~(\ref{bE}).
In these asymmetries an additional relative kinematical factor $\sqrt{-t/4 M_p^2}$ appears.
The middle row in Fig.~\ref{Fig-ALU_AUT_AUL} shows the $\sin(\phi-\phi_S)$
projection of the transverse proton beam spin asymmetry, which can also be rather large
over a wide kinematical range. As in the case of the unpolarized cross section, discussed
in the preceding section, this is caused by the fact that at smaller values of $\xB$ the
``pomeron'' behavior in ${\cal H}$ overtakes the kinematical suppression factors, see
dashed and dotted curves. We may assume that such a ``pomeron'' behavior is also
contained in ${\cal E}$. For our choice of $\kappa^{\rm sea}= 1.5$ the $\cal E$
contribution will mostly cancel the ${\cal H}$ contribution, see (\ref{A^sin_UT})
where $F_2(t=0) \approx 1.79$. In contrast to the electron beam spin asymmetry, for a
neutron target the asymmetry is now more sensitive to the helicity conserving
CFF $\cal H$. For the $\cos(\phi-\phi_S)$
projection of the transverse proton beam spin asymmetry
(\ref{A^cos_UT}) the common expectation is that the
parity-odd CFFs $\widetilde{\cal H}$ and
$\overline{\cal E}$ behave more gently at small $\xB$ and, hence, we expect that
this observable is small in the EIC kinematics (not shown).

Finally, we consider the longitudinally polarized proton beam spin asymmetry.
Its projection on the dominant ${\sin\phi}$ harmonic reads
\begin{equation}
\label{A_UL}
A^{\sin\phi}_{\rm UL} \propto
\frac{\sqrt{1-y}}{2-y} \sqrt{\frac{-t}{y^2 \Q^2}} \times \xB\im
\left[ F_1 \widetilde{\cal H}- \left(\!\frac{\xB}{2} F_1  +\frac{t}{4M_p^2} F_2\!\right)\overline{\cal E}
+\frac{\xB }{2}(F_1 + F_2) {\cal H} \right]+\cdots.
\end{equation}
It is sensitive to the imaginary part of CFF $\widetilde{\cal H}$ and
$\overline{\cal E}$, and other CFFs might contribute as well.
As already noted, one expects that here the dominant CFF $\widetilde{\cal H}$ behaves
gently
at small $\xB$ and models that incorporate such a behavior predict a rather tiny asymmetry
(dotted and solid lines). In contrast, in the \KM model (dashed line), a
rather big GPD $\widetilde H$ has been incorporated with a generic $1/\sqrt{\xB}$ behavior
at small $\xB$.  Hence, we get a sizable asymmetry for $5\times100 \GeV^2$ beam energies
which is getting smaller at higher beam energies $20\times250 \GeV^2$, see lower row
on Fig.~\ref{Fig-ALU_AUT_AUL}.  We emphasize again that not much is known about the
small-$\xB$ behavior of CFF $\widetilde {\cal H}$. We add that for a neutron target
the asymmetry becomes sensitive to the CFF $\overline{\cal E}$.

Let us summarize the lessons from the approximated equations
(\ref{A_{LU}-approx}--\ref{A_UL}), quantified by numerics. The experimentally established
`pomeron' behavior of the CFF $\cal H$ predicts a large single beam spin and a large
$\cos(\phi-\phi_S)$  projection of the transverse proton beam spin asymmetry for the EIC
kinematics. If $\cal E$ contains also a `pomeron' behavior, the latter asymmetry can be
weakened (amplified) for a positive (negative) imaginary part of $\cal E$. The remaining
two single spin asymmetries cannot be predicted easily; however, based on common
phenomenological/theoretical wisdom they are probably small.
Let us note that the normalization of these asymmetries obviously depends
also on the real part of the twist-two associated CFFs and the remaining eight ones.
As advocated above, a  measurement of cross section differences are not affected by
this normalization uncertainty.

\subsection{Further EIC opportunities}
\label{sec:opportunities}

An EIC machine provides further opportunities for DVCS studies:
\begin{itemize}
\item Double spin flip experiments provide a handle on the real part of CFFs,
however, in such measurements the spin-dependent BH cross section contributes.

\item As demonstrated by the HERMES collaboration, having a positron beam at hand
allows also to separate the interference and DVCS harmonics in single spin target
experiments.
Measuring spin-dependent cross sections in the charge odd sector (interference term)
and the charge even sector allows to extract CFFs from experimental measurements, based
on minimal assumptions.

\item The large kinematical coverage of the proposed high-luminosity EIC
(see Fig.~\ref{x-q2-dvcs}) and the partial
overlap with JLAB 12GeV kinematics raises the question: Can one utilize evolution, even
at moderate $\xB$ values, to access GPDs away from their cross-over line?

\item Photon electroproduction off the neutron offers the possibility for a flavor separation.

\item Photon electroproduction off nuclei is a mostly unexplored experimental field.
\end{itemize}
Below we will discuss a minimalistic version of the second point in more detail,
namely, having an {\it un}polarized positron beam at hand.
Let us mention here that a study of GPD evolution was presented in
Ref.~\cite{Kumericki:2011zc}, however, we may conclude here that a wide coverage in
$\Q^2$ is extremely helpful in getting constraints on GPDs away from the cross-over line,
however, a ``measurement'' of the GPD in the outer region certainly cannot be reached.
DVCS on a ``neutron target'' is certainly needed for a GPD flavor decomposition. However,
this program is more complicated than in DIS, since in the interference term the various
CFFs are accompanied by nucleon form factors, see short discussions in the previous
section.  We will not discuss DVCS off nuclei, which is interesting in itself.  It has been
worked out theoretically for a spin-zero target, where one can adopt the equation from \cite{Belitsky:2000vk},
and to some extent also for spin-one target \cite{Berger:2001zb,Kirchner:2003wt,Cano:2003ju}, while the
formalism for spin-1/2 nuclei can be adopted from the proton.

We should also emphasize the EIC opportunities for Compton scattering measurements below
the deeply virtual regime.
\begin{itemize}
\item Quasi-real Compton scattering can be measured  over a rather wide energy range in anti-tagged
electron scattering experiments, where  the VCS cross section is peaked at $\Q^2\sim 0$.
\item
We  expect that at stage I binning of low photon virtualities, i.e., $\Q^2 < 1\GeV^2$, will be possible.
\end{itemize}
Such measurements will provide understanding on the transition from the deeply virtual to
the quasi-real regime. This, in turn, is needed if radiative electromagnetic corrections to
photon electroproduction are to be elaborated in a more complete manner than they presently are.

Finally, we should remind that other exclusive channels can be measured at EIC:
\begin{itemize}
\item Deeply virtual production of light vector mesons can be employed for a partial
flavor separation of quark GPDs.
\item $J/\Psi$ production  gives naturally access to the gluon GPD.
\item Also, experimental studies on deeply virtual production  of pseudo scalar mesons,
the production of two final meson states,  time-like DVCS, and
double DVCS may turn out to be feasible.
\end{itemize}
We would like to add that deeply virtual
production of light vector mesons and DVCS measurements
at HERA collider experiments can be simultaneously described with a GPD formalisms
\cite{Meskauskas:2011aa,Kroll:2012sm}.
Whether the measurements, listed in the last item above, are
actually feasible at EIC, can only be stated in terms of models. Thereby, based
on phenomenological knowledge of the dominant GPD $H$, cross sections for
time-like \cite{Berger:2001xd, Moutarde:2013qs} and/or double
\cite{Guidal:2002kt,Belitsky:2002tf,Belitsky:2003fj} DVCS might be more or less
realistically estimated, however, were not part of our studies.

\subsubsection{Uses of an unpolarized positron beam}
\label{sec:positron}

The isolation of the interference term, which contains the most valuable information on
CFFs, is most easily done by forming charge asymmetries, which require
a positron beam. We emphasize again, that the alternative Rosenbluth
separation is expected to be more intricate and has not been so far either considered
theoretically or explored experimentally (e.g., by the use of
approximated expressions).
Forming differences and sums of spin-dependent cross section
measurements with both kinds of lepton beams allows to extract the pure interference and
DVCS squared terms and might allow to quantify twist-three and gluon transversity effects.
{}From such experiments one can extract the imaginary part of CFFs. Note that only an
unpolarized positron beam is needed to perform such a program for the
single proton spin asymmetries -- of course, for the projection of the single
electron spin asymmetry a polarized positron beam would be needed. In double spin flip
measurements one can use the same procedure to access the real part of the CFFs.
Although existing data indicate that twist-three effects are small, as it is expected
based on kinematic factors, the twist-three related CFFs are not necessarily small.
Surely, one needs very high precision data to extract non-dominant twist-two CFFs or
twist-three related ones. However, even obtaining only an upper limit is important for
a determination of the systematic uncertainties of the (dominant) twist-two CFFs.

Let us consider here only the lepton beam charge asymmetry (\ref{A_C}) for an unpolarized
proton. Its first harmonic is dominated by the real part of the twist-two related CFFs
$\cal H$ and $\cal E$, rather analogous to equation (\ref{A_{LU}-approx}) for the electron
beam spin asymmetry,
\begin{equation}
\label{A_C-approx}
A_{\rm C}^{\cos\phi}\propto
\frac{\sqrt{1-y}}{2-y}\sqrt{\frac{-t}{y^2 \Q^2}}\times \xB\,\re \left[
 F_1 {\cal H} -\frac{t}{4M_p^2} F_2 {\cal E} + \frac{\xB}{2} (F_1 + F_2) \widetilde {\cal H}
\right](\xB,t,\Q^2)  +\cdots\,.
\end{equation}
\begin{figure}[ht]
\begin{center}
\includegraphics[width=1.00\textwidth]{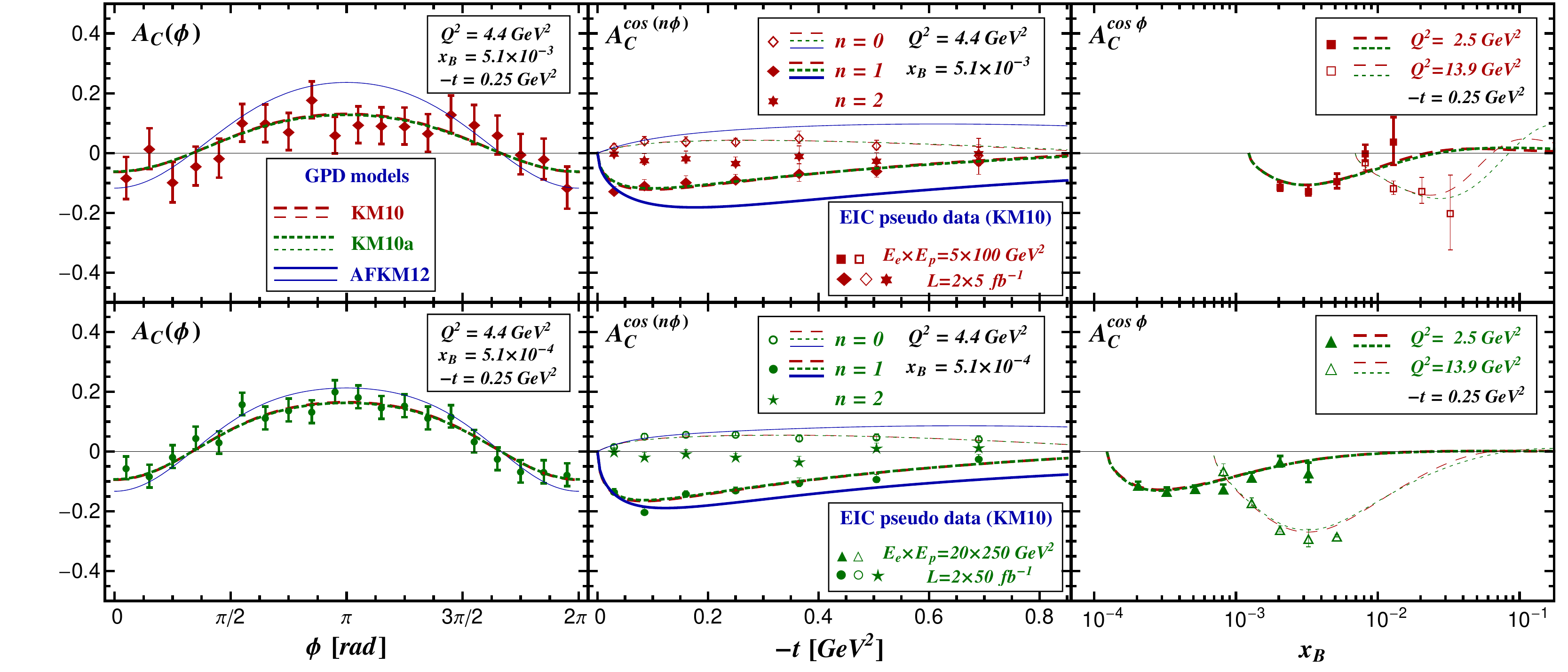}
\end{center}
\vspace{-0.5cm}
\caption{\small \KM (dashed), \KMa (dotted), and \our (solid) model predictions for
the DVCS lepton beam charge asymmetry (\ref{A_C}, \ref{A_C-approx}) with
$E_e\times E_p = 5\times 100 \GeV^2$ (upper row) and
$E_e\times E_p =20\times 250 \GeV^2$ (lower row).
Left column: $A_{\rm C}$ versus $\phi$ for $\xB=5.1\times 10^{-3}$,
${\cal Q}^2 = 4.4~{\rm GeV}^2$, and $t=-0.25~{\rm GeV}^2$ (upper panel) and
 $\xB=5.1\times 10^{-4}$, ${\cal Q}^2 = 4.4~{\rm GeV}^2$, and
$t=-0.25~{\rm GeV}^2$ (lower panel).
Middle column: $A_{\rm C}^{\cos(n \phi)}$ amplitudes versus
$-t$ at same $\xB$ and $\Q^2$ values for $n=1$ (filled diamonds and circles, thick curves),
$n=0$ (empty  diamonds and circles, thin curves), and $n=2$ (stars). Right column:
Dominant amplitude $A_{\rm C}^{\cos(\phi)}$ versus $\xB$ for $\Q^2 = 2.5 \GeV^2$
(thick curves) and $\Q^2 = 13.9 \GeV^2$ (thin curves) at $t=-0.25\GeV^2$.
\label{Fig-A_C}}
\end{figure}
It is shown in Fig.~\ref{Fig-A_C} that the GPD models predict a rather sizable
lepton beam charge asymmetries for both $5\times100 \GeV^2$ (upper row) and
$20\times250 \GeV^2$ (lower row) beam energies. As for the electron beam spin
asymmetry (upper row on Fig.~\ref{Fig-ALU_AUT_AUL}), the predictions from the \KM
(dashed curves) and \KMa (dotted curves) models are almost the same,
illustrating that the CFF $\widetilde{\cal H}$ contribution is rather unimportant in
this observable ($\widetilde{\cal E}$ drops exactly out here), while the
small deviation of \our model (solid line) indicate some
sensitivity to CFF $\cal E$. The sign of this asymmetry is governed by
the effective `pomeron' trajectory $\alpha^\mathbb{P} \gtrsim 1$. It has in
the transition from the valence to the sea quark region a node (see upper right panel)
with its position depending on $-t$. In the middle column we show besides the
projection on the first even harmonic (filled diamonds and triangles) also the
projection on the zeroth harmonic (empty diamonds and triangles), which is dominated by
twist-two associated CFFs and, thus, (anti)correlated with the first harmonic.
The second harmonic (stars) is sensitive to twist-three associated CFFs, which are
set here to zero, and also depend on the twist-two associated CFFs. Therefore, the latter
induce only a small deviation, compatible with zero within one standard deviation.

\section{Partonic interpretation at small $\xB$}
\label{sec:interpretation}

Intensive GPD studies (up to NNLO accuracy) of small-$\xB$ DVCS data  measured by H1 and ZEUS
collaborations have been performed, where it
turned out that the functional form of the $t$-dependence cannot be pinned down and
an access to the CFF $\cal E$ is not feasible when having only unpolarized DVCS
cross section and the lepton beam charge asymmetry measurements \cite{Aaron:2009ac}
available \cite{Kumericki:2009uq}.
A high-luminosity EIC experiment with transversely polarized protons certainly provides
the opportunity for precise measurements of CFFs and to explore their partonic
interpretation in the small-$\xB$ region, i.e., $\xB < 0.01$. As we argued in
Sect.~\ref{sec-pred}, the set of relevant twist-two associated
CFFs is then reduced  to $\cal H$ and  $\cal E$ only, and,
moreover, valence quark contributions can be safely neglected. From our discussion there
it is also obvious that the former assumption, which is used now,
can be experimentally cross-checked. We  will also use the fact that the real parts
of the remaining two  CFFs is locally tied to their imaginary parts, see (\ref{cffHcffE-smallx}),
which is implemented in our GPD model and is in fact a  more general consequence
of the dispersion relation and the effective `pomeron' behavior.
Hence, we can restrict ourselves to two observables, namely, the unpolarized DVCS
cross section (\ref{XDVCS}) and the single transverse proton
beam asymmetry (\ref{A^sin_UT}), which now simplify to
\begin{eqnarray}
\label{KLMSPM-Def-CroSec}
\frac{d\sigma^{\rm DVCS}}{dt}(\xB,t,{\cal Q}^2) & \approx &
\frac{\pi \alpha^2 \xB^2}{{\cal Q}^4}
\left[\left| {\cal H} \right|^2  - \frac{t}{4 M^2_{p}}
\left| {\cal E} \right|^2 \right]
\left(\xB,t,{\cal Q}^2\right) \,,
\\
\label{A^sin_UT-1}
A^{\sin(\phi-\phi_S)\cos\phi}_{\rm UT}&\propto &
\frac{\sqrt{1-y}}{2-y} \frac{-t }{2y\,M_p\Q} \times \xB\im \left[F_2 {\cal H}- F_1 {\cal E} \right](\xB,t,\Q^2).
\end{eqnarray}
In the partonic interpretation of DVCS data we are in the first place interested in
the transverse distribution of sea quarks and gluons at small $\xB$ for an unpolarized and
for a transversely polarized proton.
In Sec.~\ref{sec:extraction} we explore by least-squares fitting the extraction of both GPD $H$ and $E$
from the aforementioned observables at stage II of an EIC.  In Sect.~\ref{sec:imaging} we present a
detailed study of the extraction of transverse polarized  parton distributions.  We also perform
there the Fourier transform of our GPD model fit results to the impact space, where
experimental uncertainties are propagated and extrapolation errors are taken into account.
Finally, in Sect.~\ref{sec:sum-rule} we discuss the importance
of such a measurement for the qualitative understanding of the proton spin decomposition.

\subsection{Extraction of  GPDs $H$ and $E$ from high energy EIC pseudo data}
\label{sec:extraction}

To illuminate how GPDs are experimentally constrained in the small-$x$ region at present,
we will present here also new fits to the world DVCS data set at large $W$ (small $\xB$)
that includes the propagation of experimental uncertainties.
\begin{figure}[ht]
\begin{center}
\includegraphics[width=1.00\textwidth]{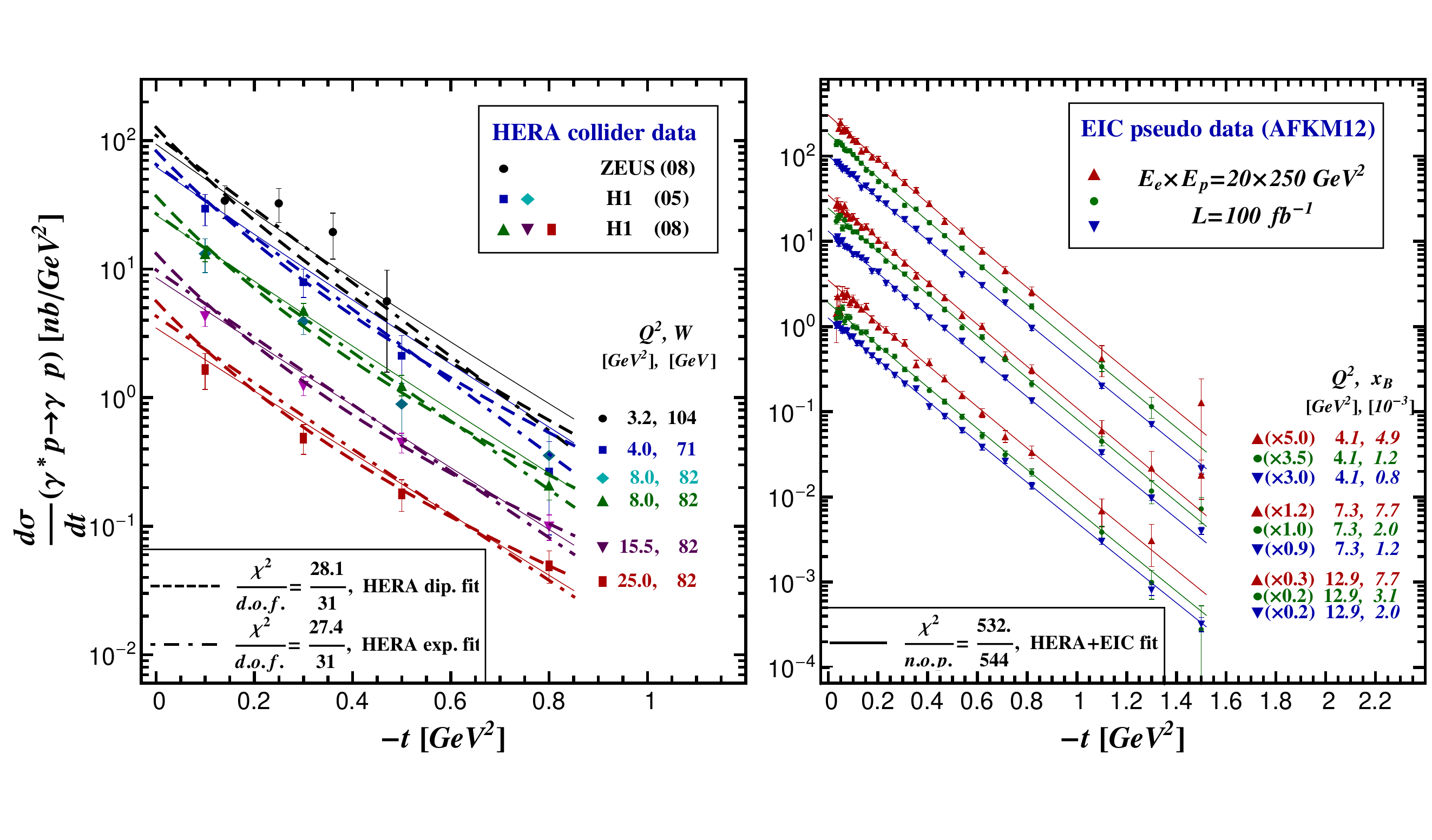}
\end{center}
\vspace{-1.5cm}
\caption{\small A model dependent extraction of GPD $H$ from cross section measurements
of the H1/ZEUS collaborations (left) and from a combined fit that includes EIC pseudo
data (right) with beam energies $E_e\times E_p = 20\times250 \GeV^2$. The HERA collider
data are taken from \cite{Chekanov:2008vy} (circle), \cite{Aktas:2005ty} (squares,diamonds),
 and \cite{Aaron:2009ac} (triangle-up, triangle-down, rectangle).
\label{fig:X}}
\end{figure}
At the HERA collider experiments H1 
\cite{Adloff:2001cn,Aktas:2005ty,Aaron:2007ab,Aaron:2009ac}
and ZEUS \cite{Chekanov:2003ya,Chekanov:2008vy} 
the unpolarized DVCS cross section could be measured
at large $W$, $-t < 1\,\GeV^2$, and with a large lever arm in $\Q^2$. Although
$\sim 200$ data points were published, we only consider 35 as statistically independent.
Apart from those for the differential cross section that are displayed on the left panel
on Fig.~\ref{fig:X} versus $-t$ we also included the following $t$-integrated cross
section measurements from ZEUS
\begin{eqnarray}
\label{meaZEUS}
7.5\GeV^2 \le \left<\Q^2 \right> \le 85 \GeV^2 && \left< W \right> = \phantom{0}89 \GeV \;
\mbox{\cite{Chekanov:2003ya}}, \\
7.5\GeV^2 \le \left<\Q^2 \right> \le 70 \GeV^2  && \left< W \right> = 104 \GeV\;\mbox{\cite{Chekanov:2008vy}}.
\nonumber
\end{eqnarray}
In our LO fits to these data we can only ask for the sea quark and gluonic components of GPD $H$,
where both components can be separated to some extent due to the large $\Q^2$ lever arm
\cite{Kumericki:2009uq}.
Since the experimental uncertainties are large, the functional form of
the $-t$ dependence for sea quark (and gluon) GPD $H$ cannot be determined by
$\chi^2/{\rm d.o.f.} \approx 1$ model fits,
done here at LO with a dipole (dashed) or an exponential (dash-dotted) residual $t$-dependence.
Thereby, also the ``pomeron'' slope parameter $\alpha^\prime$ for the sea quark (and gluon)
content cannot be determined.

\begin{figure}[ht]
\begin{center}
\includegraphics[width=1.00\textwidth]{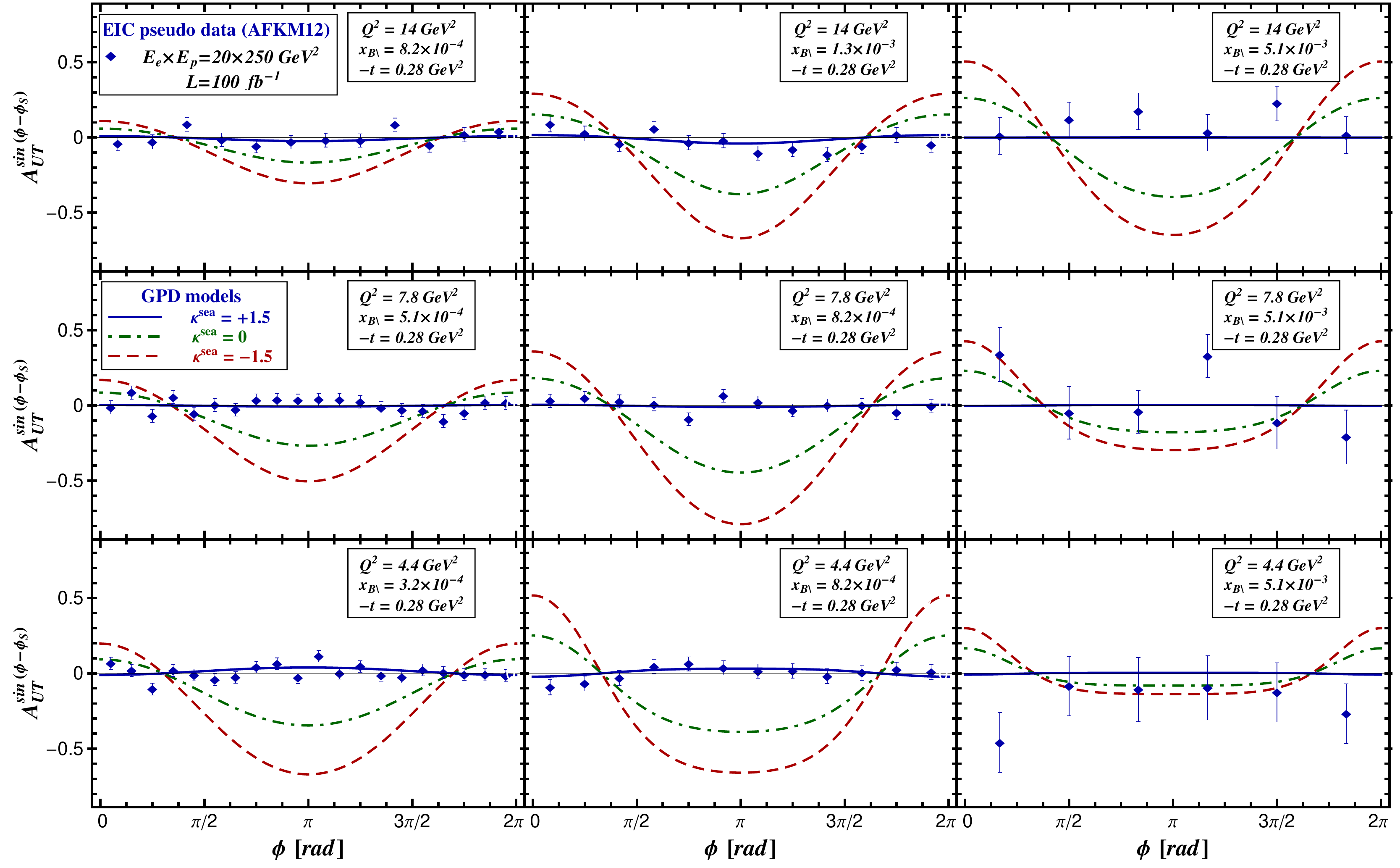}
\end{center}
\vspace{-0.7cm}
\caption{\small EIC pseudo data (diamonds) for the transverse target spin asymmetry
(\ref{A^sin_UT-1}) at beam energies $E_e\times E_p = 20\times 250 \GeV^2$ are shown
together with \our GPD model predictions, where GPD $E^{\rm sea}$ is taken as large
positive (solid), vanishing (dot-dashed), and large negative (dashed), respectively.
\label{Fig_TSA} }
\end{figure}

\begin{table}
\begin{center}
\small
\begin{tabular}{|c|cccccc|}
\hline
$H$                  & $N$ & $\alpha$ & $\alpha^\prime\; [\GeV^{-2}]$  & $b\;  [\GeV^{-2}]$ & $s_2$ & $s_4$  \\ \hline\hline
$p_0^{\rm sea}$      & 0.152  & 1.158 & 0.100 & 2.800 & $\phantom{-}0.513$  & $-0.210$ \\
$p^{\rm sea}$        & -      & -     & 0.090 & 2.858 & $\phantom{-}0.508$  & $-0.208$ \\
$\delta p^{\rm sea}$ & -      & -     & 0.009 & 0.035 & $\phantom{-}0.038$  & $\phantom{-}0.011$  \\ \hline
$p_0^{\rm G}$        &(0.448) & 1.247 & 0.100 & 2.000 & $-4.806$            & $\phantom{-}1.864$ \\
$p^{\rm G}$          & -      & -     & 0.063 & 2.086 & $-4.739$            & $\phantom{-}1.835$  \\
$\delta p^{\rm G}$   & -      & -     & 0.088 & 0.163 & $\phantom{-}0.212$  & $\phantom{-}0.106$ \\ \hline\hline
$E$ & $\kappa$       & $\alpha$ &  $\alpha^\prime\;  [\GeV^{-2}]$ & $b\;  [\GeV^{-2}]$ & $s_2$ & $s_4$ \\ \hline \hline
$p_0^{\rm sea}$      & 1.500 & 1.158 & 0.020 & 2.800     & $\phantom{-}0.513$  & $-0.210$   \\
$p^{\rm sea}$        & 1.451 & 1.164 & 0.023 & 2.779     & $\phantom{-}0.524$  & $-0.213$   \\
$\delta p^{\rm sea}$ & 0.307 & 0.005 & 0.012 & 0.049     & $\phantom{-} 0.104$ & $\phantom{-}0.026$ \\ \hline
$p_0^{\rm G}$        & $(-0.51)$     & 1.247 & 0.050 & 2.000  & $ -4.806$           & $\phantom{-}1.864$ \\
$p^{\rm G}$          & $(-0.49)$     & 1.295 & 0.001 & 1.961   & $ -4.687$           & $\phantom{-}1.803$  \\
$\delta p^{\rm G}$   & $(0.06)$      & 0.216 & 0.252 & 1.092 & $\phantom{-}0.048$  & $\phantom{-}0.077$ \\ \hline
\end{tabular}
\end{center}
\caption{ \small \our model parameters ($p_0$) and their fitted values ($p$) together with
standard uncertainties ($\delta p$) for sea quarks (superscript $^{\rm sea}$)
and gluon (superscript $^{\rm G}$) components of GPDs $H$ and $E$ at the input scale
$\Q^2_0=4 \GeV^2$.  The values in parentheses are fixed by sum rules.}
\label{tab:fit}
\end{table}
To explore the potential of the EIC measurements at stage II, we use in the following pseudo data for
the unpolarized DVCS cross section (\ref{KLMSPM-Def-CroSec}) and the transverse target spin asymmetry
(\ref{A^sin_UT-1}) for the beam energies $20\times250 \GeV^2$, as specified in Sect.~\ref{sec:simulation}.
Thereby, we utilized the flexible \our model, introduced in Sect.~\ref{sec-pred}, in which GPD
$H$ and $E$ have a different ``pomeron'' slope parameters, see $p_0$ values in
Tab.~\ref{tab:fit}, but the same residual $t$-dependencies. This choice guarantees that
positivity conditions for  GPDs at zero skewness are mostly satisfied
\cite{EIC11Die}.  The experimental uncertainties were estimated as before
(statistical uncertainties from the MILOU simulation, which are rescaled for the DVCS cross section,
$5\%$ systematic uncertainty on cross section level,
$3\%$ uncertainty of the BH cross section in the subtraction procedure
(\ref{dsigma^{DVCS}-sub}), and $5\%$ beam polarization uncertainty).
The exponential $t$-dependence of the CFFs drastically increases the
subtraction uncertainty at large $-t$, and the net uncertainty in this region can
become very large at larger $y$ values (lower $\xB$ values in particular at low $\Q^2$),
see right panel on Fig.~\ref{fig:X}. Since for the GPD $E$ we took
a model  with a positive $\kappa^{\rm sea}=1.5$, the transverse target asymmetry is becoming small.
Fig.~\ref{Fig_TSA} shows the pseudo data for this asymmetry together with the
model curve (solid), used to generate the asymmetry. Also shown is the prediction from an
analogous model which, however, has a negative $\kappa^{\rm sea}=-1.5$ value (dashed
curves) and one with vanishing CFF $\cal E$ (dash-dotted curves).  Certainly, the
predictions of all these three models are experimentally distinguishable.

We performed a combined least-squares fit to the EIC pseudo data for the unpolarized DVCS
cross section and the single transverse proton beam asymmetry together with the HERA
collider measurements, shown in Fig.~\ref{fig:X} and (\ref{meaZEUS}). Altogether we
included $2732$ data points, where the EIC pseudo data were generated as specified
above and contain 509 data points for the unpolarized cross section, obtained from
21 $\{\Q^2,\xB\}$-bins:
\begin{eqnarray}
\label{messXEIC}
\phantom{0}3.16\GeV^2 \le \Q^2 < \phantom{0}5.62\GeV^2\,, && 2.5\times 10^{-4} \le \xB \le 1.0\times 10^{-2}\;
(8\; \mbox{bins}), \nonumber\\
\phantom{0}5.62\GeV^2 \le \Q^2 < 10.00\GeV^2\,, && 4.0\times 10^{-4} \le \xB \le 1.0\times 10^{-2}\;
(7\; \mbox{bins}), \\
10.00\GeV^2 \le \Q^2 \le  17.78\GeV^2\,, && 6.3\times 10^{-4} \le \xB \le 1.0\times 10^{-2}\;
(6\; \mbox{bins}), \nonumber
\end{eqnarray}
and 2188 data points for the single transverse beam spin asymmetry as function of
$\phi$, obtained from 24 $\{\Q^2,\xB\}$-bins:
\begin{eqnarray}
\label{messAUTEIC}
\phantom{0}3.16\GeV^2 \le \Q^2 < \phantom{0}5.62\GeV^2\,, && 1.58\times 10^{-4} \le \xB \le 1.0\times 10^{-2}\;
(9\; \mbox{bins}), \nonumber\\
\phantom{0}5.62\GeV^2 \le \Q^2 <  10.00\GeV^2\,, && 2.51\times 10^{-4} \le \xB \le \; 1.0\times 10^{-2}\;
(8\; \mbox{bins}), \\
10.00\GeV^2 \le \Q^2 \le  17.78\GeV^2\,, && 3.98\times 10^{-4} \le \xB \le \; 1.0\times 10^{-2}\; (7\;
\mbox{bins}).  \nonumber
\nonumber
\end{eqnarray}
In the fit we released all 19 model parameters, which are partially correlated. In
particular, the normalization factor $\kappa^{\rm sea}$ of GPD $E$ is strongly correlated
to the skewness parameters $s_i$.
Obviously, the hypothesis of a dipole $t$-dependence  yields an unacceptably
large $\chi^2/{\rm d.o.f.}$ value, while the exponential ansatz provided, as it should,
almost the textbook value of one, $\chi^2/{\rm d.o.f.} = 0.97$. The extracted parameters
and their standard uncertainties are listed in Tab.~\ref{tab:fit}. The slope parameter of
$H^{\rm sea}$ can be well extracted with less than two standard deviations away from the
input model parameter value. The normalization of this GPD for fixed PDF
parameters is also rather robust, as indicated by small deviations of extracted
skewness parameters from the model parameters.
Since the pseudo data constrain the $t$-dependence, the correlation of
normalization parameters and $t$-slope (or dipole mass) parameters is much less pronounced
than in the fits to the HERA collider data.
For the GPD $H^{\rm G}$ the uncertainty for $\alpha^\prime$ is of the order of its model
parameter value $0.1$ and the relative uncertainty of the residual $t$-dependence is now
of the order of $7\%$ rather than $1\%$ as for sea quarks.
The relative size of skewness parameter uncertainties for gluons is on the same
$5\%$ level as for quarks.
For GPD $E^{\rm sea}$ the ``pomeron'' intercept, normalization $\kappa^{\rm sea}$ and
skewness parameters are well reproduced by the fit, where the $\kappa^{\rm sea}$
uncertainty is of the order of 20\%. The moderate size of this uncertainty also reflects
the correlation of the normalization with the skewness parameters, where the latter is now
more than twice larger than for GPD $H^{\rm sea}$. The uncertainties for the $t$-slope
parameters are only about $40\%$ larger than for GPD $H^{\rm sea}$ and are still
reasonably small. For $E^{\rm G}$ already the ``pomeron'' intercept parameter has a
very large uncertainty,
which will induce a huge normalization uncertainty. Note also the $t$-slope parameters
have big uncertainties, and they are also correlated with the remaining ones. In general
we found that with our conservative fitting strategy it is impossible to access the gluonic
component of GPD $E$ from the employed set of DVCS pseudo data. It is a standard procedure
to reduce the set of parameters to those that are not strongly correlated. This will
also reduce the size of uncertainties, however, certainly one should bear in mind that
this procedure increases the theoretical bias.

\begin{figure}[ht]
\begin{center}
\includegraphics[width=1.00\textwidth]{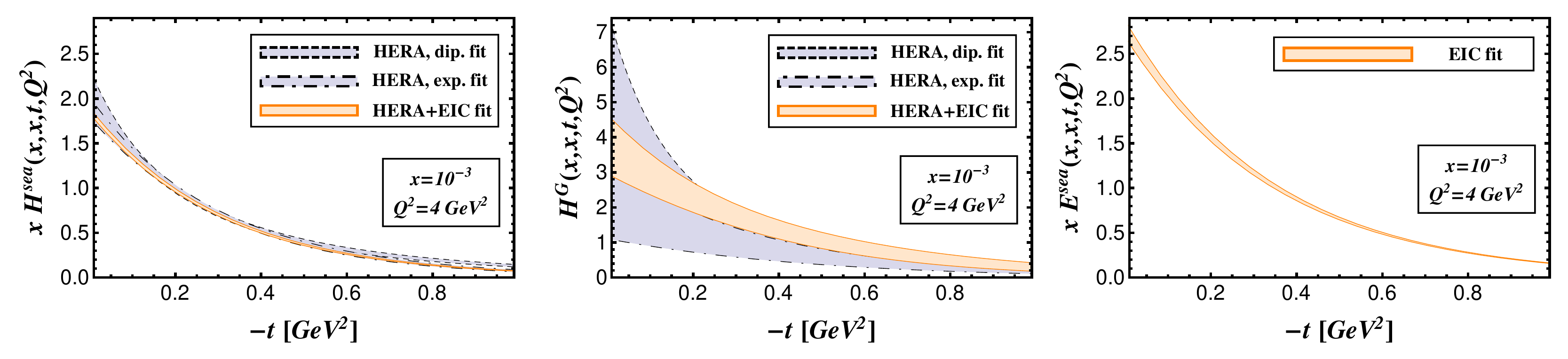}
\end{center}
\vspace{-0.5cm}
\caption{\small Least-squares fit extraction of sea quark  GPD $H^{\rm sea}$ (left) and
gluon GPD $H^G$  (middle) from a dipole ansatz (gray area surrounded by dashed curves)
and an exponential ansatz (gray area surrounded by dashed-dotted curves) using only the
HERA collider data. The results of a combined HERA/EIC fit including pseudo data for
the unpolarized DVCS cross section, c.f.~Fig.~\ref{fig:X}, and the transverse target spin
asymmetry $A^{\sin(\phi-\phi_S)}_{\rm UT}$, c.f.~Fig.~\ref{Fig_TSA}, using an exponential
ansatz are shown as light orange area (surrounded by solid curves). In addition for the
first time the sea quark GPD $E^{\rm sea}$ could be extracted (right panel).
\label{fig:Xfit1}}
\end{figure}
In Fig.~\ref{fig:Xfit1} we compare the resulting GPDs from fits to the HERA data alone
and to the combined HERA+EIC data at $\Q^2=4\,\GeV^2$, $\xB=10^{-3}$, and variable
$-t$ (covering the HERA region).
In the right panel one realizes that the uncertainty of the sea quark GPD
$H^{\rm sea}$, which is to certain extent constrained by HERA data, can be strongly
improved in particular at smaller $-t$ values. The gluon GPD $H^G$, displayed in the
middle panel, is extracted by means of the $\Q^2$ evolution and it is rather weakly
constrained by HERA DVCS data only. Here the inclusion of stage II EIC data yields a large
improvement, even if the used lever arm in $\Q^2$, compared to HERA kinematics, is still rather
limited. As emphasized above, information on the GPD $E$ can only be obtained from
a new lepton-proton scattering experiment with a transversely polarized
proton beam. In the right panel it is clearly demonstrated that the sea quark component
of this GPD can be extracted with relatively small uncertainties. As explained above,
from the utilized pseudo DVCS data the gluonic component of GPD $E$ (not shown) cannot
be reliably accessed using our flexible GPD models.

\subsection{Transverse spatial imaging}
\label{sec:imaging}

One of the main goals of GPD phenomenology is to provide the transverse spatial
distributions of partons as function of the momentum fraction $x$.
The simplest proposal to obtain a rough idea of such parton distributions
is based on the Fourier transform of the amplitude \cite{Ralston:2001xs}. Following the common
experimental procedure one would extract the $t$-dependence from a fit to a given
$(\xB,\Q^2)$ bin.
Utilizing the HERA data for DVCS and exclusive $J/\psi$ production and saying that
the former process is quark dominated while the latter is gluon dominated, one
immediately concludes from the experimental findings that the exponential $t$-slope
parameter for DVCS cross section is larger than for the $J/\psi$ cross section,
$$
B_{\rm DVCS} \approx 6 \GeV^{-2} > B_{J/\psi} \approx 4 \GeV^{-2}\,,
$$
meaning that sea quarks are more spread out in transverse space than gluons. However, we may
note that
this rather generic interpretation is based on the assumption that the proton helicity
non-conserved CFFs and/or amplitudes play no important role and that skewness effects
are unimportant. We also
emphasize that in a partonic interpretation the accessible lever arm in $-t$ is
restricted by the DVCS requirement $-t \ll \Q^2$, which ensure that possible higher
twist contributions, twist-four and higher, are small.

To quantify possible differences between a GPD interpretation and the aforementioned
procedure, our GPD fit result from the preceding section is compared with (half of) the
exponential $t$-slope of the differential DVCS cross section.
The latter is extracted by fits to the $t$-dependence in a given $\{\xB,\Q^2\}$-bin of
the pseudo data (\ref{messXEIC}) by means of the exponential model
\begin{eqnarray}
\label{dsigma/dt-b}
\frac{d\sigma_i^{\mbox{\tiny DVCS}}(t)}{dt} =  n_i\; \exp\left\{2 b_i\, t\right\}\,,
\end{eqnarray}
where $n_i$ and $b_i$ are the two fitting parameters used in bin $i$.
The $\chi^2/{\rm d.o.f}$ value in these fits is usually around one, where the propagated
standard error can be rather large for the lowest $\xB$-bins, due to the BH
subtraction procedure.  From our GPD fit, we employ both
the (sea) quark GPD $H$ on the cross-over line and the square root of the
predicted differential DVCS cross section, the latter containing also additional contribution due to
the non-vanishing CFF ${\cal E}$.
>From both of these quantities we calculate an {\it effective} exponential $t$-slopes
in the same manner.  E.g., for the GPD on the cross-over line such a $t$-slope reads
\begin{eqnarray}
\label{eq:b^eff}
b^{\rm eff}(\xB,\Q^2)=\frac{1}{t_2-t_1} \ln\frac{H(x,x,t_2,\Q^2)}{H(x,x,t_1,\Q^2)} \quad
\end{eqnarray}
with $t_1=-0.03\,\GeV^2\,,\; \;  t_2=-1.5\,\GeV^2\,,$ and $x=\xB/(2-\xB)$.

\begin{figure}[ht]
\begin{center}
\includegraphics[width=0.95\textwidth]{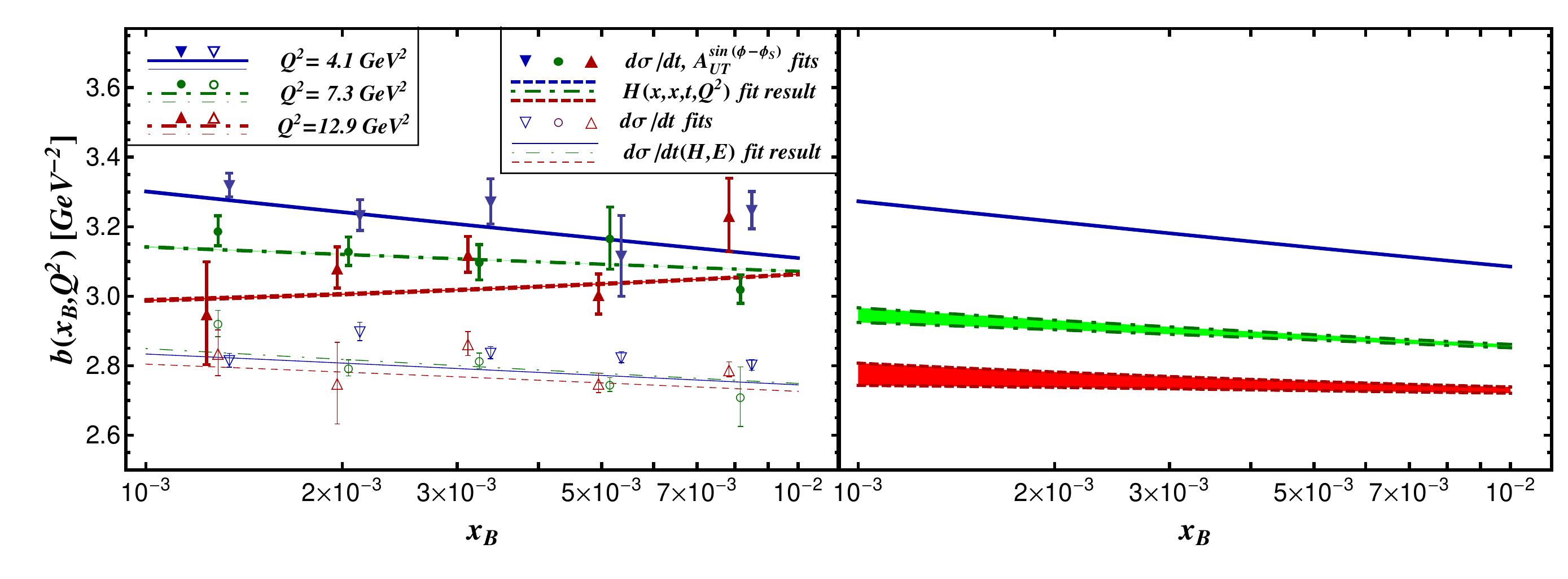}
\end{center}
\vspace{-0.7cm}
\caption{\small
Exponential $t$-slope parameters from the  model fit (\ref{dsigma/dt-b}) to the DVCS
cross section (empty symbols) and
of $\im {\cal H}$  from the model fit (\ref{|cffH|-b}) to cross section and asymmetry data
(filled symbols) as well as effective $t$-slope parameters (\ref{eq:b^eff}), obtained from our GPD fit,
for the GPD $H$ on the cross-over line (thick lines), the DVCS cross section prediction (thin lines),
and the zero-skewness GPD $H$ (error bands in the right panel) are displayed as function of $\xB$
for three different $\Q^2$ values: $4.1 \GeV^2$ (triangles-down, solid curves, blue band),
$7.3 \GeV^2$ (circles, dash-dotted curves, green band), and $12.9 \GeV^2$
(triangles-up, dashed curves, red band). Uncertainty bars arise from local fits
to EIC pseudo data, while curves and uncertainty bands originate from a combined GPD
model fit to HERA collider and EIC pseudo data. }
\label{fig:Xfit-0}
\end{figure}
In the left panel of Fig.~\ref{fig:Xfit-0} we show the results of the local
exponential model (\ref{dsigma/dt-b}) fits to the
DVCS cross sections versus $\xB$ for $\Q^2 = 4.1 \GeV^2 $, $\Q^2 = 7.3 \GeV^2 $, and
$\Q^2 = 12.9 \GeV^2 $ as empty triangles-down, circles, and triangles-up, respectively.
The effective $t$-slopes (\ref{eq:b^eff}) of the GPD on the cross-over line,
extracted from the combined GPD fit, are shown as thick solid, dash-dotted, and
dashed curves, respectively, while effective $t$-slopes of GPD-fit-predicted
cross sections are plotted as thin lines. Note that thick curves contain the uncertainty bands,
arising from the propagation of pseudo data uncertainties. As one realizes, in our large
$E$ scenario the $t$-slope of the DVCS cross section (\ref{dsigma/dt-b}) is relatively flat
w.r.t.~both $\xB$- and $\Q^2$-dependence, see empty symbols. This behavior differs drastically
from that of the extracted GPD $H$ (thick curves), which has in addition also larger slope values.
Nevertheless, the $b$-slope values extracted via exponential fit (\ref{dsigma/dt-b}) are consistent
with the effective DVCS cross section $t$-slope (thin curves),
obtained from our GPD fit, and evaluated analogously to (\ref{eq:b^eff}).
In fact, the DVCS cross section
(\ref{KLMSPM-Def-CroSec}) is in our model given as linear combination of two
exponentials with different slope parameters. The behavior of GPD $H$ (left panel: thick
curves) is partially compensated by the appearance of CFF ${\cal E}$.
Note that for a small/vanishing GPD $E$ scenario these differences in the $b$-slope
values would die out and the differences in the $t$-dependence of the CFF modulus
$|{\cal H}|$ and $\im {\cal H}$ can be considered to be small in the studied kinematical
region. Consequently, under these circumstances and restricting the fits to LO accuracy,
one can extract the $t$-dependence of GPD $H(x,x,t,\Q^2)$ directly from the DVCS cross section
measurements.

To go beyond the ${\cal H}$ dominance hypothesis in such cross section fits, one can utilize
measurements of the single transverse proton spin asymmetry (\ref{A^sin_UT-1}).
We recall that this asymmetry is sensitive to CFF  ${\cal E}$ and so the appropriate
strategy is to use cross section and asymmetry data simultaneously in an analysis. To
perform  local fits  for fixed $\xB$ and $\Q^2$ one may set the real part of CFFs to zero.
Alternatively,  one can utilize a Regge-inspired ansatz, e.g., rather analogous as in  (\ref{cffHcffE-smallx}),
and perform fits for given $\Q^2$, see, e.g., Ref.~\cite{FazFioJenLav11}. We performed such
$2\times4$ parameter ($n$, $\alpha(0)$, $\alpha^\prime$, $b$ for CFFs $\cal H$ and
$\cal E$) fits in the three $\Q^2$ bins of the pseudo data sets (\ref{messXEIC}) and
(\ref{messAUTEIC}). To compare the propagation of uncertainties with our local DVCS
cross section fits, we used the extracted values for the two Regge trajectories
$\alpha+\alpha^\prime t$ in local $2\times 2$ parameter fits to the EIC pseudo data.
The $b$-slope parameters of the imaginary parts are defined as
\begin{eqnarray}
\label{|cffH|-b}
\im {\cal H}_i(t) =  n_i\; \exp\left\{b_i\, t\right\} \quad \mbox{and}\quad \im {\cal E}_i(t) =
\overline{n}_i\; \exp\left\{ \overline{b}_i\, t\right\}\,,
\end{eqnarray}
and the slightly $t$-dependent phase for a given $\{\xB,\Q^2\}$-bin $i$ was considered to be
known.
Furthermore, we restricted the set of asymmetry data (\ref{messAUTEIC}) to those of the DVCS
cross section (\ref{messXEIC}).
The results for the exponential $t$-slope parameter of $\im {\cal H}$
are presented on the left panel in
Fig.~\ref{fig:Xfit-0} as filled symbols for $\Q^2 = 4.1 \GeV^2 $ (triangle-down),
$\Q^2 = 7.3 \GeV^2 $ (circle), and  $\Q^2 = 12.9 \GeV^2 $ (triangle-up) and, as expected,
they are compatible with the effective $t$-slope of the  GPD $H(x,x,t,\Q^2)$,
extracted from our GPD fit (thick curves). However, the propagated uncertainties in these
local fits are larger than in the previous ones, reflecting the fact that, particularly at
larger $\xB$, the asymmetry uncertainty can get large, see Fig.~\ref{Fig_TSA}.
Surely, assuming that both the assumed uncertainty distribution (Gaussian) and the model
is correct, the uncertainty propagation in global fits, e.g., with a Regge-inspired
ansatz, provides a much smaller error. It should be noted that both of these
assumptions are only true with a certain probability.

Next we consider the effective $t$-slope, analogously defined as in (\ref{eq:b^eff}), of the
unpolarized quark  GPD without skewness dependence,
\begin{eqnarray}
\label{eq:H2q}
q(x,t, \Q^2) =  H(x,\eta=0,t,\Q^2),
\end{eqnarray}
which is of great interest with regard to the transverse distribution of quarks. The result of
our model is presented on the right panel of Fig.~\ref{fig:Xfit-0} and it can be
compared to slopes of the GPD $H(x,\eta=x,t,\Q^2)$ on the cross-over line in the left panel (thick lines).
First it is observed that with growing $\Q^2$  not only the $x$-slope%
\footnote{Since $x \approx \xB/2$ is valid at small $\xB$,
twice of the $\xB$-slope that can be read off from Fig.~\ref{fig:Xfit-0} can be equivalently
considered as $x$-slope.}
of the effective $t$-slope  decreases as it is also seen for GPD $H(x,x,t,\Q^2)$, but
also its intercept drops. Loosely spoken, such a behavior means that both the ``pomeron'' slope parameter
$\alpha^\prime$ and the value of the residual $t$-slope parameter decrease.
This behavior of the zero-skewness GPD looks more natural and it is naively expected from the double log
asymptotic behavior also for the GPD on the cross-over line \cite{Mue06}.
However, with our \our model we illustrate by the thick curves  in the left panel of Fig.~\ref{fig:Xfit-0}
that specific choices of skewness parameters
combined with $t$-slope parameters can provide also
a rather flat effective residue dependence and a stronger decrease of the ``pomeron''
slope parameter $\alpha^\prime$.  Since the $\eta\to 0$ limit
commutes with the $\Q^2$-evolution, we conclude that evolution entangles the $t$-dependence with
skewness dependence. Hence, the factorization of the $t$ and skewness dependence, assumed
in our model at the initial scale, does not hold true under evolution (otherwise the
effective slope parameters should evolve similarly). The reader might be also surprised
that the uncertainty bands of the effective slope for the GPD on the cross over line remain
tiny (thick curves on the left panel), while those in the forward (zero-skewness) case get sizable
with increasing $\Q^2$ and decreasing $x$. This is caused by a naive truncation
of the covariance matrix, i.e., removing rows and columns belonging to the skewness
parameters $s_2$ and $s_4$ in the forward limit, which also alters the (anti)correlation
of uncertainties that ensure the smallness of the uncertainties for GPD
$H(x,x,t,\Q^2)$.
It is beyond the scope of this paper to study the uncertainties that arise from the
extrapolation to $\eta\to 0$ in more depth,  however, once high precision data will
become available, one should also worry about the model bias in the extrapolation of
GPD $F(x,\eta,t,\Q^2)$  from $\eta=x$ to $\eta=0$.

Finally, we would like to illustrate that the EIC measurements of the $t$-dependence in the
region $-t \lesssim 1.5\GeV^2$ can provide a probabilistic interpretation
of the transverse distribution of sea quarks and partially also for gluons. The Fourier
transform of the zero-skewness GPD (\ref{eq:H2q}) into the impact parameter space,
\begin{eqnarray}
\label{q(x,b,mu^2)}
q(x,\vec{b},\Q^2)  & = & \int\!\!\!\!\int_{-\infty}^\infty\!
\frac{d^2\vec{\Delta}}{4\pi^2}\, e^{-i \vec{\Delta}\cdot\vec{b}} \, H(x,\eta=0,t=-\vec{\Delta}^2,\Q^2) \\
&  = & \frac{1}{4 \pi}\! \int_0^\infty\!\!\!d|t|\,
J_0\!\left(b  \sqrt{|t|}\right)\, H(x,\eta=0,t,\Q^2)\,, \nonumber
\end{eqnarray}
provides in the infinite momentum frame the probability of scattering on a quark as a function
of its momentum fraction and transverse distance $b=|\vec{b}|$ from the proton center,
where $1/\Q$ is considered as the resolution scale \cite{Burkardt:2000za}.
Since for an unpolarized struck quark and proton, no direction in the transverse plane is
preferred, the integration over
the polar angle in (\ref{q(x,b,mu^2)}) yields a Bessel transform ($J_k(x)$ denotes the
Bessel function of order $k$) and results in a parton density is symmetric under rotation
of the two-dimensional impact parameter vector $\vec{b}$. For a transversely polarized
proton, e.g., the polarization vector is pointing in the $x$ direction, one finds that
the parton density is given by the unpolarized one (\ref{q(x,b,mu^2)}) and a distortion
in $y$ direction that is governed by the strength of GPD $E$,
\begin{eqnarray}
q^\Uparrow(x,\vec{b},\Q^2)  & = &
q(x,\vec{b},\Q^2)  - \frac{1}{2 M_p} \frac{\partial}{\partial b_y} E(x,\vec{b},\Q^2) \\
& =  & \frac{1}{4 \pi}\! \int_0^\infty\!\!\!d|t|\!\left[\!
J_0\!\left(b  \sqrt{|t|}\right) H +
\frac{b_y \sqrt{|t|} }{2 b M_p} J_1\!\left(b  \sqrt{|t|}\right) E
\!\right]\!\!(x,\eta=0,t,\Q^2)\,.  \nonumber
\end{eqnarray}

Before we present the resulting parton densities (\ref{q(x,b,mu^2)}) from the combined
GPD model fit to HERA and EIC pseudo data, let us shortly discuss the peculiarities
in the uncertainty estimation. The uncertainty of the resulting parton densities is, besides
the propagated experimental uncertainties, also dictated by the possible
uncertainties caused by extrapolations from the accessible kinematical region,
namely, (i) extrapolation of the skewness parameter dependence
$\eta=x$ to $\eta=0$, discussed above, (ii) extrapolation
of $t$-dependence from the experimental minimal $-t$ value $-t_1$ to $-t=0$,
as well as (iii) from maximal accessible value $-t_2$ to $-t=\infty$.
These rather intricate extrapolations are fortunately
governed by the boundary condition,
\begin{eqnarray}
\label{q2H-boundary}
q(x,\Q^2) =  H(x,\eta=0,t=0,\Q^2) = \int\!\!\!\!\int_{-\infty}^\infty\!  d^2\vec{b}\; q(x,\vec{b},\Q^2),
\end{eqnarray}
arising from the reduction of GPD $H$ in the kinematical forward limit to the standard
unpolarized PDF $q$. Hence, the normalization of the (integrated) parton density
(\ref{q(x,b,mu^2)}) is also entirely determined by the PDF normalization. To simplify
our study, we restrict ourselves to $\Q^2 = 4\GeV^2 $, where in our model
the $t$- and skewness dependencies factorize, as discussed above and exemplified also by
the agreement of the effective slope parameters in the $\eta=x$ and $\eta=0$ case, see thick
solid curves on the left and right panels on Fig.~\ref{fig:Xfit-0}.

A model analysis studying the challenges of extrapolation in $-t$ beyond the experimentally
accessible range has been presented for the
differential cross section in \cite{AscDieFazINT2011} and we essentially agree with the
conclusion that with an EIC imaging is feasible for $0.1 \fm \lesssim b \lesssim 1.5 \fm$
(or even in a wider range). Let us add some mathematical insight and let us point
out methods to increase the quality of the extrapolations.  With our model
hypothesis the $t$-dependence of the zero-skewness GPD is essentially constrained by the
EIC pseudo data in the region $0.03 \GeV^2 \le - t \le 1.5 \GeV^2$.

The uncertainty of the extrapolation into the region $[0,-t_1)$
is associated with the contribution
\begin{eqnarray}
\label{Delta_1 q}
\Delta_1 q(x,\vec{b},\Q^2) = \frac{1}{4 \pi}\! \int_0^{|t_1|}\!\!\!d|t|\, J_0\!\left(b \sqrt{|t|}\right)\,
H(x,\eta=0,t,\Q^2)\,,
\end{eqnarray}
from which one can easily obtain estimates. Although $q(x,\Q^2)=H(x,\eta=0,t,\Q^2)$ at
$t=0$ is very well known, which makes this an interpolation problem rather then
an extrapolation one, let us here calculate $\Delta_1 q$ from the knowledge of $H$ in
the vicinity of $-t_1$ by making use of a truncated Taylor series where the uncertainty is
equated with the remainder.
Consequently, to first order accuracy we have
\begin{eqnarray}
\Delta_1 q(x,\vec{b},\Q^2) &\approx & \frac{\sqrt{|t_1|}}{2 \pi b} J_1\!\left(b|  \sqrt{|t_1|}\right)\,
H(x,0,t_1,\Q^2) + \frac{t_1}{\pi b^2}\frac{\sqrt{|t_1|}}{2 \pi b} J_2\!\left(b|  \sqrt{|t_1|}\right)\,
\frac{d}{d t_1}H(x,0,t_1,\Q^2) \,, \nonumber
\\
\delta_1 q(x,\vec{b},\Q^2) &\approx&
\frac{2 |t_1|^{3/2} }{ \pi b^3}J_3\!\left(b \sqrt{|t_1|}\right)\, \frac{d^2}{d t^2_1}H(x,0,t_1,\Q^2) \,,
\end{eqnarray}
where the derivative of $H(x,0,t_1,\Q^2)$ can be evaluated numerically.
For small $-t_1$ we can roughly estimate the value of the second order derivative in
terms of the transverse width
\begin{eqnarray}
\langle b^2\rangle(x,\Q^2)= 4 \frac{d}{dt} \ln H(x,\eta=0,t,\Q^2)\Big|_{t=0}\,,
\end{eqnarray}
e.g., for a $p$-pole ansatz we find
$$
\frac{d^2}{d t^2_1} H(x,0,t_1,\Q^2) \approx \frac{1+p}{16 p} \langle b^2\rangle^2(x,\Q^2) q(x,\Q^2)\,,
$$
where the result for an exponentially functional form in $t$ follows from the limit
$p\to \infty$.
For a realistic value of $\langle \vec{b}^2\rangle \sim 0.35\, {\rm fm}^2$ in the small
$x$ region, we find that the extrapolation uncertainty is of the order of $10^{-4}$ in units
of $q(x,\Q^2)/{\rm fm}^2$. However, this uncertainty becomes important in the large
$b \gtrsim 1/\sqrt{|t_1|} \approx 1 \fm$ region,  dominated by the contributions from
the small $-t$ region. Model analyzes provide a relative uncertainty on permill level for
$b\approx 1 \fm$, which, however, will increase to the few percent level for
$b\approx 1.5 \fm$ and will then grow fast for increasing $b$.  Hence, with our
EIC pseudo data we can resolve the transverse distribution up to a distance of
$\approx 1.5\fm$.
To estimate the uncertainty of the extrapolation into the $\{-t_2, \infty]$ region we
naively use
\begin{eqnarray}
\label{Delta_2 q}
\Delta_2 q(x,\vec{b},\Q^2) = \frac{1}{4 \pi}\! \int^\infty_{|t_2|}\!\!\!d|t|\,
J_0 (b  \sqrt{|t|})\, H(x,\eta=0,t,\Q^2)\,.
\end{eqnarray}
It is most important for the small-$b$ region, where in particular for $b=0$ we have
\begin{equation}
\label{Delta_2 q-b=0}
\Delta_2 q(x,\vec{b}=0,\Q^2) = \frac{1}{4 \pi}\! \int^\infty_{|t_2|}\!\!\!d|t|\,  H(x,\eta=0,t,\Q^2)\,.
\end{equation}
The relative uncertainty at $b=0$ is easily evaluated and we find for an exponential or
$p$-pole form
$$
\frac{\Delta_2 q(x,\vec{b}=0,\Q^2)}{q(x,\vec{b}=0,\Q^2)} = e^{\langle b^2 \rangle t_2/4}\quad \mbox{and}\quad
\frac{\Delta_2 q(x,\vec{b}=0,\Q^2)}{q(x,\vec{b}=0,\Q^2)} = \left(1- \frac{\langle b^2 \rangle t_2}{4 p}\right)^{-p+1}\,,
$$
respectively.
Assuming an exponential functional form it is with $-t_2 \approx 1.5 \GeV^2$ well
under control and results in a  $\sim 3\%$ correction, which, however, would increase for a
dipole form to $\sim 40\%$. To reach the $10\%$ accuracy level, one must increase
$-t_2 \sim 8 \GeV^2$, which requires a big $\Q^2$ value to ensure the validity of DVCS
kinematics. Fortunately, the error, e.g., for $0.1\fm \le b$, gets already on the
$10\%$ level for $-t_2 \sim 3.5 \GeV^2$.  Under these circumstances, one may rely on
extrapolation techniques, e.g., based on conformal mapping or Pad\'{e} approximation, to
minimize the uncertainty. Note also that the uncertainty of extrapolation into the
$\{-t_2,\infty]$ region may be also associated with a {\it relative} uncertainty that
grows fast with increasing $ b $.
In the following the uncertainty is calculated according to (\ref{Delta_2 q-b=0})
and estimate numerically
by assuming two alternative hypotheses, namely, that the $t$-dependence falls off
exponentially or with $1/t^2$, where for a given $b$ value always the larger uncertainty is
taken. For simplicity we will neglect the uncertainty from the extrapolation
(interpolation) into the region $\{-t_1, 0]$, which is entirely justified for
$b\le 1 \fm$ and as it would be hardly visible in the visualization of the parton densities
for $b\le 1.5 \fm$.
Finally, the uncertainty from the extrapolation into the large $-t$ region was added in
quadrature to the one propagated from the (pseudo) data.

\begin{figure}[ht]
\begin{center}
\includegraphics[width=1.00\textwidth]{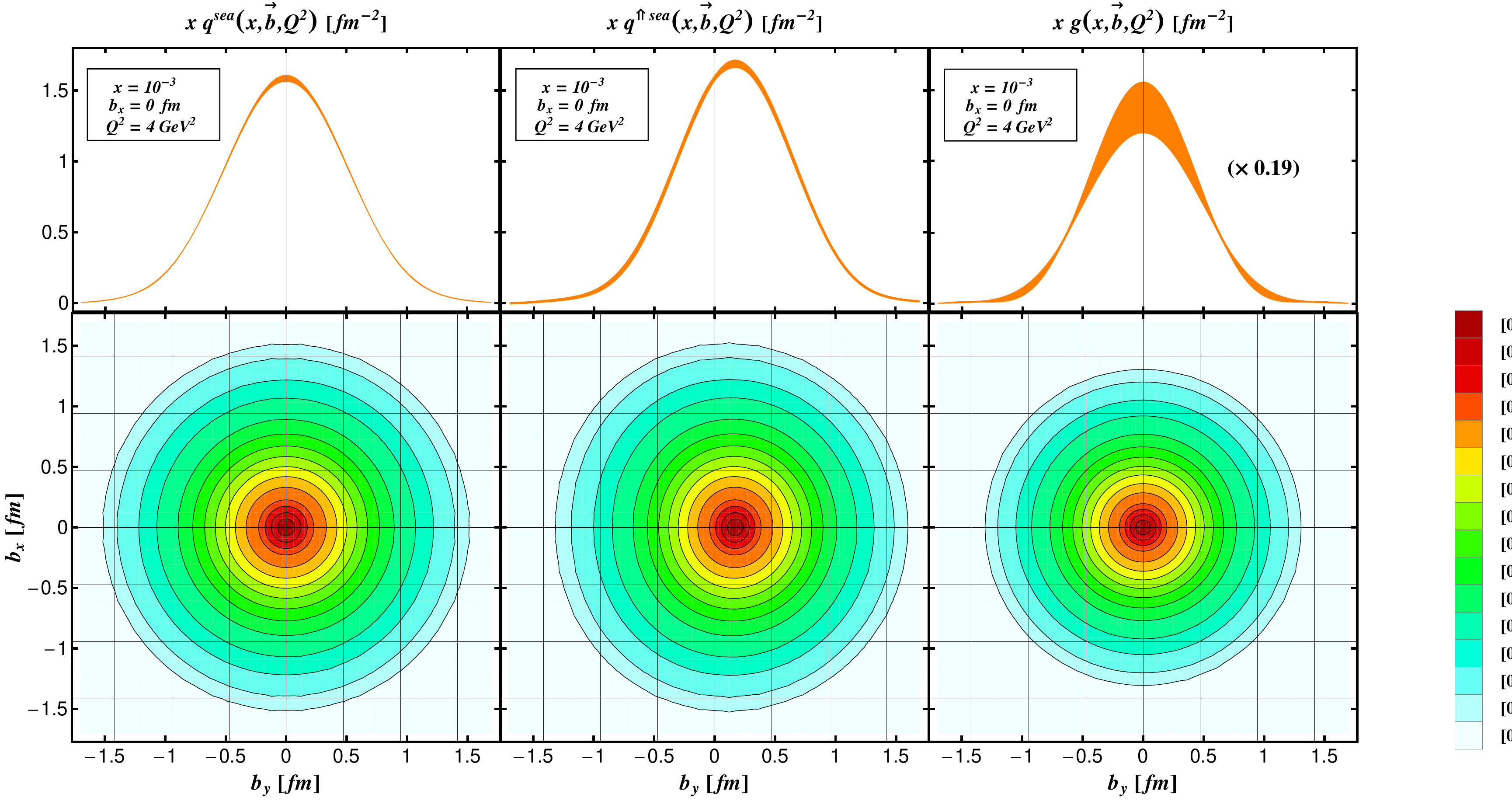}
\end{center}
\vspace{-0.5cm}
\caption{\small Parton densities at $x=0.001$ and $\Q^2=4 \GeV^2$ versus impact parameter
$b$ were obtained from a combined least-squares fit to the HERA collider and EIC pseudo
data: relative densities (lower row) and their values at $b_x=0$ for the unpolarized sea
quark parton densities of a unpolarized proton (left),
a transversely polarized proton (middle), and the unpolarized gluon parton density of
a unpolarized proton (right), its value is rescaled by a factor 0.19.
\label{fig:Xfit2} }
\end{figure}
In the left and right columns on Fig.~\ref{fig:Xfit2} the sea quark and
gluon parton densities (\ref{q(x,b,mu^2)}) at $x=10^{-3}$ and $\Q^2=4 \GeV^2$ are
shown as a relative density plot versus $b_y$ and $b_x$ (lower panels) and for $b_x=0$
as function of $b_y$ (upper panels). Note, the gluon density is rescaled
by a factor of $0.19$.
Since for sea quarks the propagated uncertainty is small and it was assumed that there is
no cross-talk between $t$- and  skewness dependencies, and the PDF uncertainties [fixed PDF
parameters, see  also boundary condition (\ref{q2H-boundary})] are neglected, the final
uncertainties are smaller than for the corresponding $t$-dependent GPD at the cross-over
line shown in the left panel of Fig.~\ref{fig:Xfit1}. Nevertheless, the increase of the
uncertainty
in the vicinity of $b=0$ due to the $\{-t_2,\infty ]$ extrapolation is  visible for
the quark density (upper left  panel) and much  more pronounced for the gluon density
(upper right panel).  Generally, the larger error for the gluon density is mainly based
on the fact that the DVCS process alone does not allow to pin down this quantity
on the same quantitative level as for sea quarks, see earlier discussions and
Fig.~\ref{fig:Xfit1}. We again emphasize that the functional form of the
$t$-dependence will influence the uncertainties related to the extrapolation error.
For instance, a power-like falloff will increase the DVCS amplitude in the
accessible large $-t$ region and therefore decrease the experimental uncertainties
in this region, however, on the other hand, the uncertainties of the extrapolation into
the $\{-t_2 ,\infty]$ region will become more important.

Apart from the uncertainties that appear in the unpolarized parton densities, we
also have the normalization uncertainty of GPD $E$, which is not protected
by a boundary condition. Nevertheless, we found in our model that this
uncertainty is not large. This is illustrated in the middle column of
Fig.~\ref{fig:Xfit2}, where we display the sea quark density for a transversely polarized
proton.

\subsection{Angular momentum sum rule}
\label{sec:sum-rule}

Finally, we shortly discuss the role of the EIC measurements in elucidation of the Ji spin sum rule
\cite{Ji:1996ek}. This rule states that the proton spin
\begin{eqnarray}
\frac{1}{2} = \sum_{q=u,d,s,\cdots} J^q(\Q^2) + J^G(\Q^2)
\end{eqnarray}
is built from quark and gluon angular momenta $J^q$ and $J^G$, which are defined via a
gauge invariant decomposition of the QCD energy momentum tensor.
Note that several other decompositions have been proposed, which are related to Ji`s
ones by reshuffling a certain amount of angular momentum fraction $\Delta J$, i.e.,
\begin{eqnarray}
 \sum_q J^q \Rightarrow  \sum_q J^q + \Delta J \quad\mbox{and}\quad J^G \Rightarrow J^G - \Delta J\,,
\end{eqnarray}
where one may take the freedom to define $\Delta J$ as the expectation value of a
gauge variant operator in order to reach a partonic interpretation of the gluon component
in terms of spin and orbital angular momentum, e.g., to arrive at the Jaffe-Manohar spin sum rule
\cite{Jaffe:1989jz}.
Ji`s decomposition implies that the partonic components of the proton spin are given by
the momentum fraction part, called here $A$, and the anomalous gravitomagnetic moment $B$,
\begin{equation}
J^i(\Q^2) = \frac{1}{2} A^i(\Q^2) + \frac{1}{2} B^i(\Q^2),\quad
\left\{ { A \atop B}\right\}^i(\Q^2) = \int_0^1\!dx\,x \left\{ { H \atop E}\right\}^i(x,\eta=0,t=0,\Q^2)\,,
\end{equation}
which are given by the first moments of GPDs $H$ and $E$, respectively. A phenomenological
quantification of this sum rule is a highly intricate task, which is often trivialized
by entirely relying on simple-minded GPD (or even transverse momentum dependent PDF) models or assumptions.

The definitions in Ji`s angular momentum sum rule allow to employ any other QCD, i.e.,
field theory based framework to quantify the quark and gluon angular momenta. Most
promising for achieving this goal are lattice gauge field simulations and once reliable
results can be obtained, for a review see \cite{Hagler:2009ni},
one may incorporate them in GPD models. There are various
systematic uncertainties in the lattice estimation of angular momentum carried by sea quarks
and gluons and hence their phenomenological determination is an important task
for the future.

As we have seen, DVCS measurements at an EIC will allow to access the
GPD $E$ at the cross-over line and allows, in a model dependent manner, to extract also its
normalization in the forward kinematics. In fact, what we called anomalous magnetic
moment of sea quarks is simply related to their angular momentum:
$$
J^{\rm sea} = \frac{1}{2}\left(1+\kappa^{\rm sea}\right) A^{\rm sea}\,,
$$
where the phenomenological value of the momentum fraction is at $\Q^2=4\GeV^2$ given by
$$A^{\rm sea}(\Q^2=4\GeV^2) \approx 0.15.$$

\section{Summary}
\label{sec:summary}

We show on some selected examples the physics case for DVCS measurements at a
proposed EIC. Pseudo data were generated by the MC program
MILOU that is tuned to HERA collider DVCS measurements. Full detector simulations have
not yet been included; it was rather assumed that the systematical uncertainty
for cross section measurements is on the 5\% level. The statistical uncertainties of
these simulations have been included in model predictions for various single spin
asymmetries, electron charge
asymmetries, and unpolarized cross sections, covering the EIC kinematics at stage I
and II.

We illustrated that present GPD models, constrained by global fits to present
DVCS data, provide a variety of EIC predictions, where in particular the
$t$-dependence of the different models
is poorly known and can be constrained to a large degree by EIC measurements.
We did not discuss in completeness the extraction of CFFs, which can be
done by having a polarized positron beam at hand. Here already a unpolarized one
would help to have a cleaner access to twist-two associated CFFs. In particular, it
can be used to isolate the interference term, which contains the most valuable
information.
This also provides an experimental cross-check for the smallness of the
$\phi$-integrated interference term in cross section
measurements.
Rosenbluth separation, as it was worked out in Sect.~\ref{sec:Rosenbluth},
provides another handle
on the isolation of BH and DVCS cross sections. This technique has to be
explored further for the access of higher harmonics in the interference term. At present
it is not known to what extent this method can be employed
in a model independent manner, however, certainly it looks more intricate than
in the case of unpolarized DIS or elastic form factor measurements.

While the access to CFFs and GPDs at lower beam energies requires the measurement of
many observables the situation becomes simpler at higher energies. Here, we can
assume that only two twist-two associated CFFs $\cal H$ and $\cal E$ show a
``pomeron'' behavior and are as such accessible in these kinematics.
Moreover, their real parts are small compared to the imaginary and their phases
are approximately given by an effective ``pomeron'' trajectory.
Therefore, they can be accessed by a measurement of the DVCS cross section and the
single transverse proton spin asymmetry. Thereby, at large electron energy loss
$y$ and $-t$ the DVCS cross section may drop drastically and perhaps cannot be measured.
In such circumstances one can use the single electron beam spin asymmetry measurements,
which are predicted to be sizable at large $y$.

For parton imaging, it was illustrated that in a large GPD $E$ scenario its
$t$-dependence extracted from the unpolarized cross section does not necessarily match
the $t$-dependence of the sea quark GPD $H^{\rm sea}$.
To extract in such a situation the
$t$-dependence of CFFs $\cal H$ and $\cal E$ one may neglect the real part or use
global fits, e.g., with a Regge-inspired ansatz in a given $\Q^2$ bin.
Taking EIC pseudo data generated from MILOU and propagated to the predictions of \our
model, we studied the error propagation to the sea quark and gluonic components of
GPDs $H$ and $E$ by means of least-squares fits. Thereby, it turned out that with our
rather flexible model the sea quark component of both GPDs can be pinned down quite
precisely, while the knowledge of the gluon GPD $H^{\rm G}$ can also be substantially improved.
However, the gluon GPD
$E^{\rm G}$ is without further assumptions not accessible. To obtain a probabilistic
interpretation, a model dependent extrapolation to the zero-skewness GPD has to be
performed. We adopt in our studies the popular GPD model hypothesis that the skewness effect is
$t$-independent, however, we also pointed out that this hypotheses may not hold
under evolution. Certainly, this extrapolation
may be considered as the largest
theoretical uncertainty. Concerning the extrapolation uncertainties in the unmeasured $-t$
region, we found
that the extrapolation into the small $-t$ region is well under control if the
transverse resolution is of the order of $200\, {\rm MeV}$. This allows to resolve the
transverse distribution of partons  up to $1.5 \fm$.  In the case that the CFFs decrease
strongly with increasing $-t$, e.g., exponentially, the experimentally accessible range,
which overlaps with DVCS kinematics, is sufficient to provide an image of the sea quark
GPDs $H$ and $E$ and also for the gluon GPD $H^{\rm G}$. If this will not be the case,
the imaging procedure may not be under control if one simply neglects the
non-accessible large $-t$ region. Under these circumstances one may increase
experimentally the $-t$ range together with $\Q^2$ or employ mathematical
extrapolation methods, which we did not explore here.

Let us also emphasize that the revealing of the CFF $\cal E$ in the small-$\xB$
region is of more general interest, since it is loosely related to the problem
whether  the ``pomeron''  coupling can flip the spin of the proton. In the partonic
language it is related to the question of whether sea quarks and gluons carry a non-vanishing
gravitomagnetic moment or their angular momentum is simply given by half of their
momentum fraction.  Certainly the phenomenological access to this problem suffers from
the uncertainties of extrapolation to the forward kinematics; however, the experimental
measurement of CFF $\cal E$ can shed light on these questions, which at present can be
hardly addressed with lattice gauge field simulations.

In summary, the proposed high-luminosity Electron Ion Collider, combined with its
designated detector, would be an ideal apparatus for precise measurements of
exclusive channels in both electron-proton and electron-nuclei scattering.
Besides hard exclusive vector meson and photon electroproduction,
one might address other exclusive channels, too. In particular,  utilizing
Monte-Carlo simulations and GPD fitting routines, we have shown the potential of
such experiments for GPD phenomenology and the ability to
obtain from such measurements the spatial distributions of sea quarks and gluons.

\acknowledgments

We are grateful to M.~Diehl for many useful discussions. D.M. and K.K. thank the
Nuclear Physics group at Brookhaven National Laboratory for the warm
hospitality during their stay, where this project has been staged and mostly completed.
This work was supported in part by the U.S. Department of Energy under
contract number DE-AC02-98CH10886, by Croatian Ministry of Science,
Education and Sport, contract no.
119-0982930-1016, and the Joint Research Activity
\emph{Study of Strongly Interacting Matter} (acronym HadronPhysics3,
Grant Agreement No.~283286) under the Seventh Framework Program of the European Community.

\newpage
\appendix

\section{Updates to the MILOU code}
\label{appendix:MILOU}

The MILOU code has been modified from its original version and it is currently
maintained at BNL
\footnote{All information can be found at the dedicated page:
https://wiki.bnl.gov/eic/index.php/MILOU\label{foot:MILOU2}}. The updates to the code
mainly include bug fixing together with an improved output. The most relevant
updates are the following:

\begin{itemize}

\item {\bf Bug fixed in the FORTRAN common blocks.} Now they preserve the random
seeds set in the cards and there is no need for recalculating the integral
every event generation.

\item {\bf The correct ALLM parametrization for the $F_2$ structure function has
been implemented.} This is relevant when running MILOU using the option for
the Frankfurt-Freund-Strikman (FFS)\cite{FFS} model, which computes the complex DVCS
amplitude to LO and is not based on GPDs. Formerly, a wrong implementation of
the ALLM parametrization caused a disagreement between the NLO GPD based and the
FFS models. The correct ALLM is now taken from \cite{ALLM}, and the agreement
between the two models and with the predictions from the GenDVCS \cite{GenDVCS}
Monte Carlo (also using FFS) at HERA energies is satisfactory.

\item {\bf A new output format.} Beside the original output in the form of a PAW
$n$-tuple \cite{PAW}, a new output has been implemented in the form of a
Pythia-like ascii format text file, in the same standard as other MCs used at
EIC. A detailed description of the new output can be found on the web-page in
footnote \ref{foot:MILOU2}.

\item {\bf Simulation of harmonics.} In calculating the beam charge asymmetry, a
functional form for the $cos(\phi)$ harmonic was formerly hard coded. Now the
code points to the correct values from Freund/McDermott model.

\item {\bf Simulation of the interference term.} it is now properly set to the
values expected from the Freund/McDermott model at NLO, without the twist-three
contribution.
\end{itemize}

\newpage

\end{document}